\documentclass{ws-ijmpe}  

\usepackage[square]{ws-rv-van}
\usepackage{amsmath}
\usepackage{amssymb}

\newcommand{\beq}{\begin{equation}}
\newcommand{\eeq}{\end{equation}}
\newcommand{\be}{\begin{equation}}
\newcommand{\ee}{\end{equation}}
\newcommand{\ber}{\begin{eqnarray}}
\newcommand{\eer}{\end{eqnarray}}
\newcommand{\berr}{\begin{eqnarray*}}
\newcommand{\eerr}{\end{eqnarray*}}

\newcommand{\sg}{\sigma(\omega,\vec{p})}
\newcommand{\sgh}{\sigma_H(p_0,\vec{p})}

\newcommand{\tc}{T_c}

\newcommand{\addnew}[1]{#1}
\newcommand{\old}[1]{}
\newcommand{\nn}{\nonumber}
\newcommand{\non}{\nonumber}

\def\als{\alpha_{\rm s}}

\def\siml{{\ \lower-1.2pt\vbox{\hbox{\rlap{$<$}\lower6pt\vbox{\hbox{$\sim$}}}}\ }}
\def\simg{{\ \lower-1.2pt\vbox{\hbox{\rlap{$>$}\lower6pt\vbox{\hbox{$\sim$}}}}\ }}

\begin{document}

\title{QUARKONIUM AT FINITE TEMPERATURE}

\author{Alexei Bazavov}
\address{Department of Physics, University of Arizona,
         Tucson, AZ 85721, USA}
\author{P\'eter Petreczky}
\address{RIKEN-BNL Research Center and Physics Department,\\ 
         Brookhaven National Laboratory, Upton NY 11973, USA}
\author{Alexander Velytsky\footnote{current address: Physics Department,
         Brookhaven National Laboratory, Upton NY 11973, USA}}
\address{Enrico Fermi Institute, University of Chicago,
         5640 S. Ellis Ave.,\\ Chicago, IL 60637, USA \\ and \\
         HEP Division and Physics Division, 
         Argonne National Laboratory,\\ 9700 Cass Ave., Argonne, IL 60439, USA}

\maketitle

\begin{abstract}
We discuss properties of heavy quarkonium states at high temperatures
based on lattice QCD and potential models. We review recent progress made
in lattice calculations of spatial static quark anti-quark correlators
as well as quarkonium correlators in Euclidean time. Recent developments
in effective field theory approach and potential models are also discussed.
\end{abstract}


\section{Introduction}

There was considerable interest in the properties and the fate of
heavy quarkonium states at finite temperature since the famous conjecture 
by Matsui and Satz \cite{Matsui:1986dk}. It has been argued that color screening in
medium will lead to  quarkonium dissociation above deconfinement, which in turn
can signal quark gluon plasma formation in heavy ion collisions. The basic
assumption behind the conjecture by Matsui and Satz was the fact that 
medium effects can be understood in terms of a temperature dependent 
heavy quark potential. Color screening makes the potential 
exponentially suppressed at distances larger than the Debye radius and it 
therefore cannot bind the heavy quark and anti-quark once the temperature is
sufficiently high.
Based on this idea 
potential models at finite temperature with different temperature dependent
potentials have been used over the last two decades to study quarkonium properties at finite
temperature (see Ref. \cite{Mocsy:2008eg} for a recent review). 
It was not until recently that effective field theory
approach, the so-called thermal pNRQCD, has been developed to justify 
the use of potential models at finite
temperature \cite{Brambilla:2008cx}. 
This approach, however, is 
based on the weak coupling techniques and will be discussed in the next section. 
To understand the non-perturbative aspects of color screening lattice calculations 
of the spatial correlation functions of static quarks  are needed. 
Recently a lot of progress has been made in this direction which will be the topic of 
section \ref{sec_static_corrs}. To prepare the reader for this 
in section \ref{sec_lgt_basics} we review the basics of
lattice gauge theory.

In principle it is possible to study the problem of quarkonium dissolution without any use
of potential models. In-medium properties of different quarkonium states and/or their
dissolution are encoded in spectral functions. Spectral functions are related to Euclidean 
meson correlation functions which can be calculated on the lattice.
Reconstruction of the spectral functions from the lattice meson correlators turns out to be
very difficult, and despite several attempts its outcome  
still remains inconclusive. One remarkable feature of the studies of the lattice meson
correlators is their small temperature dependence despite the expected color screening. 
This seems to be puzzling.
We will discuss the possible resolution of  
this puzzle in section \ref{sec_pot_model}, while the current status 
of the lattice calculations of the Euclidean
correlators and the corresponding meson spectral functions will be presented in section 
\ref{sec_spec_fun}. 
The summary and outlook will be given in section \ref{sec_concl}.

\section{pNRQCD at finite temperature}
\label{sec_pnrqcd}

There are different scales in the heavy quark bound state problem related to the heavy quark mass $m$,
the inverse size $\sim m v$ and the binding energy $~m v^2$. Here $v$ is the typical heavy quark velocity
in the bound state and is considered to be a small parameter.
Therefore it is possible to derive a sequence of effective field theories using this 
separation of scales (see Refs. \cite{Brambilla:2004wf,Brambilla:2004jw} for  recent reviews). 
Integrating out modes at the highest energy scale $\sim m$ leads to
an effective field theory called non-relativistic QCD or NRQCD, where the pair creation of heavy quarks is
suppressed by powers of the inverse mass and the heavy quarks are described by non-relativistic Pauli spinors \cite{Caswell:1985ui}.
At the next step, when the large scale related to the inverse size is integrated out, the potential NRQCD
or pNRQCD appears. In this effective theory the dynamical fields include the singlet 
$\rm S(r,R)$ and octet $\rm O(r,R)$ fields 
corresponding to the heavy quark anti-quark pair in singlet and octet states respectively, as well as light quarks and gluon fields
at the lowest scale $\sim mv^2$. The Lagrangian of this effective field theory has the form
\begin{eqnarray}
{\cal L} =
&&
- \frac{1}{4} F^a_{\mu \nu} F^{a\,\mu \nu}
+ \sum_{i=1}^{n_f}\bar{q}_i\,iD\!\!\!\!/\,q_i
+ \int d^3r \; {\rm Tr} \,
\Biggl\{ {\rm S}^\dagger \left[ i\partial_0 + \frac{\nabla_r^2}{m}-V_s(r) \right] {\rm S}\nonumber\\
&&
+ {\rm O}^\dagger \left[ iD_0 + \frac{\nabla_r^2}{m}- V_o(r) \right] {\rm O} \Biggr\}
+ V_A\, {\rm Tr} \left\{  {\rm O}^\dagger {\vec r} \cdot g{\vec E} \,{\rm S}
+ {\rm S}^\dagger {\vec r} \cdot g{\vec E} \,{\rm O} \right\}
\nonumber\\
&&
+ \frac{V_B}{2} {\rm Tr} \left\{  {\rm O}^\dagger {\vec r} \cdot g{\vec E} \, {\rm O}
+ {\rm O}^\dagger {\rm O} {\vec r} \cdot g{\vec E}  \right\}  + \dots\;.
\label{pNRQCD}
\end{eqnarray}
Here the dots correspond to terms which are higher order in the multipole expansion \cite{Brambilla:2004jw}.
The relative distance $r$ between the heavy quark and anti-quark plays a role of a label, the light
quark and gluon fields depend only on the center-of-mass coordinate $R$. The singlet $V_s(r)$ and octet $V_o(r)$ 
heavy quark potentials
appear as matching coefficients in the Lagrangian of the effective field theory
and therefore can be rigorously defined in QCD at any order of the perturbative expansion.
At leading order 
\be
V_s(r)=-\frac{N^2-1}{2 N} \frac{\alpha_s}{r},~V_o(r)=\frac{1}{2 N}\frac{\alpha_s}{r}
\ee
and $V_A=V_B=1$.

The free field equation for the singlet field is 
\be
\left[ i\partial_0 + \frac{\nabla_r^2}{m}-V_s(r) \right] {\rm S(r,R)}=0,
\ee
i.e. has the form of a Schr\"odinger equation with the potential $V_s(r)$. 
In this sense, potential models emerge from
the pNRQCD. Note, however, that pNRQCD also accounts for interaction of the soft gluons which cannot be included
in potential models, i.e. it can describe retardation effects.

One can generalize this approach to finite temperature. However, the presence of additional 
scales makes the analysis more complicated \cite{Brambilla:2008cx}. The effective Lagrangian will have the same form as above,
but the matching coefficients may be temperature dependent. In the weak coupling regime there are three different thermal scales :
$T$, $g T$ and $g^2 T$. The calculations of the matching coefficients depend on the relation of these thermal scales to
the heavy quark bound state scales \cite{Brambilla:2008cx}. To simplify the analysis the static approximation has been used, in which
case the scale $m v$ is replaced by the inverse distance $1/r$ between the static quark and anti-quark. The binding energy
in the static limit becomes $V_o-V_s \simeq N \alpha_s/(2 r)$. When the binding energy is larger than the temperature the derivation of pNRQCD proceeds
in the same way as at zero temperature and there is no medium modifications of the heavy quark potential \cite{Brambilla:2008cx}. 
But bound state
properties will be affected by the medium through interactions with ultra-soft gluons, in particular, the binding energy will be reduced
and a finite thermal width will appear due to medium induced singlet-octet transitions arising from the dipole interactions in
the pNRQCD Lagrangian \cite{Brambilla:2008cx} (c.f. Eq. (\ref{pNRQCD})).
When the binding energy is smaller than one of the thermal scales the singlet
and octet potential will be temperature dependent and will acquire an imaginary part \cite{Brambilla:2008cx}. 
The imaginary part of the potential arises because of the singlet-octet transitions induced by the dipole vertex as well as
due to the Landau damping in the plasma, i.e. scattering of the gluons with space-like momentum off the thermal excitations in
the plasma. 
In general, the thermal corrections
to the potential go like $(r T)^n$ and $(m_D r)^n$ \cite{Brambilla:2008cx}, where $m_D$ denotes the Debye mass. 
Only for distances $r>1/m_D$ there is an exponential screening. 
In this region the singlet potential has a simple form
\begin{eqnarray}
&
\displaystyle
V_s(r)=
 -C_F\,\frac{\als}{r}\,e^{-m_Dr}
+ iC_F\,\als\, T\,\frac{2}{rm_D}\int_0^\infty dx \,\frac{\sin(m_Dr\,x)}{(x^2+1)^2}- C_F\, \als \left( m_D + i T \right),\nn\\
&
\displaystyle
C_F=(N^2-1)/(2N)
\label{Vs}
\end{eqnarray}
The real part of the singlet potential coincides with the leading order 
result of the so-called singlet free energy \cite{Petreczky:2005bd}.
The imaginary part of the singlet potential in this limit has been first calculated in \cite{Laine:2006ns}.
For small distances the imaginary part vanishes, while at large distances it is twice the damping rate of a
heavy quark \cite{Pisarski:1993rf}. This fact was first noted in Ref. \cite{Beraudo:2007ky} for thermal QED.

The effective field theory at finite temperature
has been derived in the weak coupling regime assuming the separation of 
different thermal scales as well as $\Lambda_{QCD}$.
In practice the separation of these scales is not evident and 
one needs lattice techniques to test the approach. Therefore
section \ref{sec_static_corrs} will be dedicated to the study of 
static quarks at finite temperature on the lattice. To prepare the reader for
this discussion some basics of the lattice gauge theory will be given in the next section.

\section{Basics of lattice gauge theory}
\label{sec_lgt_basics}

To study non-perturbative aspects of QCD we use lattice gauge theory
\cite{Wilson:1974sk}. In this formalism a field theory is defined in a
gauge-invariant way on a discrete space-time domain. This serves
at least two purposes: a) to provide an ultra-violet cut-off for the 
theory, restricting highest momentum to $\pi/a$ ($a$ being the lattice
spacing), and b) to evaluate the path integrals in the Euclidean
formulation stochastically using importance sampling.

On the lattice the fundamental degrees of freedom of a theory
with local $SU(N)$ gauge symmetry are fermion fields $\psi_x$ that
reside on the sites of the lattice and carry flavor, color and Dirac
indeces, which we suppress through the most of this paper, and
gauge, bosonic degrees of freedom that in the form of $SU(N)$ matrices 
$U_{x,\mu}$ reside on links. Sites on a four-dimensional lattice 
are labeled with $x\equiv(\vec{x},t)$.

The theory is defined by the partition function
\begin{equation}\label{partfun}
  Z=\int DU D\bar\psi D\psi \exp(-S)
\end{equation}
where the action
\begin{equation}\label{Saction}
  S=S_g+S_f
\end{equation}
contains gauge, $S_g$ and fermionic, $S_f$ parts. The latter part is bi-linear
in fields and has the form
\begin{equation}\label{Sfaction}
  S_f=\bar\psi M \psi
\end{equation}
where $M$ is the fermion matrix. In the simplest formulation the lattice gauge action can
be written as 
\be
S_g=\beta \sum_P \left(1-\frac{1}{N}{\rm Tr} U_P\right),
\ee
where $U_P$ is the so-called plaquette, a product of link variables along the elementary 
square and $\beta=2 N/g^2$ with $g^2$ being the bare gauge coupling. This is the Wilson gauge
action \cite{Wilson:1974sk}. The explicit form of the fermion action that is often 
used in lattice QCD calculations will
be discussed in section \ref{sec:charm_lat}.

The expectation value of an operator $\hat O$ is given then by
\begin{equation}\label{expectO}
  \langle \hat O\rangle=\frac{1}{Z}
  \int DU D\bar\psi D\psi\hat O \exp(-S).
\end{equation}
Integration over the fermion fields (which are Grasmann variables)
can be carried out explicitly:
\begin{equation}\label{partfun_eff}
  Z=\int DU \det M[U]\exp(-S_g)\equiv
  \int DU \exp(-S_{eff}),
\end{equation}
where $S_{eff}=S_g-\ln\det M[U]$ is the effective action.
The fermion determinant $\det M[U]$ describes the vacuum polarization effects due to the dynamical
quarks and makes the effective action non-local in gauge variables.
For this reason simulations with dynamical quarks are very resource demanding
and the quenched approximation is often employed, where $\det M[U]$ is set to 1.

To evaluate the path integral (\ref{expectO}) stochastically,
an ensemble of $N_U$ gauge configurations, weighted with $\exp(-S_{eff})$,
is generated using Monte Carlo or Molecular Dynamics techniques.
The expectation value of the operator is then approximated by the
ensemble average:
\begin{equation}\label{expectO_appr}
  \langle\hat O\rangle \simeq \frac{1}{N_U} \sum_{i=1}^{N_U}O_i(U),
\end{equation}
where $O_i(U)$ is the value of the operator $\hat O$, 
calculated on $i$-th configuration. (When the operator
$\hat O$ depends explicitly on the quark fields extra factors
of $M^{-1}$ appear as shown for a meson correlator below.)

Consider a meson (quark anti-quark pair) operator of a general form
\begin{equation}\label{Jgen}
  J(\vec{x},\vec{y};t)=
  \bar\psi(\vec{x},t)\Gamma{\cal U}(\vec{x},\vec{y};t)\psi(\vec{y},t),
\end{equation}
where $\Gamma$ determines the spin structure and ${\cal U}$ is a gauge connection
that corresponds to the excitations of the gluonic field. Dirac and color 
indeces in (\ref{Jgen}) are suppressed.

The propagation of such meson from time $t=0$ to $t$ is described 
by the correlation function 
\begin{multline}\label{Jprop}
  \langle J(\vec{x}_1,\vec{y}_1;0) J(\vec{x}_2,\vec{y}_2;t)\rangle =
  \frac{1}{Z}\int DU D\bar\psi D\psi\exp(-S) \\
  \times\bar\psi(\vec{x}_1,0)\Gamma{\cal U}(\vec{x}_1,\vec{y}_1;0)\psi(\vec{y}_1,0)
  \bar\psi(\vec{y}_2,t)\Gamma^\dagger{\cal U}^\dagger(\vec{x}_2,\vec{y}_2;t)
  \psi(\vec{x}_2,t).
\end{multline}
Again, integration over the quark fields can be carried out resulting in
\begin{multline}\label{JJplus}
  \langle J(\vec{x}_1,\vec{y}_1;0) J(\vec{x}_2,\vec{y}_2;t)\rangle=\\
  \langle {\rm Tr}\left[M^{-1}(\vec{x}_2,t;\vec{x}_1,0)
  \Gamma {\cal U}(\vec{x}_1,\vec{y}_1;0)
  M^{-1}(\vec{y}_1,0;\vec{y}_2,t)\Gamma^\dagger
  {\cal U}^\dagger(\vec{x}_2,\vec{y}_2;t)\right]\rangle\\
  -\langle {\rm Tr}\left[M^{-1}(\vec{y}_1,0;\vec{x}_1,0)
  \Gamma {\cal U}(\vec{x}_1,\vec{y}_1;0)
  \right]\rangle\\
  \times\langle {\rm Tr}\left[M^{-1}(\vec{x}_2,t;\vec{y}_2,t)
  \Gamma^\dagger
  {\cal U}^\dagger(\vec{x}_2,\vec{y}_2;t)\right]\rangle.
\end{multline}
The inverse of the fermion matrix, $M^{-1}$ has meaning of a fermion
propagator.

Gauge transporters ${\cal U}(\vec{x},\vec{y};t)$ can be taken as weighted sums
of different paths connecting points $\vec{x}$ and $\vec{y}$. By choosing
paths of certain shape or combinations of different paths it is possible to
achieve a better overlap of the meson operator with a given state.
One of the possibilities is 
to construct the gauge transporters by using APE smearing
\cite{Albanese:1987ds} on spatial links: 
a link variable $U_{x,\mu}$ is replaced by a weighted average
of itself and a sum of the 3-link paths connecting the same sites as $U_{x,\mu}$:
\begin{equation}
  U_{x,\mu}\to U'_{x,\mu}=(1-6c)U_{x,\mu}+c\sum_{\nu\neq\mu}
  U_{x,\mu}U_{x+\hat\nu,\mu}U^\dagger_{x+\hat\mu,\nu}.
\end{equation}
This procedure can be applied iteratively. Then a gauge transporter 
${\cal U}(\vec{x},\vec{y};t)$,
taken as a product of smeared links, is equivalent to a weighted sum of
differently shaped paths connecting the sites $\vec{x}$ and $\vec{y}$.

In the following we will consider meson correlators at finite temperature.
The finite temperature is introduced by compactifying the the Eucliden time
direction, i.e. $T=1/(N_{\tau} a)$ with $N_{\tau}$ being the number of temporal
time slices. Gauge fields and fermion fields obey periodic and anti-periodic 
boundary conditions in the temporal direction.
In the next section we consider a spinless static quark 
anti-quark pair, $\Gamma=\mathbf{I}$ that can propagate only in time.
In this case second term in (\ref{JJplus}) vanishes. In 
Sec.~\ref{sec_spec_fun} we consider local meson operators with
$\vec{x}_i=\vec{y}_i$, $i=1,2$ which means 
${\cal U}(\vec{x},\vec{y};t)=\mathbf{I}$.

\section{Correlation functions of static quarks in lattice gauge theory}
\label{sec_static_corrs}

\subsection{Static meson correlators}
\label{subsec_meson_cor}

Consider static (infinitely heavy) quarks. The position
of heavy quark anti-quark pair is fixed in space and propagation happens only
along the time direction. In this limit the second term on the right hand side of
(\ref{JJplus}) vanishes. We are interested in a spinless state and set
$\Gamma=\mathbf{I}$ in this section. With respect to the color the meson
can be in a singlet or adjoint state. These states are described by the following
gauge connections
\begin{eqnarray}
  {\cal U}(\vec{x},\vec{y};t)&=&U(\vec{x},\vec{y};t),\label{calUs}\\
  {\cal U}^a(\vec{x},\vec{y};t)&=&U(\vec{x},\vec{x}_0;t)
  T^a U(\vec{x}_0,\vec{y};t),\label{calUas}
\end{eqnarray}
where $U(\vec{x},\vec{y})$ is a spatial gauge transporter, -- 
the product of the gauge variables along the path 
connecting $\vec{x}$ and $\vec{y}$, $\vec{x}_0$ is the coordinate
of the center of mass of the meson and $T^a$ are the $SU(N)$ group generators.

The meson operators are given then by Eq. (\ref{Jgen}) with 
$\vec{x}_1=\vec{x}_2$, $\vec{y}_1=\vec{y}_2$ for static quarks
\begin{eqnarray}
  \label{Jmesstat}
  J(\vec{x},\vec{y};t)&=&
  \bar\psi(\vec{x},t)U(\vec{x},\vec{y};t)\psi(\vec{y},t),\\
  \label{Jamesstat}
  J^a(\vec{x},\vec{y};t)&=&
  \bar\psi(\vec{x},t)U(\vec{x},\vec{x}_0;t)
  T^a U(\vec{x}_0,\vec{y};t)\psi(\vec{y},t).
\end{eqnarray}
Substituting expressions (\ref{Jmesstat}) into eq. (\ref{JJplus})
and noting that for a static quark the propagator
$M^{-1}(\vec{x},0;\vec{x},t)\sim L(\vec{x})$, where 
the temporal Wilson line $L(\vec{x})=\prod_{t=0}^{N_\tau-1} U_{(\vec{x},t),0}$
with $U_{(\vec{x},t),0}$ being the temporal links, 
we get for the meson correlators at $t=1/T$:
\begin{eqnarray}
\displaystyle
G_1(r,T)&\equiv&\frac{1}{N}
\langle J(\vec{x},\vec{y};0)\bar J(\vec{x},\vec{y};1/T)\rangle\nonumber\\
\displaystyle
&=&\frac{1}{N} 
\langle {\rm Tr}\left[
L^{\dagger}(\vec{x}) U(\vec{x},\vec{y};0) 
L(\vec{y}) U^{\dagger}(\vec{x},\vec{y},1/T)\right]\rangle, \label{defg1}\\
\displaystyle
G_a(r,T)&\equiv&\frac{1}{N^2-1}\sum_{a=1}^{N^2-1}
\langle J^a(\vec{x},\vec{y};0)\bar J^a(\vec{x},\vec{y};1/T)\rangle
\nonumber\\
\displaystyle
&=&\frac{1}{N^2-1}
\langle {\rm Tr} L^{\dagger}(x)  {\rm Tr} L(y) \rangle\nonumber\\
\displaystyle
&-&\frac{1}{N (N^2-1)} 
\langle {\rm Tr}\left[
L^{\dagger}(x) U(x,y;0) L(y) U^{\dagger}(x,y,1/T)\right] \rangle,
\label{defg3}\\
&&r=|\vec{x}-\vec{y}|. \nonumber
\end{eqnarray}

The correlators depend on the choice of the spatial transporters
$U(\vec{x},\vec{y};t)$. Typically, a straight line connecting points 
$\vec{x}$ and $\vec{y}$ is used as a path in the gauge transporters, 
i.e. one deals with time-like rectangular cyclic Wilson loops.
This object has been calculated at finite temperature in hard thermal loop (HTL) 
petrurbation theory in context of perturbative 
calculations of the singlet potential introduced in the previous section and quarkonium 
spectral functions \cite{Laine:2006ns,Laine:2007qy,Burnier:2007qm}.
In the special gauge, where $U(\vec{x},\vec{y};t)=1$ the above
correlators give the standard definition of the singlet and
adjoint free energies of a static $Q\bar Q$ pair 
\begin{eqnarray}
\displaystyle
\exp(-F_1(r,T)/T)&=&
\frac{1}{N} \langle {\rm Tr}[L^{\dagger}(x)  L(y)]\rangle,\label{F1def}\\
\displaystyle
\exp(-F_a(r,T)/T)&=&
\frac{1}{N^2-1}
\langle {\rm Tr} L^{\dagger}(x)  {\rm Tr} L(y) \rangle \nonumber\\
\displaystyle
&-&\frac{1}{N (N^2-1)} \langle {\rm Tr}
\left[L^{\dagger}(x)  L(y)\right]\rangle.\label{Fadef}
\end{eqnarray}
The singlet and adjoint free energies can be calculated at high temperature in
leading order HTL approximation \cite{Petreczky:2005bd} resulting in
\begin{eqnarray}
\displaystyle
F_1(r,T)&=&-\frac{N^2-1}{2 N} \frac{\alpha_s}{r} \exp(-m_D r)-\frac{(N^2-1)\alpha_s m_D }{2 N},
\label{f1p}\\
\displaystyle
F_a(r,T)&=&\frac{1}{2 N} \frac{\alpha_s}{r} \exp(-m_D r)-\frac{(N^2-1) \alpha_s m_D}{2N},
\label{f3p}
\end{eqnarray}
with $m_D=g T \sqrt{(N/3+N_f/6)} $ being the leading order Debye mass and $N_f$ is the number of quark flavors.
At this order $F_1$ and $F_a$ are gauge independent or, in other words, do not depend
on the choice of
the parallel transporters $U(\vec{x},\vec{y};t)$.
Note that at small distances ($r m_D\ll 1$) the singlet free energy 
\begin{equation}
F_1(r,T) \simeq -\frac{N^2-1}{2 N} \frac{\alpha_s}{r}
\end{equation}
is temperature independent and coincides with the zero temperature potential, 
while the adjoint free energy 
\begin{equation}
F_a(r,T) \simeq \frac{1}{2 N} \frac{\alpha_s}{r}-\frac{N}{2} \alpha_s m_D 
\end{equation}
depends on the temperature.

The physical free energy of a static 
quark anti-quark pair, i.e. the one related to the work that has to be done to
separate the static charges by certain distance is given by the thermal average
of the singlet and adjoint free energies \cite{McLerran:1981pb}
\begin{eqnarray}
\displaystyle
\exp(-F(r,T)/T)&=&\frac{1}{N^2} \exp(-F_1(r,T)/T) 
+ \frac{N^2-1}{N^2} \exp(-F_a(r,T)/T)\nonumber\\
\displaystyle
&=&\frac{1}{N^2} \langle {\rm Tr}
\left[L(x) {\rm Tr} L(y)\right]\rangle \equiv \frac{1}{N^2} G(r,T) \label{defg}.
\end{eqnarray}
This quantity is explicitly gauge independent. In leading order HTL approximation
the free energy is 
\begin{equation}
F(r,T)=-\frac{(N^2-1)}{8 N^2} \frac{\alpha_s^2}{r^2 T} \exp(-2 m_D r).
\end{equation}
The $1/r^2$ behavior is due to partial cancellation between the singlet and adjoint contribution
\cite{McLerran:1981pb,Nadkarni:1986cz} and has been confirmed by lattice calculations 
in the intermediate distance regime above the  deconfinement transition \cite{Petreczky:2001pd,Digal:2003jc}.

Using the transfer matrix one can show that in the confined phase 
\begin{eqnarray}
\displaystyle
G_1(r,T)&=&\sum_{n=1}^{\infty} c_n(r) e^{-E_n(r,T)/T},\label{g1}\\
\displaystyle
G(r,T)&=&\sum_{n=1}^{\infty} e^{-E_n(r,T)/T},
\label{g}
\end{eqnarray}
where $E_n$ are the energy levels of static quark and anti-quark pair \cite{Jahn:2004qr}. The 
coefficients $c_n(r)$ depend on the choice of $U(x,y;t)$ entering the static meson
operator in Eqs.~(\ref{Jmesstat}-\ref{Jamesstat}). 
Since the color averaged correlator $G(r,T)$ corresponds to a 
gauge invariant measurable quantity 
it does not contain $c_n$.
The lowest energy level is the usual static quark anti-quark
potential, while the higher energy levels correspond to hybrid potentials 
\cite{Bali:2000gf,Morningstar:1998da,Juge:2002br,Michael:1990az}. 
Using multi-pole  expansion in pNRQCD one can
show that at short distances the hybrid potential corresponds to the adjoint potential
up to non-perturbative constants \cite{Brambilla:1999xf}. 
Indeed, lattice calculations of the hybrid potentials
indicate a repulsive short distance part \cite{Bali:2000gf,Morningstar:1998da,Juge:2002br,Michael:1990az}.
Furthermore, the gap between the static potential and the first hybrid potential can be estimated
fairly well at short distances in perturbation theory \cite{Bali:2003jq}.

If $c_1=1$ the dominant contribution to
$G_a$ would be the first excited state $E_2$, i.e. the lowest hybrid potential which
at short distances is related to the adjoint potential. In this sense $G_a$ is related 
to static mesons with quark anti-quark in adjoint state. 
Numerical calculations show, 
however, that $c_1$ is $r$-dependent and in general $c_1(r) \ne 1$. Thus $G_a$ also
receives contribution from $E_1$ \cite{Jahn:2004qr}. The lattice data suggests that $c_1$ 
approaches unity at short distances \cite{Jahn:2004qr} in accord with expectations based on 
perturbation theory, where $c_1=1$  up to ${\cal O} (\alpha_s^3)$ corrections  \cite{Brambilla:1999xf}. 
Therefore  at short distances, $r \ll 1/T$ the color singlet and color averaged free energy
are related $F(r,T)=F_1(r,T) + T \ln (N^2-1)$.

In the following we  consider $SU(2)$ and $SU(3)$ gauge theories and refer to the adjoint state as
triplet and octet, correspondingly.

\subsection{Lattice results on static meson correlators}\label{sec_ana_cor}

Correlation function of static quarks have been extensively studied  
on the lattice since the pioneering work by Mclerran and Svetistky \cite{McLerran:1981pb}.
Most of these studies, however, considered only the color averaged correlator, i.e.
the correlation function of two Polyakov loops (for the most complete analysis see 
Ref. \cite{Kaczmarek:1999mm} and references therein). Since both color singlet and color octet
degrees of freedom contribute to the Polyakov loop 
correlator it has large temperature dependence even at short distances and it is not a
very useful quantity if we want to learn something about quarkonium properties at high
temperatures. As this has been pointed out in Refs. \cite{Digal:2001iu,Digal:2001ue} we need
to know the quark anti-quark interactions in the singlet channel in order to learn about in-medium quarkonium
properties. Therefore in recent years the singlet correlator has been computed on the lattice
in $SU(2)$ and $SU(3)$ gauge theories 
\cite{Kaczmarek:2002mc,Philipsen:2002az,Digal:2003jc,Kaczmarek:2003dp,Kaczmarek:2004gv}
as well as in full QCD
with 3 and 2 flavors of dynamical quarks \cite{Petreczky:2004pz,Kaczmarek:2005ui}. Preliminary results
also exist for 2+1 flavor QCD with physical value of the strange quark mass and light $u,d$-quark masses
corresponding to pion mass of about $220MeV$ \cite{Kaczmarek:2007pb,rbc_progres}.
All these calculations use Coulomb gauge definition of the singlet correlator, i.e. the
definition (\ref{F1def}) with Coulomb gauge fixing.

\begin{figure}
\centering
\includegraphics[width=0.9\columnwidth]{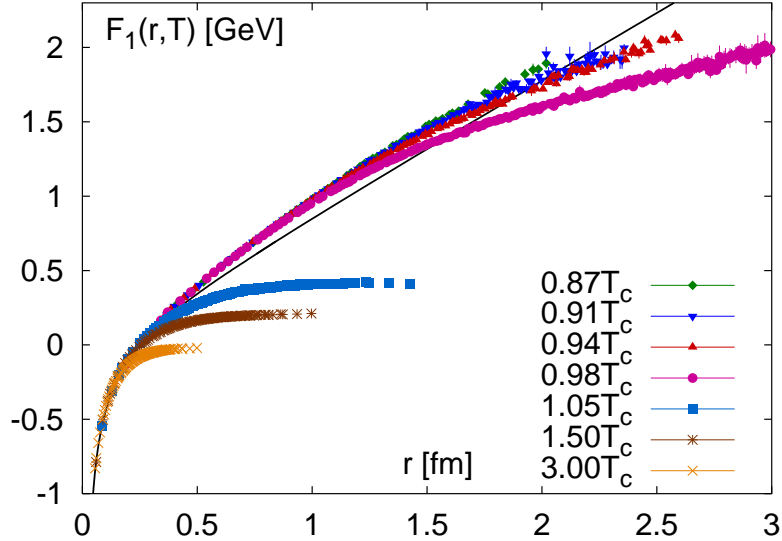}
\caption{The color singlet free energy singlet free energy in quenched QCD
(SU(3) gauge theory) calculated in Coulomb gauge \cite{Kaczmarek:2002mc,Kaczmarek:2003dp,Kaczmarek:2004gv}.
The solid line shows the zero temperature potential.} 
\label{fig:fes}
\end{figure}

The numerical results for $SU(3)$ gauge theory are presented in Figure \ref{fig:fes}
and compared with the zero temperature potential. Here we used the 
string Ansatz for the zero temperature potential 
\be
V(r)=-\frac{\pi}{12 r}+\sigma r,
\label{vstring}
\ee
since this form gives a very good description of the lattice data in $SU(3)$ gauge theory \cite{Necco:2001xg}.
To convert lattice units to physical units the value $\sqrt{\sigma}=420$MeV has been used for the
string tension. To remove the additive renormalization in the singlet free energy the lattice data
have been normalized to $V(r)$ given by Eq. (\ref{vstring}) at the shortest
distance available. As one can see from  Figure \ref{fig:fes} the singlet free energy is temperature independent
at short distances and coincides with the zero temperature potential. Below the deconfinement transition, $T<T_c$, it rises
linearly, indicating confinement. The string tension at finite temperature is smaller than the zero temperature
string tension $\sigma$. At large distances the singlet free energy is the same as the free energy determined from
the Polyakov loop correlators and therefore the finite temperature string tension agrees with findings of 
Ref. \cite{Kaczmarek:1999mm}. 
Medium effects set in at distance $r_{med} \simeq 0.4\mbox{ fm}/(T/T_c)$ and we see exponential screening 
at distance $r > 1/T$ \cite{Kaczmarek:2004gv}.
In the intermediate distance regime $0.5\mbox{ fm}<r<1.5\mbox{ fm}$ we see that the singlet free energy is enhanced relative
to the zero temperature potential. This is not a real physical effect but an artifact of the calculations. 
Similar effect has also been seen in $SU(2)$ gauge theory \cite{Philipsen:2002az,Digal:2003jc,Jahn:2004qr}.
At these distances
the color singlet correlator is sensitive to the value of the coefficients $c_n(r)$ in Eq.~(\ref{g1}). 
Below we will show that the
enhancement of the singlet free energy is due to the fact that $c_1(r)<1$ in this region.

\begin{figure}
\centering
\includegraphics[width=0.9\columnwidth]{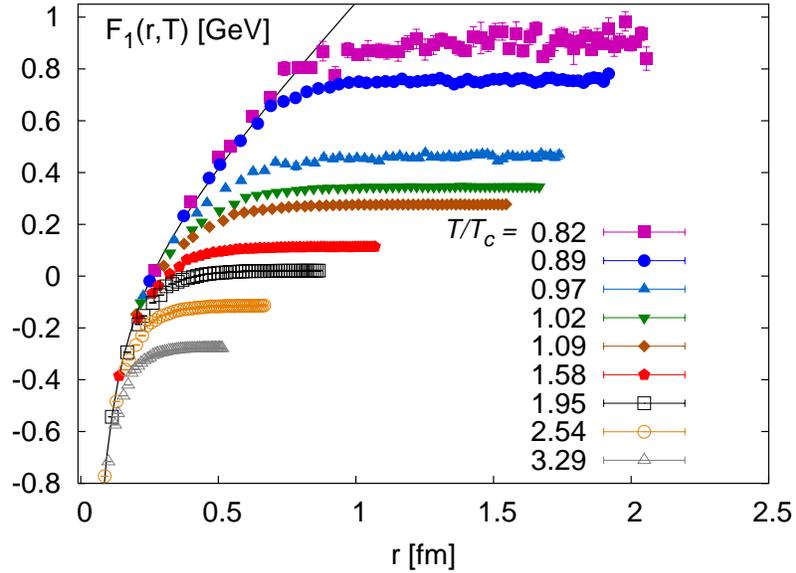}
\caption{The color singlet free energy in 2+1 flavor QCD \cite{Kaczmarek:2007pb,rbc_progres}. The solid line
is the parametrization of the lattice data on the zero temperature potential from Ref. \cite{Cheng:2007jq}. }  
\label{fig:fes_2+1}
\end{figure}

Numerical results for the singlet free energy in 2+1 flavor QCD are shown in Figure \ref{fig:fes_2+1}.
In this case a different normalization procedure has been used. 
The additive renormalization has been determined at zero temperature for each value of the lattice
spacing used in the finite temperature calculations by matching the zero temperature potential to the
string Ansatz (\ref{vstring}) at distance $r=r_0$, with $r_0$ being the Sommer scale \cite{Sommer:1993ce}.
This additive renormalization then has been used for the singlet free energy.
For detailed discussion of the calculation of the static potential and its renormalization see Ref. \cite{Cheng:2007jq}.
As in quenched QCD the singlet free energy is temperature independent at short distances and
coincides with the zero temperature potential. At distances larger than the inverse temperature
it is exponentially screened.
The novel feature of the singlet free energy in full QCD
is the string breaking, i.e. the fact that it approaches a constant at large separation.
This happens because if the energy in the string exceeds the binding energy of a heavy-light
meson the static quark and anti-quark are screened due to pair creation from the vacuum. This happens at
distances of about $0.8$ fm according to the figure. As temperature increases the distance where the singlet
free energy flattens out becomes smaller and for sufficiently high temperatures turns out to be inversely proportional 
to the temperature. Thus string breaking smoothly turns into color screening as temperature increases. 

\subsection{Color singlet correlator in SU(2) gauge theory at low temperatures}
\label{color_singlet_cor}
For a better understanding of the temperature dependence of the singlet correlator
and its physical interpretation the values of the overlap factors $c_n(r)$ should be estimated.
To extract the overlap factors we need to calculate the singlet correlators in the low
temperature region, where only few energy levels contribute to the correlator. 
The calculations of the singlet correlator at low temperatures is difficult because of the
rapidly deceasing signal to noise ratio. To overcome this difficulty it has been suggested
to calculate Wilson loops with multilevel Luescher-Weisz algorithm \cite{Luscher:2001up}
instead of the Coulomb gauge fixing \cite{Bazavov:2008rw}.
In this case the color singlet 
correlator defined by Eq.~(\ref{defg1}) is an expectation
value of the gauge-invariant Wilson loop with gauge
transporter $U(\vec{x},\vec{y};t)$ being a product of the iteratively smeared spatial links.

\begin{figure}
\centering
\includegraphics[width=0.9\columnwidth]{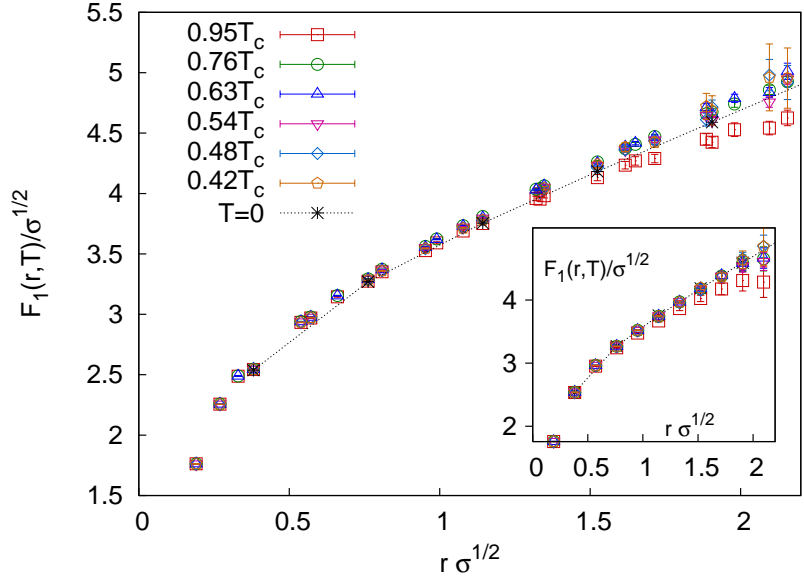}
\caption{The color singlet free energy in 
$SU(2)$ gauge theory below the deconfinement
temperature at $\beta=2.5$ calculated on $32^3 \times N_{\tau}$ lattices.
Also shown is the $T=0$ potential. The inset shows the color singlet free energy
from which the contribution from the matrix element $T \ln c_1$ has been subtracted. }  
\label{fig:f125}
\end{figure}

Numerical calculations have been performed in $SU(2)$ gauge theory using standard Wilson gauge action \cite{Bazavov:2008rw}.
With the use of the multilevel algorithm it was possible to go down to temperatures as low
as $0.32T_c$ not accessible in the previous studies. 
It is well known that smearing increases the overlap with the ground state by removing 
the short distance fluctuations in the spatial links \cite{Booth:1991kk}.
For this reason smearing also reduces the breaking of the rotational invariance to the
level expected in the free theory.
When no smearing is used the color singlet
free energy, $-T \ln G_1(r,T)$ shows a small but visible temperature dependence.
The temperature dependence of the singlet free energy is significantly
reduced when APE smearing is applied.
The color singlet free energy for $\beta=2.5$
and 10 APE smearings is shown in Fig.~\ref{fig:f125}. 
As one can see from the figure
the color singlet free energy shows very mild temperature dependence in the confined phase
with noticeable temperature effects appearing only at $T=0.95T_c$.

\begin{figure}
\centering
\includegraphics[width=0.7\columnwidth]{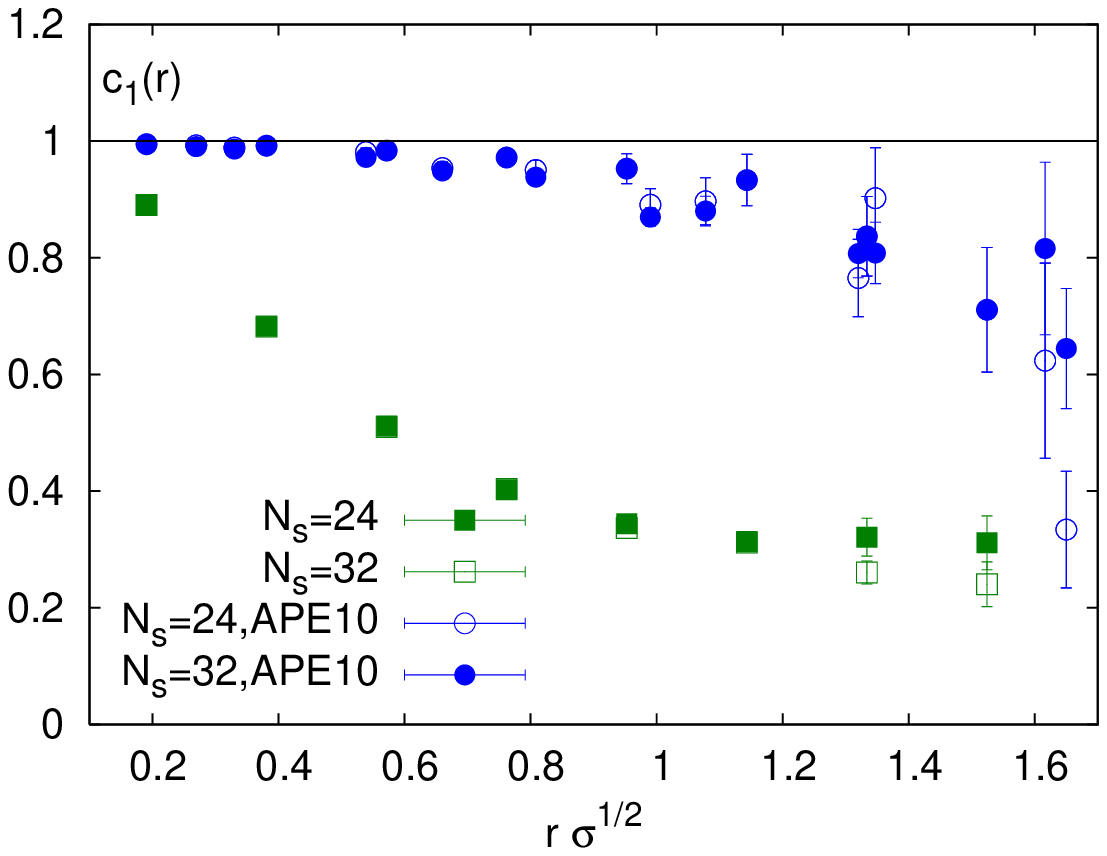}
\includegraphics[width=0.7\columnwidth]{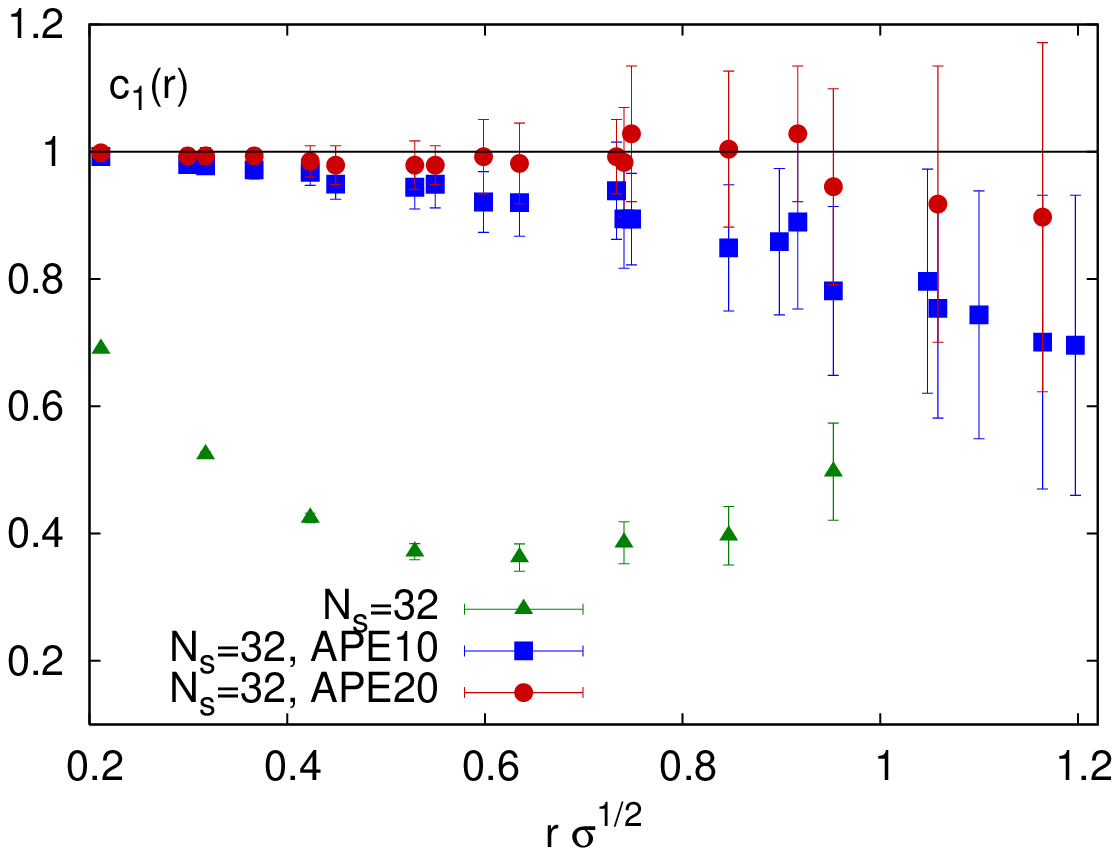}
\caption{The pre-exponential factor of the color singlet correlators as function of
distance $r$ for $\beta=2.5$ (top) and $\beta=2.7$ (bottom). Shown are results
for unsmeared spatial links and 10 and 20 steps of APE smearing.}
\label{fig:c1}
\end{figure}

Consider the expansion (\ref{g1}). The dominant contribution comes from the
ground state, so it is reasonable to fit the singlet correlator to the form
\begin{equation}\label{G1expfit}
  G_1(r,T)=c_1(r) \exp(-E_1(r)/T).
\end{equation}
This allows to extract the matrix element $c_1(r)$ using a simple exponential fit, which is shown in
Fig.~\ref{fig:c1}.
When no APE smearing is used the value of $c_1(r)$ strongly depends on the separation $r$.
At small distances it shows a tendency of approaching unity as one would expect in perturbation theory.
However, $c_1(r)$ decreases with increasing distance $r$. 
At large distance its value is around $0.3-0.5$.
Similar results for $c_1(r)$ have been obtained in the study of $SU(2)$ gauge theory 
in 3 dimensions \cite{Jahn:2004qr}.
When APE smearing is applied the $r$-dependence of the amplitude $c_1(r)$ is largely reduced
and its value is close to unity both for $\beta=2.5$ and $\beta=2.7$. For $\beta=2.7$ we also see
that increasing the number of smearing steps from 10 to 20 reduces the deviation of $c_1(r)$ from unity.

As discussed in section \ref{subsec_meson_cor} perturbation theory 
predicts that the deviation of $c_1(r)$ from unity is
of order $\alpha_s^3$. Therefore it can be made arbitrarily small by going to sufficiently small
distances. It is known, however, that lattice perturbation theory converges very
poorly. The main reason for this has been identified with the short distance fluctuations of the link variables,
which makes their mean value very different from unity \cite{Lepage:1992xa}. Smearing removes these short
distance fluctuations and this is the reason why $c_1(r)$ is much closer to unity when APE smearing is
applied.
Thus, almost the entire temperature dependence 
of the singlet free energy at distances $0.5<r \sqrt{\sigma}<2$ is due to the deviation 
of $c_1$ from unity and can be largely reduced by applying APE smearing to
the links in the spatial gauge connections. To further demonstrate this point in the inset of
Figure \ref{fig:f125} we show the results for $F_1(r,T)+T \ln c_1(r)$. Clearly no temperature dependence
can be seen in this quantity up to $0.95T_c$, where we see temperature dependence at 
distances $r\sqrt{\sigma} \ge 1.5$ 
corresponding to the expected drop of the effective string tension.

\subsection{Color singlet free energy in the deconfined phase}
\label{sec:deconf_f1}

The behavior of the color singlet free energy
in the deconfined phase has been studied in Coulomb gauge
\cite{Kaczmarek:2002mc,Digal:2003jc,Kaczmarek:2004gv,Petreczky:2004pz,Kaczmarek:2005ui}
and from cyclic Wilson loops \cite{Bazavov:2008rw}.
As discussed above at short distances it is
temperature independent and coincides with the zero temperature potential.
At large distances it approaches a constant $F_{\infty}(T)$, which monotonically
decreases with the temperature. The constant $F_{\infty}(T)$  is the free energy of 
two isolated static quarks, or equivalently of a quark anti-quark pair at infinite separation. 
Its value is 
therefore independent of the definition 
of the singlet correlator $G_1(r,T)$ and is related to the renormalized
Polyakov loop $L_{ren}(T)=\exp(-F_{\infty}(T)/(2 T))$ \cite{Kaczmarek:2002mc}. 

At leading order  $F_1(r,T)-F_{\infty}(T)$ is of Yukawa form (c.f. Eq. (\ref{f1p})). 
Therefore it is useful to define a quantity called the screening function 
\begin{equation}
S(r,T)=r \cdot (  F_1(r,T)-F_{\infty}(T)).
\end{equation}
This quantity shows exponential decay at distances $r>1/T$ in the deconfined phase 
both in pure gauge theories \cite{Digal:2003jc,Kaczmarek:2004gv} and full QCD \cite{Kaczmarek:2005ui}.
From its exponential decay the Debye screening mass has been determined and turns out to be about 
$40\%$ larger than the leading order perturbative value.

\begin{figure}
\centering
\includegraphics[width=0.7\columnwidth]{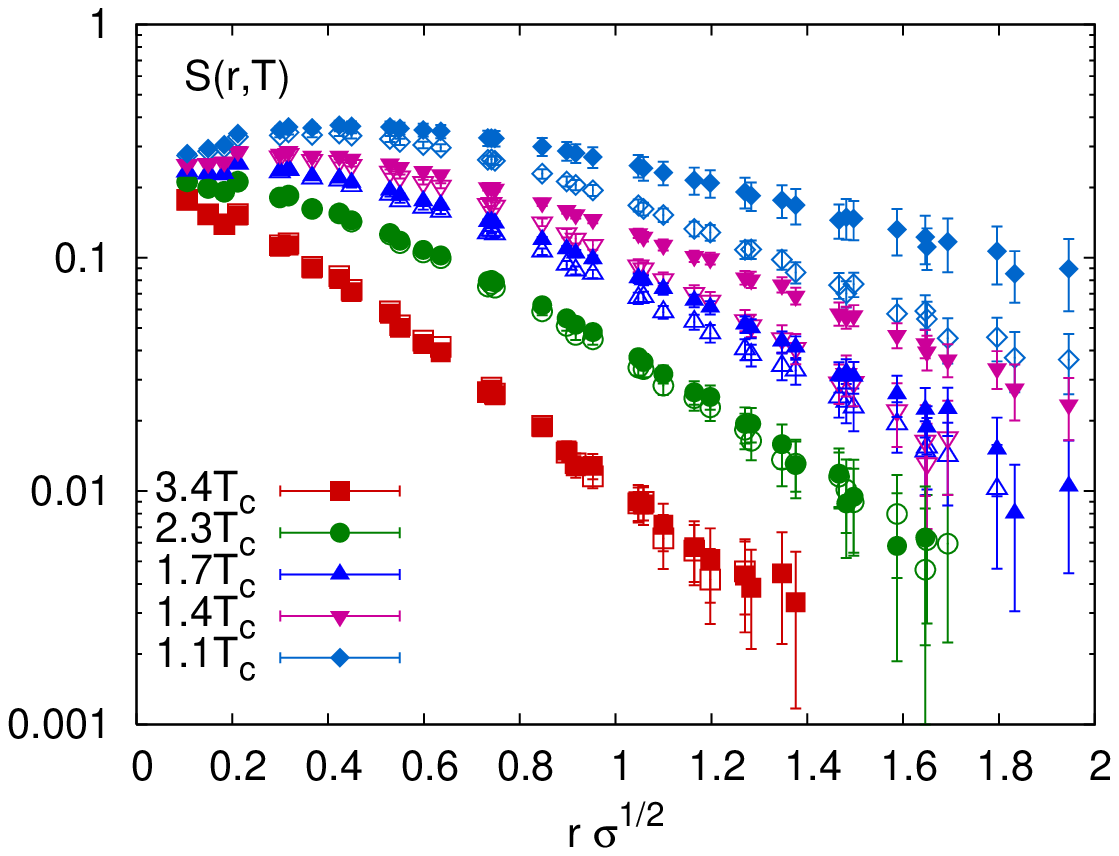}
\includegraphics[width=0.7\columnwidth]{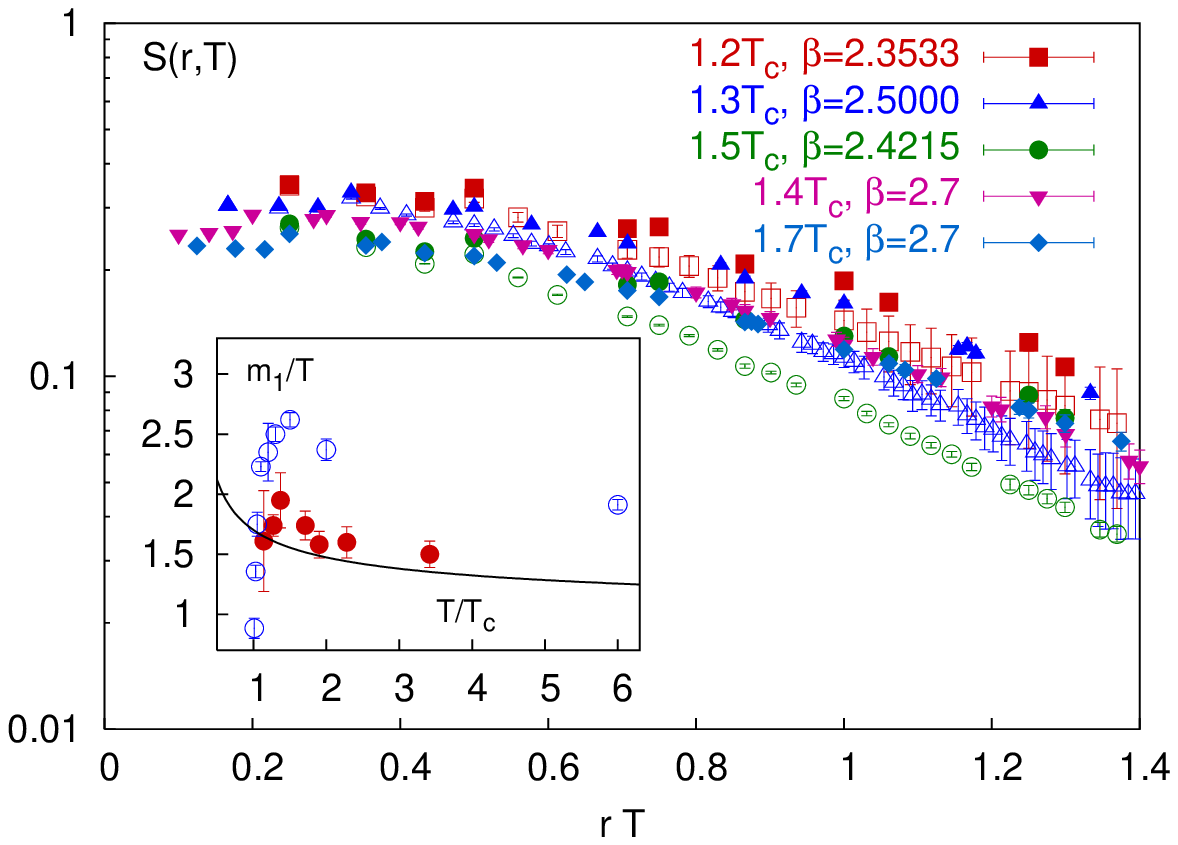}
\caption{The screening function $S(r,T)=r (  F_1(r,T)-F_{\infty}(T))$ in $SU(2)$ gauge theory at different
temperatures calculated for $\beta=2.7$ as function of $r \sqrt{\sigma}$ (top) and as function
of $r T$ (bottom).
}
\label{fig:s1}
\end{figure}

It is interesting to study the screening function using 
the free energy determined from cyclic Wilson loops and make comparison with Coulomb gauge results. This analysis
has been recently done in Ref. \cite{Bazavov:2008rw} for $SU(2)$ gauge theory.
The behavior of the screening function at different temperatures
is shown in Fig.~\ref{fig:s1}.
At short distances ($r T<0.5$) the singlet free energy 
does not depend on the smearing level. Furthermore, it is very close to the free energy
calculated in Coulomb gauge. 
At large distances the screening function $S(r,T)$ shows an exponential 
decay determined by a temperature dependent screening mass $m_1(T)$, which is equal to the leading order
Debye mass up to the non-perturbative $g^2$ corrections: $m_1=m_D + {\cal O}(g^2)$ \cite{Rebhan:1993az, Rebhan:1994mx}.
There is some dependence on the smearing level at larger distances
which, however, disappears
at high temperatures and with increasing the smearing level. In particular, for 
$\beta \leqslant 2.5$
it turns out that there is no dependence on smearing level for 5 or more smearing steps.
For $\beta=2.7$ 10--20 steps are needed, depending on the temperature.
Fitting the large distance behavior of the screening function by 
an exponential form $\exp(-m_1(T) r)$ allows to
determine the screening mass $m_1(T)$.
In the inset of Fig.~\ref{fig:s1} the color singlet screening masses 
extracted from the fits are shown in comparison with 
the results obtained in Coulomb gauge in Ref. \cite{Digal:2003jc}.
The solid line is the leading order Debye mass calculated using 2-loop gauge
coupling $g(\mu=2 \pi T)$ in $\overline{MS}$-scheme. 
As we see from the figure the screening
masses are smaller than those calculated in Coulomb gauge and agree well with the leading order
perturbative prediction.

\subsection{Color adjoint free energy}
\label{subsec_adj}

The structure of Eqs.~(\ref{defg1}) and (\ref{defg3}) shows
that color adjoint correlator is given by a difference
of the color averaged and singlet correlators.
The color adjoint correlator in Coulomb gauge has been studied in pure gauge theory \cite{Digal:2003jc,Petreczky:2005bd},
3-flavor QCD \cite{Petreczky:2004pz} and 2-flavor QCD \cite{Kaczmarek:2005ui}. At low temperatures the color adjoint free energy
turned out to be significantly smaller than the first hybrid potential contrary to the expectations. In fact at sufficiently
large distance it was found to be identical to the singlet free energy. As has been pointed out in Ref. \cite{Jahn:2004qr}
this is due to the non-trivial $r$-dependence of the overlap factor $c_1(r)$ and its deviation from unity. 

\begin{figure}
\centering
\includegraphics[width=0.97\columnwidth]{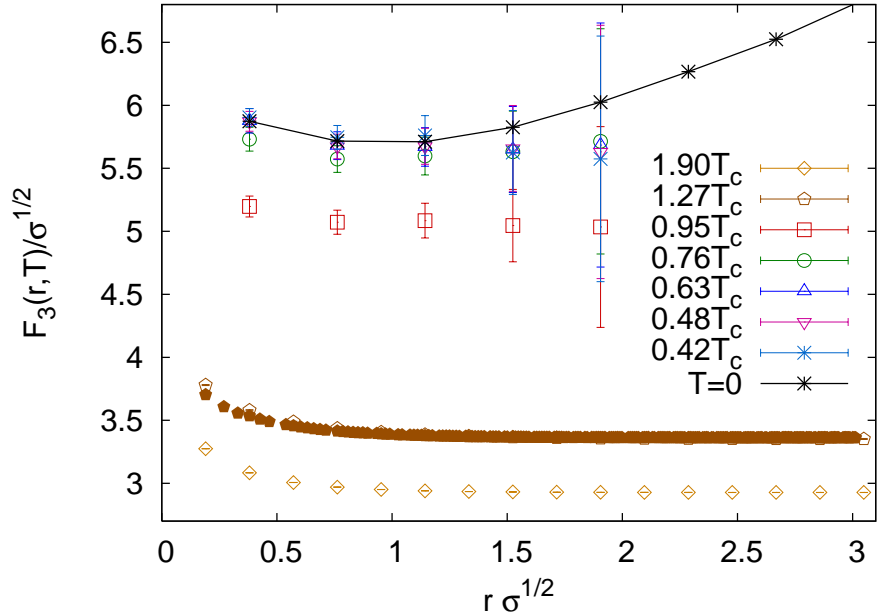}
\caption{The triplet free energy at different temperatures calculated
at $\beta=2.5$. 
The filled symbols correspond to calculations in Coulomb gauge.
Also shown is the first hybrid potential calculated in \cite{Michael:1990az}.
}
\label{fig:f3corr}
\end{figure}

We have shown in section \ref{color_singlet_cor} that deviations of the overlap factor $c_1(r)$ from unity can be greatly reduced when one uses
Wilson loops with the smeared spatial gauge connection. Therefore it is interesting to see how the 
adjoint free energy behaves in this approach. The numerical analysis has been done in $SU(2)$
gauge theory \cite{Bazavov:2008rw} therefore below
we will refer to the adjoint free energy as the triplet free energy.
If we assume that only two states contribute to the Eqs. (\ref{g1}) and (\ref{g}),
then from Eq. (\ref{defg3}) it follows that
\begin{equation}
F_3(r,T)=E_2(r)-T\ln\left(1-c_2(r)+\frac{1}{3}(1-c_1(r)) e^{\Delta E(r)/T}
\right),
\label{f3_2comp}
\end{equation}
with $\Delta E(r)=E_2(r)-E_1(r)$. 
We have seen in section 
\ref{color_singlet_cor} that the temperature dependence of the 
singlet free energy is quite small. In any case it is considerably smaller than the temperature
dependence of the averaged free energy. Therefore the contribution of the excited states
to $G_1(r,T)$ is quite small and it is reasonable to assume that $c_2(r) \ll 1$.
From the analysis of the multipole expansion we also expect that at small distances, $c_2(r) \sim (r \Lambda_{QCD})^4$ \cite{vairo_priv}. 
Thus, the temperature dependence of $F_3(r,T)$ and its deviation from the hybrid state $E_2(r)$ is
due to small deviation of $c_1(r)$ from unity. At low temperatures, when $\Delta E \gg T$ these small
deviations are amplified by the exponential factor.
This can be easily verified by
subtracting the correction $T \ln (1+\frac13(1-c_1) e^{\Delta E/T})$ from the triplet
free energy and assuming that $E_1(r)$ is given 
by the ground state potential and $E_2(r)$ is given by the
first hybrid potential as calculated in Ref. \cite{Michael:1990az}.
The numerical results are summarized in Fig. \ref{fig:f3corr} which shows that after this 
correction is accounted for in the confined phase the triplet
free energy at low temperatures agrees reasonably well with the first hybrid potential. As temperature
increases more excited states contribute. In particular,  at $0.76T_c$ the value of the triplet
free energy can be accounted for by including the next hybrid state \cite{Michael:1990az}.
However, at  $0.95T_c$ there are large temperature effects, 
which cannot be explained by including the contribution from only few excited states.

In Fig. \ref{fig:f3corr} we also show the triplet free energy above the deconfinement temperature
compared to the calculations in Coulomb gauge \cite{Digal:2003jc}. It turns out to be much smaller than in the
confined phase and agrees well with Coulomb gauge results. This means that the small deviation of
the overlap factor $c_1(r)$ from unity is unimportant in this case. The triplet free energy monotonically
decreases with increasing temperature as expected in HTL perturbation theory (c.f. Eq.~(\ref{f3p})). 
In the limit of high temperatures and short distances, $r \ll 1/T$ we have 
$E_2(r)=\alpha_s/(4 r),~\Delta E(r)=\alpha_s/r,c_2(r) \simeq 0$ and $c_1(r)=1+{\cal O}(\alpha_s^3)$.
Therefore we can expand the logarithm in Eq. (\ref{f3_2comp}) to get 
\begin{equation}
F_3(r,T)=+\frac{1}{4} \frac{\alpha_s}{r}+{\cal O}(\alpha_s^3 T)+{\cal O}(\alpha_s m_D).
\end{equation}
Thus the correction due to $c_1(r) \ne 0$ is much smaller than the expected leading order
thermal effects in the triplet free energy.

The strong temperature dependence of the adjoint free energy has been also observed in $SU(3)$ gauge theory \cite{Petreczky:2005bd}
and in full QCD \cite{Petreczky:2004pz,Kaczmarek:2005ui}.
Although the adjoint free energy is strongly temperature dependent even at short distances its $r$-dependence is almost
the same as  $r$-dependence of the singlet free energy.
The derivative of the adjoint free energy with respect to $r$ is identical withing errors 
to that of the singlet free energy at short distances up to the Casimir
factor, i.e. $F_{a}'(r,T)/F_1'(r,T)=-1/(N^2-1)$ \cite{Petreczky:2005bd}.

\section{Quarkonium spectral functions}
\label{sec_spec_fun}

\subsection{Meson correlators and spectral functions}
\label{sec.corr}

Now we focus the discussion on the relation between the
Euclidean meson correlators and spectral functions at finite
temperature. The zero temperature limit is straightforward.
Most dynamic properties of the finite temperature system are incorporated 
in the spectral function. The spectral function $\sgh$ for a given 
mesonic channel $H$ in a system at temperature $T$ can be defined 
through the Fourier transform of the real time two point functions
$D^{>}$ and $D^{<}$ or equivalently as the imaginary part of 
the Fourier transformed retarded 
correlation function \cite{LeBellac:375551},
\ber
\sgh &=& \frac{1}{2 \pi} (D^{>}_H(p_0, \vec{p})-D^{<}_H(p_0, \vec{p}))
\nonumber\\
&&
=\frac{1}{\pi} Im D^R_H(p_0, \vec{p}) \nonumber \\
 D^{>(<)}_H(p_0, \vec{p}) &=& \int{d^4 p \over (2
\pi)^4} e^{i p \cdot x} D^{>(<)}_H(x_0,\vec{x}) \label{eq.defspect} \\
D^{>}_H(x_0,\vec{x}) &=& \langle
J_H(x_0, \vec{x}), J_H(0, \vec{0}) \rangle \nonumber\\
D^{<}_H(x_0,\vec{x}) &=& 
\langle J_H(0, \vec{0}), J_H(x_0,\vec{x}) \rangle , x_0>0 \
\eer
In essence $\sigma_H$ is the Fourier transformation of the thermal average of the commutator $[J(x),J(0)]$.

In the present paper we study local meson operators of the form (c.f. (\ref{Jgen}) with ${\cal U}=I$)
\beq
J_H(t,x)=\bar q(t,x) \Gamma_H q(t,x)
\label{cont_current}
\eeq
with $q$ a continuous real-time fermion position operator and
\beq
\Gamma_H=1,\gamma_5, \gamma_{\mu}, \gamma_5 \gamma_{\mu}, \gamma_{\mu} \gamma_{\nu}
\eeq
for scalar, pseudo-scalar, vector, axial-vector and tensor channels. 
The relation of these quantum number channels to different meson states is given
in Tab. \ref{tab.channels}.

\begin{table*}
\begin{tabular}
[c]{||c|c|c||c|}\hline
$\Gamma$ & $^{2S+1}L_{J}$ & $J^{PC}$ & $u\overline{u}$\\\hline
$\gamma_{5}$ & $^{1}S_{0}$ & $0^{-+}$ & $\pi$\\
$\gamma_{s}$ & $^{3}S_{1}$ & $1^{--}$ & $\rho$\\
$\gamma_{s}\gamma_{s^{\prime}}$ & $^{1}P_{1}$ & $1^{+-}$ & $b_{1}$\\
$1$ & $^{3}P_{0}$ & $0^{++}$ & $a_{0}$\\
$\gamma_{5}\gamma_{s}$ & $^{3}P_{1}$ & $1^{++}$ & $a_{1}$\\
&&$2^{++}$&\\\hline
\end{tabular}%
\begin{tabular}
[c]{|cc|}\hline
$c\overline{c}(n=1)$ & $c\overline{c}(n=2)$\\\hline
$\eta_{c}$ & $\eta_{c}^{^{\prime}}$\\
$J/\psi$ & $\psi^{\prime}$\\
$h_{c}$ & \\
$\chi_{c0}$ & \\
$\chi_{c1}$ & \\
$\chi_{c2}$ & \\\hline
\end{tabular}
\begin{tabular}[c]{|cc|}\hline
$b\overline{b}(n=1)$ & $b\overline{b}(n=2)$\\
\hline
$\eta_b$ & $\eta_b'$ \\
$\Upsilon(1S)$ & $\Upsilon(2S)$\\
$h_b$ & \\
$\chi_{b0}(1P)$& $\chi_{b0}(2P)$\\
$\chi_{b1}(1P)$& $\chi_{b1}(2P)$\\
$\chi_{b2}(1P)$&  $\chi_{b2}(2P)$\\ 
\hline                                                                   
\end{tabular}
\caption{Meson states in different channels}
\label{tab.channels}
\end{table*}

The correlators $D^{>(<)}_H(x_0,\vec{x})$ satisfy the 
well-known Kubo-Martin-Schwinger
(KMS) condition \cite{LeBellac:375551}
\beq
D^{>}_H(x_0,\vec{x})=D^{<}_H(x_0+i/T,\vec{x}).
\label{kms}
\eeq
Inserting a complete set of
states and using Eq. (\ref{kms}), one gets the expansion
\ber
&
\sgh = {(2 \pi)^2 \over Z} \sum_{m,n} (e^{-E_n / T} \pm e^{-E_m / T})\times \nonumber\\ 
&
\langle n | J_H(0) | m \rangle|^2 \delta^4(p_\mu - k^n_\mu + k^m_\mu) 
\label{eq.specdef}
\eer
where $Z$ is the partition function, and 
$k^{n(m)}$ refers to the four-momenta of the state $| n (m) \rangle $.

A stable mesonic state contributes a $\delta$ function-like
peak to the spectral function:
\beq
\sgh = | \langle 0 | J_H | H \rangle |^2 \epsilon(p_0)
\delta(p^2 - m_H^2),
\label{eq.stable}
\eeq
where $m_H$ is the mass of the state and $\epsilon(p_0)$ is the sign function. For 
a quasi-particle in the medium one gets a smeared peak, with the width
being the  thermal width. 
As one increases the temperature the width increases 
and at sufficiently high
temperatures, the contribution from the meson state in the spectral function may 
be sufficiently broad so that it is not very meaningful to speak of it
as a well defined state any more. 
The spectral function as defined in
Eq. (\ref{eq.specdef}) can be directly accessible by high energy
heavy ion experiments. For example, the spectral function for the vector 
current is directly related to the differential thermal cross section 
for the production of dilepton pairs \cite{Braaten:1990wp}:
\beq
\left.{dW \over dp_0 d^3p}\right|_{\vec{p}=0} = {5 \alpha_{em}^2 \over 27 \pi^2} 
{1 \over p_0^2 (e^{p_0/T}-1)} \sigma_V(p_0, \vec{p}).
\label{eq.dilepton} \eeq
Then presence or absence of a bound state in the spectral function
will manifest itself in the peak structure of the differential 
dilepton rate.

In finite temperature lattice calculations, one calculates
Euclidean time propagators, usually
projected to a given spatial momentum:
\beq
G_H(\tau, \vec{p}) = \int d^3x e^{i \vec{p}.\vec{x}} 
\langle T_{\tau} J_H(\tau, \vec{x}) J_H(0,
\vec{0}) \rangle
\eeq
This quantity is an analytical continuation
of $D^{>}_H(x_0,\vec{p})$
\beq
G_H(\tau,\vec{p})=D^{>}_H(-i\tau,\vec{p}).
\eeq
Using this equation and the KMS condition           one can
easily show that $G_H(\tau,\vec{p})$ is related to the 
spectral
function, Eq. (\ref{eq.defspect}), by an integral equation
(see e.g. appendix B of Ref. \cite{Mocsy:2005qw}):
\ber
G_H(\tau, \vec{p}) &=& \int_0^{\infty} d \omega
\sg_H K(\omega, \tau) \label{eq.spect} \non\\
K(\omega, \tau) &=& \frac{\cosh(\omega(\tau-1/2
T))}{\sinh(\omega/2 T)}.
\label{eq.kernel}
\eer
This equation is the basic equation for extracting the spectral
function from meson correlators. 
Equation (\ref{eq.kernel})
is valid in the continuum. 
Formally the same spectral representation can be written for
the Euclidean correlator calculated on the lattice $G^{lat}_H(\tau,\vec{p})$.
The corresponding spectral function, however, will be distorted by the effect
of the finite lattice spacing. These distortions have been calculated in the
free theory \cite{Karsch:2003wy,Aarts:2005hg}. When discussing the numerical results in
following sections the subscript $H$ denoting different channels for meson correlators and spectral functions 
will be omitted.

\subsection{Lattice formulations for charmonium physics}
\label{sec:charm_lat}

The quarkonium system at zero and finite temperature
was studied in \cite{Umeda:2002vr,Datta:2003ww,Asakawa:2003re,Jakovac:2006sf} using Wilson-type fermions
\begin{equation}
S_q^{Wilson}=\sum_x {\bar{\psi}}(x) \left[
m_0 +
       \!\not\! {D}^{\rm Wilson}-\frac{c_{SW}}{2} \sum_{\mu,\nu} \sigma_{\mu \nu} F_{\mu \nu}
      \right]{\psi}(x),
\label{sqw}
\end{equation}
where the Dirac operator is defined as 
\begin{equation}
D_{\mu}^{\rm Wilson} =  \nabla_\mu - {1 \over 2} \gamma_\mu \Delta_\mu
\end{equation}
with
\begin{eqnarray}
 \nabla_\mu \psi(x) & =&
{1\over 2}\, \biggl[ U_\mu(x) \psi(x+\mu) - U_{\mu}^{\dagger}(x-\mu)
           \psi(x-\mu)\biggr] \non\\
 \Delta_\mu \psi(x)  &=&
  \, \biggl[ U_\mu(x) \psi(x+\mu) + U_{\mu}^{\dagger}(x-\mu) \psi(x-\mu)
                                       -2 \psi(x) \biggr]   \, .\non\\
\end{eqnarray}
Furthermore, $\sigma_{\mu,\nu}=\left \{ \gamma_{\mu},  \gamma_{\nu} \right \}$ and the field strength tensor is defined as 
\ber
&
F_{\mu\nu}(x)    = 
                  - \frac{i}{2}[Q_{\mu\nu} - Q_{\mu\nu}^\dagger]  \\
&
4 Q_{\mu\nu}(x)  = 
                   U_{\mu}(x) U_{\nu}(x+\hat\mu) U_{\mu}^\dagger(x+\hat\nu)
                   U_{\nu}^\dagger(x)  \nonumber +          \non \\
               &  U_{\nu}(x) U_{\mu}^\dagger(x-\hat\mu+\hat\nu)
                   U_{\nu}^\dagger(x-\hat\mu) U_{\mu}(x-\hat\mu) \nonumber +
                                                          \non \\
               &  U_{\mu}^\dagger(x-\hat\mu) U_{\nu}^\dagger(x-\hat\mu-\hat\nu)
                   U_{\mu}(x-\hat\mu-\hat\nu) U_{\nu}(x-\hat\nu) \nonumber +
                                                          \non \\
               &  U_{\nu}^\dagger(x-\hat\nu) U_{\mu}(x-\hat\nu)
                   U_{\nu}(x+\hat\mu-\hat\nu) U_{\mu}^\dagger(x)
                                                          \non . \\
&
\eer
The last term in the brackets in Eq. (\ref{sqw}) helps to suppress ${\cal O}(a)$ lattice artifacts and is called the clover term. 
Formulations with $c_{SW} \ne 0$ usually referred to as clover action. The standard Wilson action for fermions corresponds to $c_{SW}=0$.
In the lattice literature usually the form with the hopping parameter $\kappa=1/(2am_0+8)$ is used. In this form the fields need to be
accordingly normalized $\psi(x)\rightarrow\psi(x)a^{-3/2}/\sqrt{2 \kappa}$.

In \cite{Jakovac:2006sf}  the anisotropic Fermilab formulation is used for the heavy quarks.  
The anisotropy allows to use fine temporal lattice spacings without significant increase in the computational cost.
The study uses the quenched approximation and the standard Wilson action
in the gauge sector for which the relation between the bare $\xi_0$ and the renormalized anisotropy 
$\xi=a_s/a_t$ is known in a wide range of the gauge coupling $\beta=6/g^2$ \cite{Klassen:1998ua}. 

The anisotropic clover action \cite{Chen:2000ej} is
\ber
\displaystyle
S_q^{\xi}
 &=& \sum_{x} {\bar{\psi}}(x) \left[
      {m_0} +
      \nu_t  \!\not\! {D}^{\rm Wilson}_t +
      \frac{\nu_s}{\xi_0}  \sum_s \!\not\! {D}^{\rm Wilson}_s  \right. \non\\
\displaystyle
      &-& \left.\frac{1}{2} \left(
      C_{\rm sw}^t  \sum_{s} \sigma_{ts} {F}_{ts} +  
      \frac{C_{\rm sw}^s}{\xi_0}
      \sum_{s<s\prime} \sigma_{ss\prime} {F}_{ss\prime} \right) 
      \right]  {\psi}(x) .
\label{s_fermilab}
\eer
Because the lattice spacings in space and time directions are different the spatial and temporal Dirac operators as well
as the clover terms have different coefficients.
By tuning the clover coefficients according to the Fermilab prescription \cite{Chen:2000ej,ElKhadra:1996mp}
it is possible to reduce $O(a_t m_0)$ discretization errors. We will refer to this formulation as anisotropic
Fermilab formulation.

In the following we will show results from \cite{Jakovac:2006sf}, where anisotropic Fermilab formulation was used
as well as results from the study that used isotropic clover action with non-perturbatively determined values of $c_{SW}$ \cite{Datta:2003ww}.
In the study with anisotropic Fermilab action two values for the renormalized  anisotropy 
$\xi=2,~4$ as well as  $\beta=5.7,~5.9,~6.1,~6.5$ were used. 
These parameters correspond to temporal lattice spacings $a_t^{-1}=1.905-14.12$ GeV.  
To set the scale for the lattice spacing the traditional phenomenological value $r_0=0.5$ fm for the Sommer scale \cite{Sommer:1993ce} was used.
The Sommer scale $r_0$ has also been  calculated for anisotropic
Wilson action for $\beta=5.5-6.1$ \cite{EdwardsHellerKlassen:unpub}. 
Alternatively one can estimate the lattice spacing from the difference between the mass of $^1P_1$ state and
the spin averaged $1S$ mass: $\Delta M(^1P_1-\overline{1S})$. To a very good approximation this
mass difference  is not affected by fine and hyperfine splitting and thus is
not very sensitive to quenching errors.
It was found that close to the continuum limit  the lattice spacing determined from  $\Delta M(^1P_1-\overline{1S})$
is different from that determined from $r_0$ by $10\%$ \cite{Chen:2000ej,Okamoto:2001jb} if the phenomenological
value $r=0.5$ fm is used. Using the value of $r_0=0.469(7)$ determined in full QCD \cite{Gray:2005ur} would give a value for 
 $\Delta M(^1P_1-\overline{1S})$ splitting which is closer to the experimental one, however, the $\Delta M(\overline{2S}-\overline{1S})$ splitting
would be even further away from the experimental value \cite{Okamoto:2001jb}.
This problem is  due to the quenched approximation.

In the studies with isotropic clover action the following values of the lattice gauge couplings have been used 
$\beta=6.499,~6.640$ and $7.192$ \cite{Datta:2002ck,Datta:2003ww,Datta:2004js,Petreczky:2008px}.
These correspond to lattice spacings $a^{-1}=4.04,~4.86$ and $9.72$GeV respectively if the value $\sqrt{\sigma}=420$MeV for the
string tension is used.

The continuum meson current in Eq. (\ref{cont_current}), 
$J_H$ is related to the lattice current as
\begin{equation}
J_H=Z_H a_s^3 \bar \psi \Gamma_H \psi,
\end{equation}
where $\psi$ is the lattice quark field in Eq. (\ref{sqw}).
The renormalization constant $Z_H$ can be calculated in perturbation theory
or non-perturbatively. For isotropic clover action this has been done with both methods
(see Ref. \cite{Datta:2003ww} and references therein).

\subsection{Bayesian analysis of meson correlators}
\label{sec:bayes}
The obvious difficulty in the reconstruction of the spectral function from
Eq. (\ref{eq.kernel}) is the fact that the Euclidean correlator is calculated
only at ${\cal O}(10)$ data points on the lattice, while for a reasonable discretization
of the integral in Eq. (\ref{eq.kernel}) we need ${\cal O}(100)$ degrees of freedom. The problem can be solved using Bayesian analysis
of the correlator, where one looks for a spectral function which maximizes the 
conditional probability $P[\sigma|DH]$ of having the spectral function $\sigma$ given
the data $D$ and some prior knowledge $H$  (for  reviews see \cite{Asakawa:2000tr,Lepage:2001ym}).
Different Bayesian  methods differ in the choice of the prior knowledge.
One version of this analysis which is extensively used in the literature is the 
{\em Maximum Entropy Method} (MEM) \cite{Bryan:1990,Nakahara:1999vy}.
It has been used to study different correlation functions in Quantum Field Theory 
at zero and finite temperature
\cite{Umeda:2002vr,Asakawa:2003re,Datta:2003ww,Asakawa:2000tr,Nakahara:1999vy,Yamazaki:2001er,Karsch:2001uw,Karsch:2002wv,Datta:2002ck,Asakawa:2002xj,Blum:2004zp,Petreczky:2003iz,Petreczky:2001yp,Allton:2002mv,Langfeld:2001cz,Fiebig:2002sp,Sasaki:2005ap}.
In this method the basic prior knowledge is the positivity of the spectral function and 
the prior knowledge is given by the Shannon - Janes entropy  
$$
\displaystyle
\old{S=\int d \omega \biggl [ \sigma(\omega)-m(\omega)-m(\omega)
  \ln(\frac{\sigma(\omega)}{m(\omega)}) \biggr]. }
\addnew{S=\int d \omega \biggl [ \sigma(\omega)-m(\omega)-\sigma(\omega)
  \ln(\frac{\sigma(\omega)}{m(\omega)}) \biggr]. }
$$
The real function $m(\omega)$ is called the default model and parametrizes all additional prior knowledge about the
spectral functions, e.g. such as the asymptotic behavior at high energy  \cite{Asakawa:2000tr,Nakahara:1999vy}.
For this case the conditional probability becomes
\beq
 P[\sigma|DH]=\exp(-\frac{1}{2} \chi^2 + \alpha S),
\addnew{\label{eq:PDH}}
\eeq
with $\chi^2$ being the standard likelihood function and $\alpha$ a real parameter.
Previously in the MEM analysis of the meson spectral functions the Bryan's
algorithm was used \cite{Bryan:1990}.
A new algorithm was introduced in \cite{Jakovac:2006sf}.
It is worth to make connection between this method and the Bryan
algorithm. In both cases the true problem is
number-of-data dimensional -- in more dimensions the problem would be 
under-determined. To find the relevant subspace, the Bryan algorithm
uses singular value decomposition, while the new algorithm finds the same relevant
subspace by exact mathematical transformations. Although the method of
identifying the subspace is different, the result is the same, and in
both cases one proceeds with solving the original problem in this
restricted subspace. The advantage of the new algorithm is that 
it is more stable numerically when one reconstructs quarkonium spectral functions
at zero temperature. 

The comparison of the two algorithms was done for the pseudo-scalar spectral function for $\beta=6.1$, $\xi=4$ \cite{Jakovac:2006sf}. The problem with the
Bryan algorithm is that it does not work well for charmonium correlators  if the time
extent is sufficiently large, which is the case at low temperatures; the iterative procedure
does not always converge. For instance at $\beta=6.1$, $\xi=4$ and $16^3 \times 96$ lattice 
it was possible to get the spectral functions
using the Bryan algorithm only when using $\tau_{max}=24$ data points in the time direction.
With the new algorithm there is no restriction on $\tau_{max}$ which can be as large as $N_{\tau}/2$.

\subsection{Charmonium spectral functions at zero temperature}
\label{sec.t0spf}
 
\subsubsection{Pseudo-scalar and vector spectral functions at zero temperature}

\begin{figure}
\centering
\includegraphics[width=0.7\columnwidth]{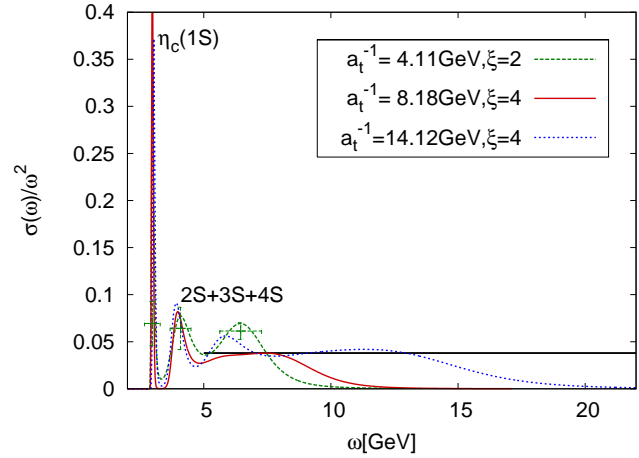}
\caption{The pseudo-scalar spectral function at zero temperature
for three finest lattice spacings. The horizontal line corresponds to the spectral function 
in the free massless limit at zero lattice spacing.}
\label{spf_ps_T0}
\end{figure}

In this subsection we discuss the results of Ref. \cite{Jakovac:2006sf} on
charmonium spectral functions obtained using MEM.
The zero temperature spectral functions for three different lattice 
spacings obtained by the new method from Ref. \cite{Jakovac:2006sf} are summarized in Fig.~\ref{spf_ps_T0}. Here the
simple default model $m(\omega)=1$ 
(in units of the spatial lattice spacing) is used.
To get a feeling for the statistical errors in the spectral functions
its mean value in some interval $I$ is calculated:
\begin{equation} 
\bar \sigma = \frac{\int_I d \omega \sigma(\omega)}{\int_I d \omega}.
\end{equation}
Then the error on $\bar \sigma$ is calculated using standard jackknife method.
These errors are shown in Fig.~\ref{spf_ps_T0}, where the length of the intervals are shown
as horizontal error bars.
As one can see from the figure, 
the $\eta_c(1S)$ can be identified very well. The second peak is likely
to correspond to excited states.
Because of the heavy quark mass the splitting between different
radial excitations is small and MEM cannot resolve different excitations
individually but rather produces a second broad peak to which all radial excitation
contribute. 
This can be seen from the fact that the amplitude of the second peak (i.e. the area under the peak)
is more than two times larger than the first one. Physical considerations tell us that it should 
be smaller than the first amplitude if it was a $2S$ state. When comparing amplitudes and 
peak positions from MEM analysis and from double exponential fits a very good agreement for the 
first peak and a fair agreement for the second peak are found. This gives confidence that at zero temperature charmonium 
properties can be reproduced
well with MEM.

For energies larger than $5$GeV one probably sees a continuum
in the spectral functions which is distorted by finite lattice spacing. In particular the
spectral function is zero above some energy which scales roughly as 
$a_s^{-1}$.
Note that for $\omega<5$GeV the spectral function does not depend 
on the lattice spacing.

\begin{figure}
\centering
\includegraphics[width=0.7\columnwidth]{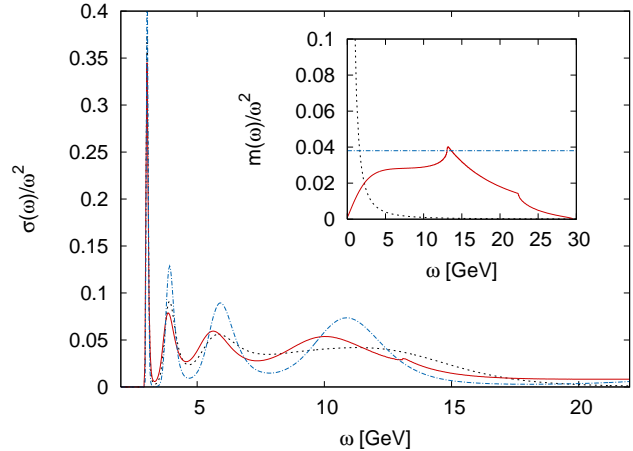} 
\caption{The default model dependence of the pseudo-scalar
spectral function at the finest lattice spacing ($\beta=6.5$).
In the inset the default models corresponding to different
spectral functions are shown.}
\label{spf_ps_dmdep}
\end{figure}

One should control how the result depends on
the default model. In Fig.~\ref{spf_ps_dmdep} we show the spectral
function for three different default models. One can see that
the default model dependence is significant only for $\omega > 5$ GeV.
This is not surprising as there are very few time slices which
are sensitive to the spectral functions at $\omega > 5$ GeV, while
the first peak is well determined by the large distance behavior
of the correlator. 

\begin{figure}
\centering
\includegraphics[width=0.7\columnwidth]{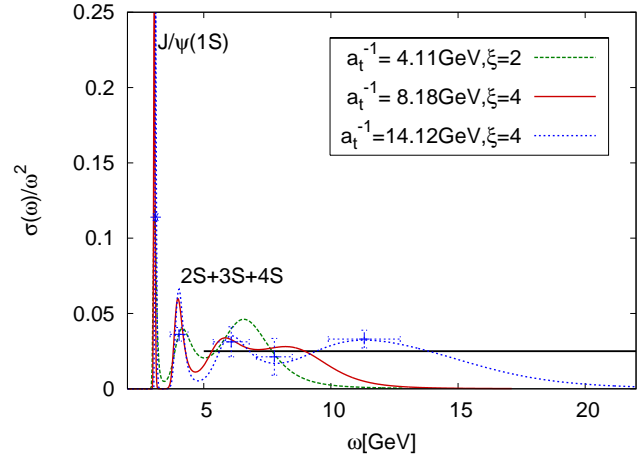}
\caption{The vector spectral function at zero temperature
for three finest lattice spacings. The horizontal line is the 
spectral function in the free massless limit.}
\label{spf_vc_T0}
\end{figure}

The spectral function in the vector channel defined
as
\be
\sigma_V(\omega)=\frac{1}{3} \sum_i \sigma_{ii} (\omega),
\ee
was also calculated and is shown 
 in Fig.~\ref{spf_vc_T0}
for the three finest lattice spacings. 
The conclusions which can be derived from this figure are
similar to the ones discussed above for the pseudo-scalar channel.
The first peak corresponds to the $J/\psi(1S)$ state, the second peak
most likely is a combination of $2S$ and higher excited states, finally
there is a continuum above $5$ GeV which is, however, distorted by
lattice artifacts. The similarity between the pseudo-scalar and vector
channel is, of course, expected. The lower lying states in theses channels
differ only by small hyperfine splitting. 

\subsubsection{Spectral functions for P states}

\begin{figure}
\centering
\includegraphics[width=0.7\columnwidth]{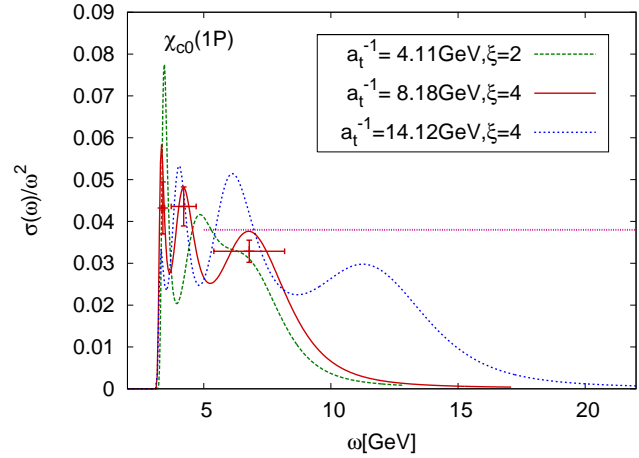}
\caption{The scalar spectral function at zero temperature
for three finest lattice spacings.}
\label{spf_sc_T0}
\end{figure}

The spectral functions in the scalar, axial-vector and
tensor channels which have the $1P$ charmonia  as the ground state were also calculated in Ref. \cite{Jakovac:2006sf}.
The scalar spectral functions reconstructed using MEM are
shown in Fig. \ref{spf_sc_T0}.
The first peak corresponds to $\chi_{c0}$ state, but it is
not resolved as well as the ground state in the pseudo-scalar
channel. 
This is due to the fact that the scalar correlator is considerably more noisy
than the pseudo-scalar or vector correlator.
This can be understood as follows. For the heavy quark mass the
contribution of the ground state in the scalar channel is suppressed
as $1/m^2$ relative to the ground state contribution in the 
pseudo-scalar and vector channels, and therefore it is considerably
smaller than the continuum contribution to the scalar correlator.
For the two finest lattice spacings there is a second peak
which may correspond to a combination of excited $P$ states.
Above $\omega>5$ GeV we see a continuum which is strongly distorted
by lattice artifacts and probably also by MEM.

\begin{figure}
\includegraphics[width=0.48\textwidth]{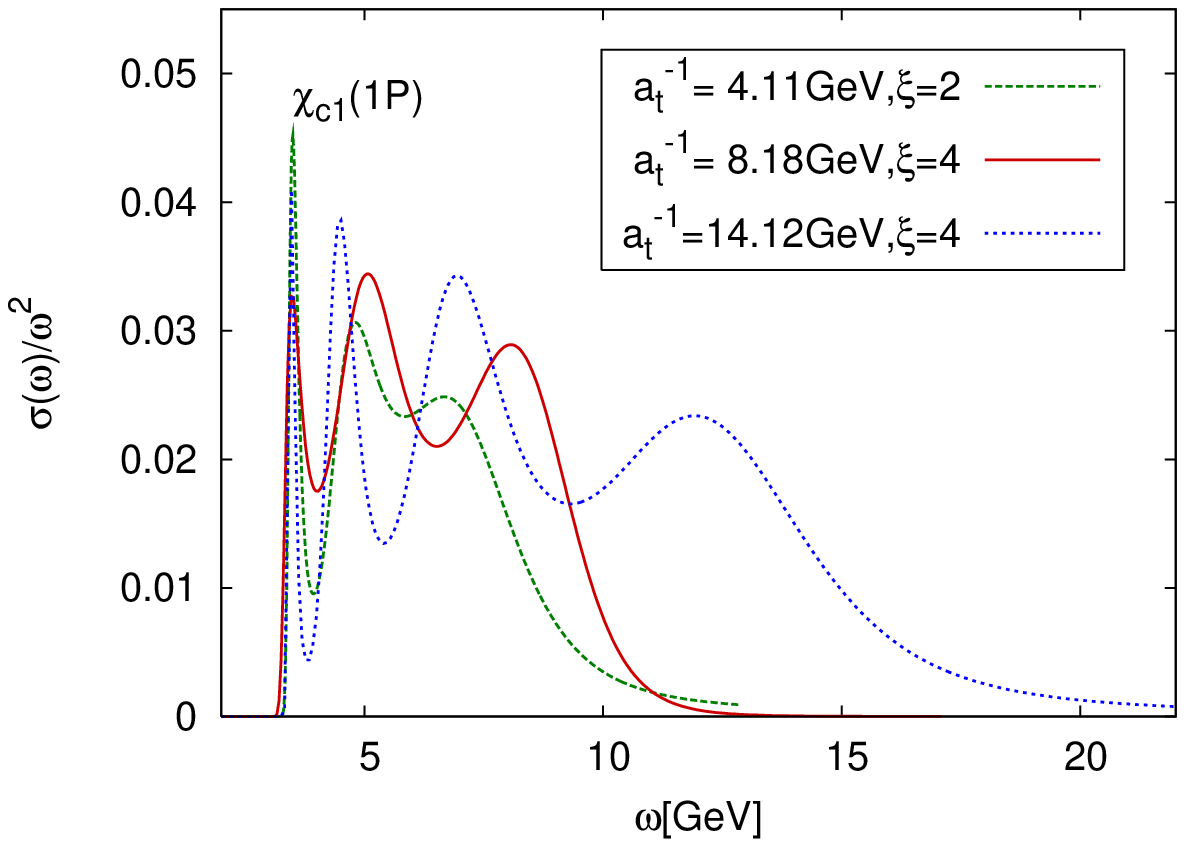}
\includegraphics[width=0.48\textwidth]{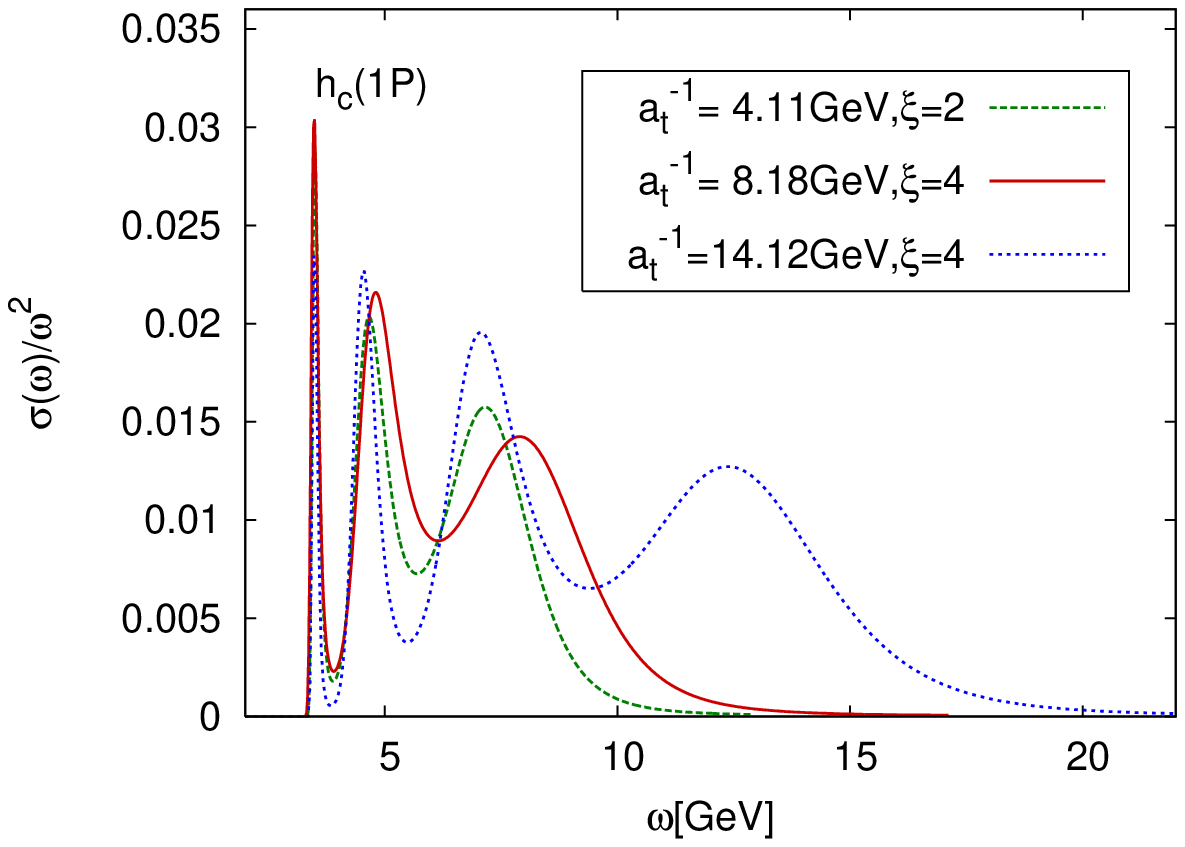}
\caption{The axial-vector (left) and tensor (right) spectral functions 
at zero temperature
for three lattice spacings.}
\label{spf_axt_T0}
\end{figure}

The spectral functions in the axial-vector
and tensor channels are shown in Fig. \ref{spf_axt_T0}.
They look similar to the scalar spectral functions. As in the scalar channel
the first peak is less pronounced than in the case of $S$-wave 
charmonium spectral functions, and it corresponds to $\chi_{c1}$ and 
$h_c$ state,  respectively. The continuum part of the spectral function is
again strongly distorted. 

\subsection{Charmonium correlators at finite temperature}

The reconstruction
of the spectral functions from lattice correlators is difficult
already at zero temperatures. At finite temperature it is even more 
difficult 
to control the systematic errors in the spectral functions reconstructed
from MEM. 
This is because with increasing temperature the maximal time extent
$\tau_{max}$ is decreasing as $1/T$. Also the number of data points available for the analysis
becomes smaller. 
Therefore other  methods which can give
some information about the change of the spectral functions as
the temperature is increasing are often used. The temperature dependence of the
spectral function will manifest itself in the temperature dependence
of the lattice correlator $G(\tau,T)$. Looking at Eq. (\ref{eq.kernel})
it is easy to see that the temperature dependence of $G(\tau,T)$ comes from the temperature
dependence of the spectral function $\sigma(\omega,T)$ and the 
temperature dependence of the kernel $K(\tau,\omega,T)$. To
separate out the trivial temperature dependence due to $K(\tau,\omega,T)$
one calculates the reconstructed correlator\cite{Datta:2003ww} 
\begin{equation}
G_{recon}(\tau,T)=\int_0^{\infty} d \omega \sigma(\omega,T=0) K(\tau,\omega,T).
\end{equation}
If the spectral function does not change with increasing temperature 
we expect $G(\tau,T)/G_{recon}(\tau,T)=1$. In Ref. \cite{Jakovac:2006sf} an extensive
study of the temperature dependence of this ratio for different channels
at different lattice spacings was carried out. 

\subsubsection{The pseudo-scalar correlators}

\begin{figure}
\centering
\includegraphics[width=0.9\columnwidth]{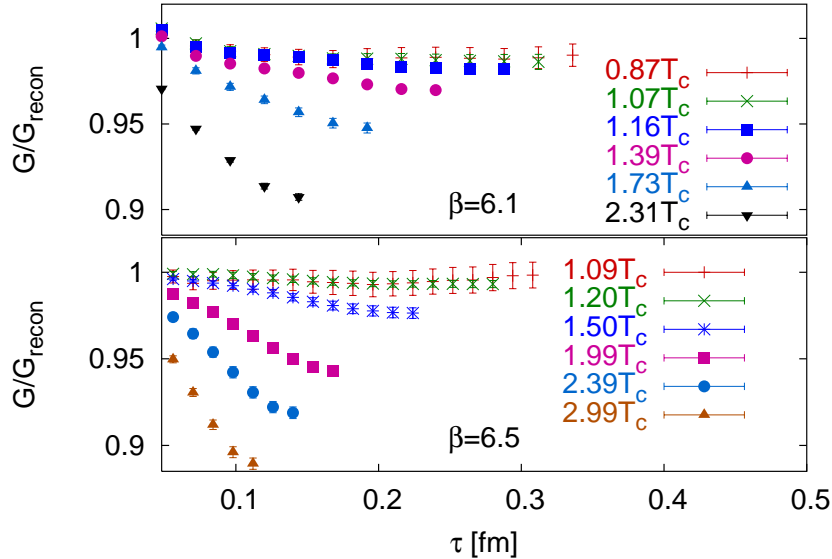} 
\caption{The ratio $G/G_{recon}$ for the pseudo-scalar channel
for the two finer $\xi=4$ lattices.}
\label{psxi4}
\end{figure}

First we present results from Ref. \cite{Jakovac:2006sf} for the temperature dependence of the pseudo-scalar
correlators. In Fig. \ref{psxi4} we show numerical results for
$G/G_{recon}$ on lattices with $\xi=4$. 
We see almost no change in the pseudo-scalar correlator till temperatures
as high as $1.2T_c$. The temperature dependence of the pseudo-scalar correlator
remains small for temperatures below $1.5T_c$. 
Medium modifications of the correlator slowly
turn on as we increase the temperature above this value.
From the figures it is clear that the temperature dependence of 
the correlators is not affected significantly by the finite lattice spacing. 
The very small temperature dependence of the pseudo-scalar correlator suggests
that the corresponding ground state $\eta_c(1S)$ may survive till
temperatures as high a $1.5T_c$. The temperature dependence of the
correlator found in this study is similar to that of Ref. \cite{Datta:2003ww}, where
isotropic lattices with very small lattice spacings, $a^{-1}=4.86,9.72$ GeV
have been used. 
However, at temperatures higher than $1.5T_c$ the deviations of $G/G_{recon}$
from unity  become slightly larger than those found in Ref. \cite{Datta:2003ww}. This is possibly due to the fact
that cutoff effects  are more important at higher temperatures. 
Thus despite similarities of the temperature dependence of the pseudo-scalar
correlator to findings of Ref. \cite{Datta:2003ww} there are quantitative differences.
One should note, however, statistical errors and systematic uncertainties are larger
in the analysis presented in Ref. \cite{Datta:2003ww} than in Ref. \cite{Jakovac:2006sf}.
In the study on isotropic lattices the  ratio $G/G_{recon}$ 
starts to depend more strongly on the temperature only around $3T_c$ \cite{Datta:2003ww}. 

\subsubsection{The P-wave correlators}

\begin{figure}
\centering
\includegraphics[width=0.9\columnwidth]{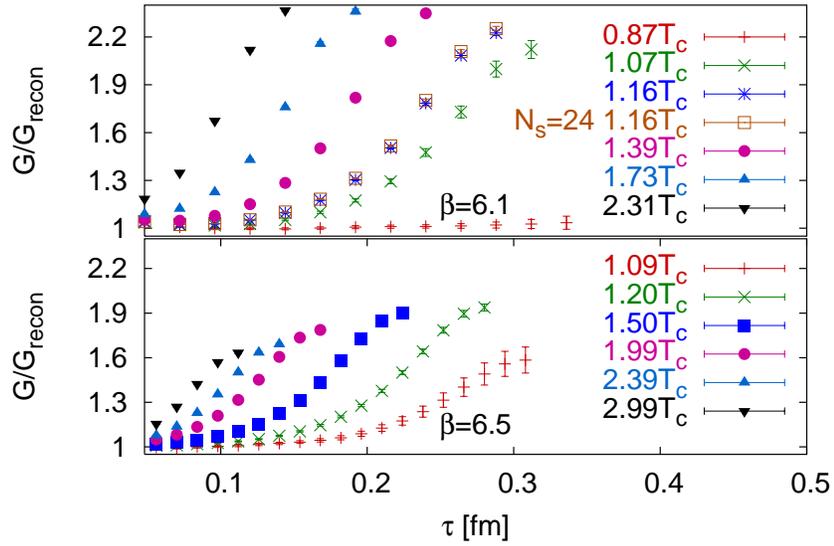}
\caption{The ratio $G/G_{recon}$ for the scalar channel
for the two finer $\xi=4$ lattices.}
\label{scxi4}
\end{figure}

Next we present the temperature dependence of
the scalar, axial-vector and tensor correlators corresponding to $P$-states. 
The $\tau$ dependence of the scalar correlator was studied on $\xi=2,4$ lattices \cite{Jakovac:2006sf}.
The numerical results on fine
lattices with $\xi=4$ are shown in Fig. \ref{scxi4}. 
We see some differences in $G/G_{recon}$ calculated at $\beta=6.1$ and
$\beta=6.5$.  Thus the cutoff dependence of $G/G_{recon}$  is larger
in the scalar channel than in the pseudo-scalar one. For $\beta=6.1$ and $\xi=4$
the calculations were done on $24^3 \times 24$ lattice to check finite volume effects.
The corresponding results are shown in Fig. \ref{scxi4} indicating that the finite
volume effects are small.
On the finest lattice
the enhancement of the scalar correlator is very similar to that
found in calculations done on isotropic lattices \cite{Datta:2003ww}, but small
quantitative differences can be identified.

\begin{figure}
\centering
\includegraphics[width=0.9\columnwidth]{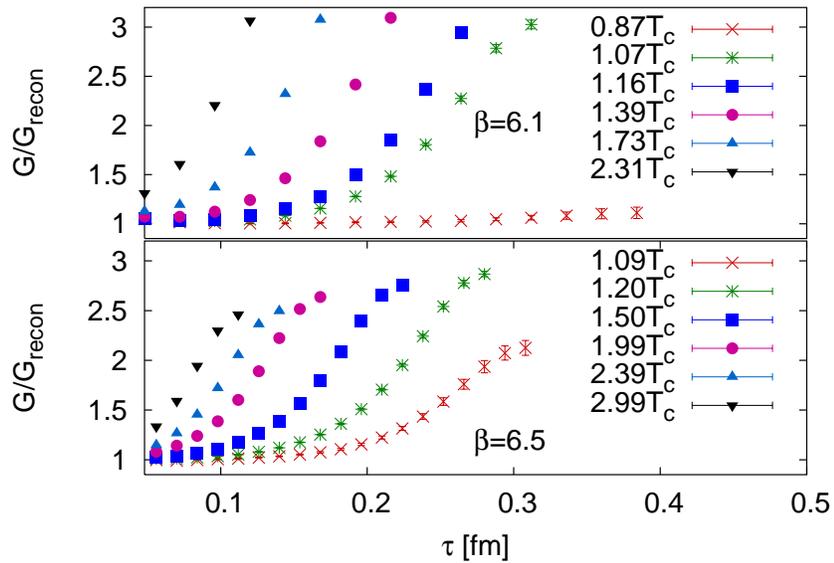}
\caption{The ratio $G/G_{recon}$ for the axial-vector channel
for $\xi=4$ lattices.}
\label{axxi4}
\end{figure}

\begin{figure}
\centering
\includegraphics[width=0.9\columnwidth]{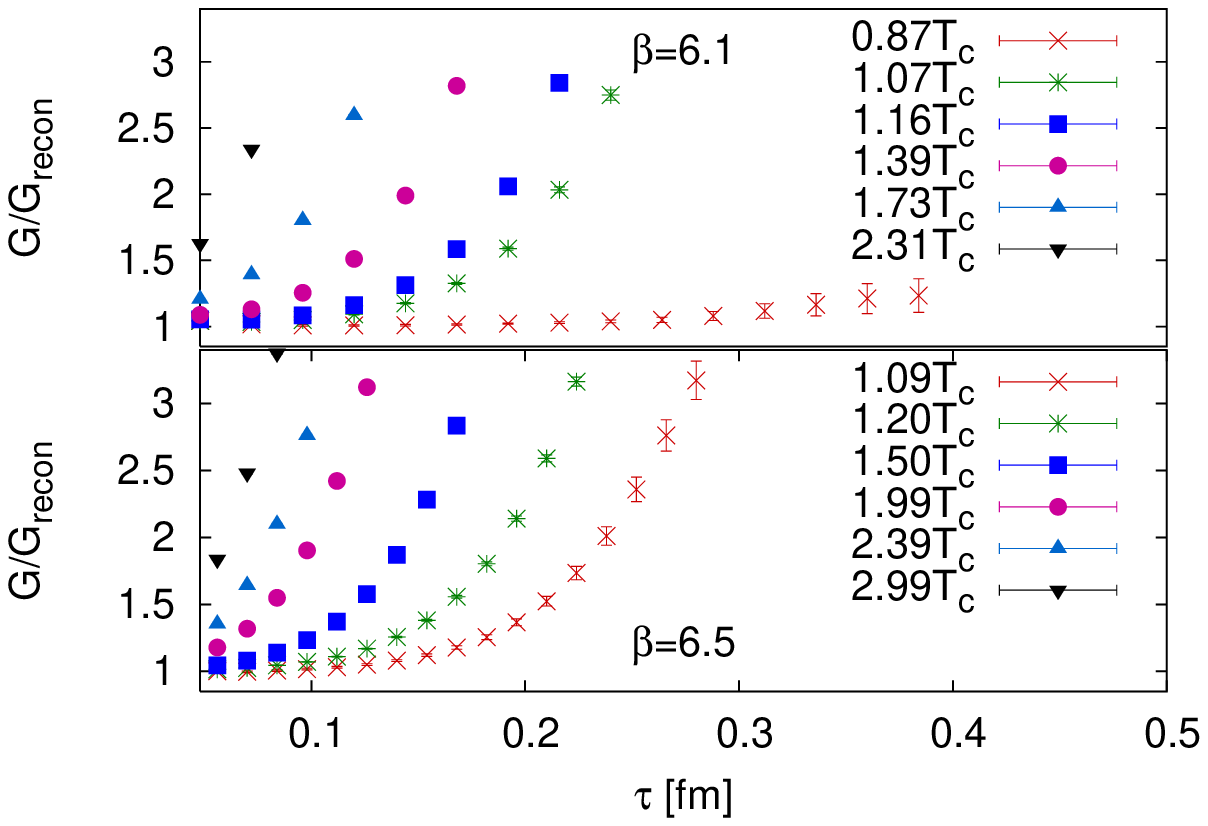}
\caption{The ratio $G/G_{recon}$ for the tensor channel
for $\xi=4$ lattices.}
\label{b1xi4}
\end{figure}

In Figs. \ref{axxi4} and \ref{b1xi4} we show the temperature dependence of the
axial-vector and tensor correlators respectively for $\xi=4$.
Qualitatively their behavior is very similar to the scalar correlator but the enhancement
over the zero temperature result is larger. The results for the axial-vector correlators 
again are very similar to those published in Ref. \cite{Datta:2003ww}.
The difference in $G/G_{recon}$ calculated  at $\beta=6.1$ and $\beta=6.5$ are
smaller than in the scalar channel.

The large increase in the scalar, axial-vector and tensor correlators 
may be interpreted as indicator of strong modification of the corresponding spectral function and, possibly the dissolution of $1P$ charmonia
states. However, as we will see in the next sections the situation is more complicated. 
It has been noticed in \cite{Umeda:2007hy} that the increase in $G/G_{recon}$ may be due to the zero mode contribution.

\subsubsection{The vector correlator}

\begin{figure}
\centering
\includegraphics[width=0.9\columnwidth]{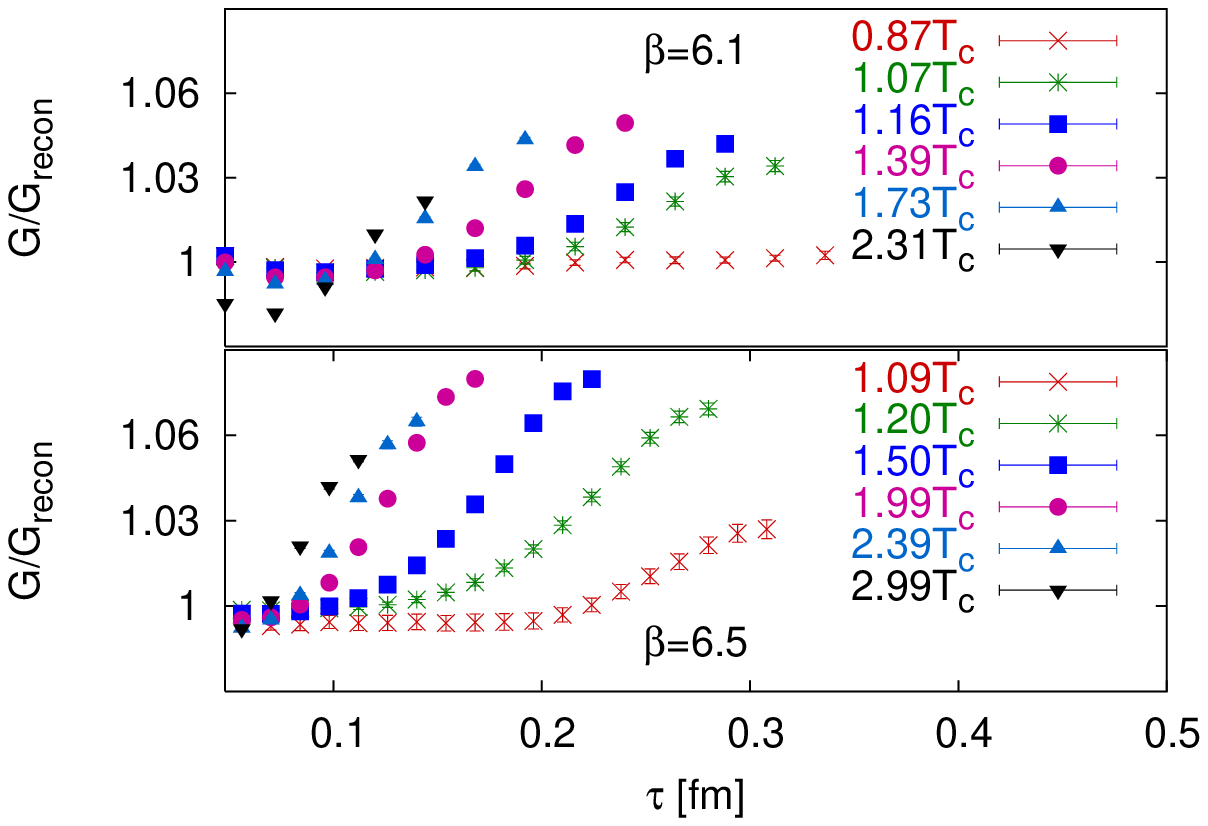}
\caption{The ratio $G/G_{recon}$ for the vector channel
for the two finer $\xi=4$ lattices.}
\label{vcxi4}
\end{figure}

The numerical results for the vector correlator of Ref. \cite{Jakovac:2006sf} are shown
in Fig. \ref{vcxi4}
for $\xi=4$. As one can see from the figures the temperature dependence of $G/G_{rec}$ is
different from the pseudo-scalar case and this ratio is larger than unity for all lattice spacings.
Similar results have been obtained on isotropic lattices \cite{Datta:2004js}. The enhancement of the vector correlator is due to the
presence of the transport contribution in the spectral function \cite{Mocsy:2005qw,Petreczky:2005nh} 
and will be discussed in section \ref{sec_zero_modes}.
Since the vector current is conserved the vector correlator also caries information about the transport of heavy quarks in the plasma. 
This shows up as a peek in the spectral function at zero energy, leading to the observed enhancement in $G/G_{recon}$.

\subsection{Charmonium spectral functions at finite temperature}
\begin{figure}
\centering
\includegraphics[width=0.7\columnwidth]{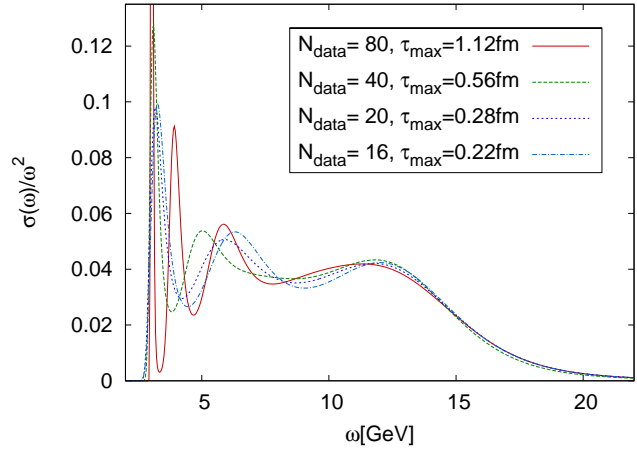}
\caption{The dependence of the reconstructed pseudo-scalar spectral function on the maximal
temporal extent for $\beta=6.5$. In the analysis the default model $m(\omega)=1$ has been
used.
}
\label{spf_ps_ntdep}
\end{figure}

In section \ref{sec.t0spf} we have seen that using MEM one can 
reconstruct well the main features of the spectral function, in particular
the ground state properties. At finite temperature the situation becomes
worse because the temporal extent is decreasing. The maximal time 
separation is $\tau_{max}=1/(2T)$. As a consequence 
it is no longer possible to isolate the ground state  well. 
Also the number of available data points becomes smaller.  While the later 
limitation can be overcome by using smaller and smaller lattice spacings in time direction
the former limitation is always present. Therefore we should investigate 
systematic effects due to limited extent of the temporal direction. It appears
that the pseudo-scalar channel is the most suitable case for this investigation
as at zero temperature it is well under control and there is no contribution from
heavy quark transport.
To estimate the effect of limited temporal extent in Ref. \cite{Jakovac:2006sf} the 
spectral functions at zero temperature is calculated considering only the first $N_{data}$ 
time-slices in the analysis for $\beta=6.5$, $\xi=4$. 
The result of this calculation is shown in Fig. \ref{spf_ps_ntdep}
where $N_{data}=80,~40,~20$ and $16$.  The last
two values correspond to the finite temperature lattices in the deconfined phase.
In this case we see the first peak quite clearly.
As one can see from the figure already for $N_{data}=40$ and $\tau_{max}=0.56$ fm
the second peak corresponding to radial excitation is no longer visible and the first peak
becomes significantly broader. The position of the first peak, however, is unchanged.
As the number of data points is further decreased ($N_{data}=20,16$) we see further broadening of the first peak
and a small shift of the peak position to higher energies.   
These systematic effects should be taken into account when analyzing the spectral functions
at finite temperature. Therefore when studying the spectral functions at finite temperature
we always compare with the zero temperature spectral functions reconstructed 
with the same number of data points and $\tau_{max}$ as available at that temperature.

\begin{figure}
\centering
\includegraphics[width=0.67\columnwidth]{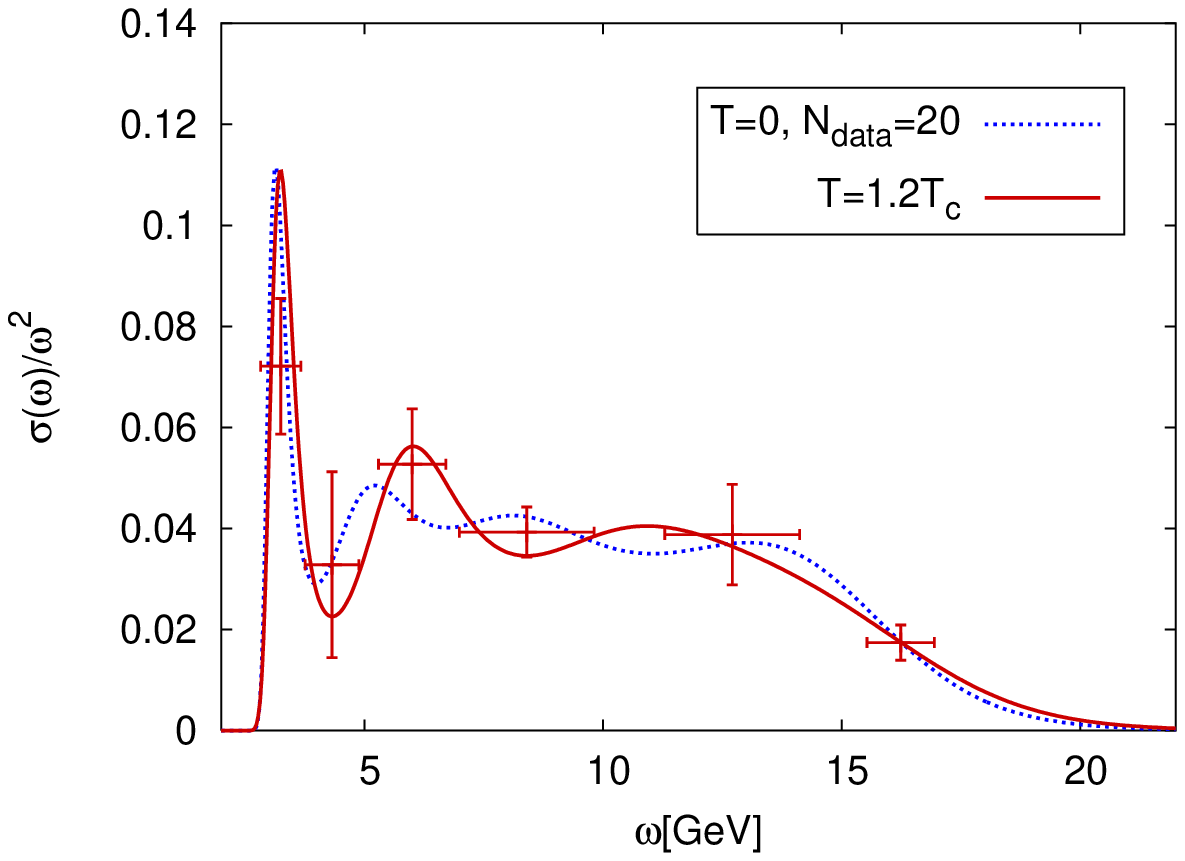}
\includegraphics[width=0.67\columnwidth]{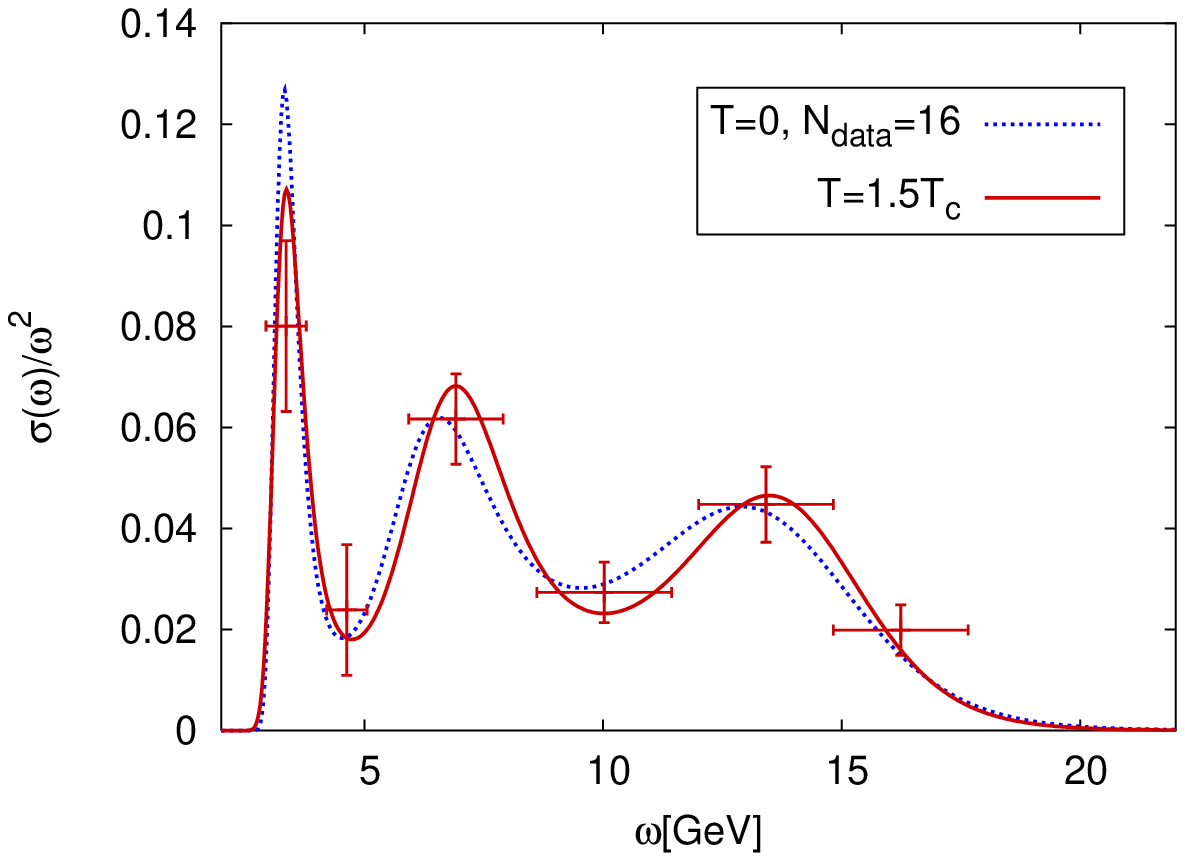}
\includegraphics[width=0.67\columnwidth]{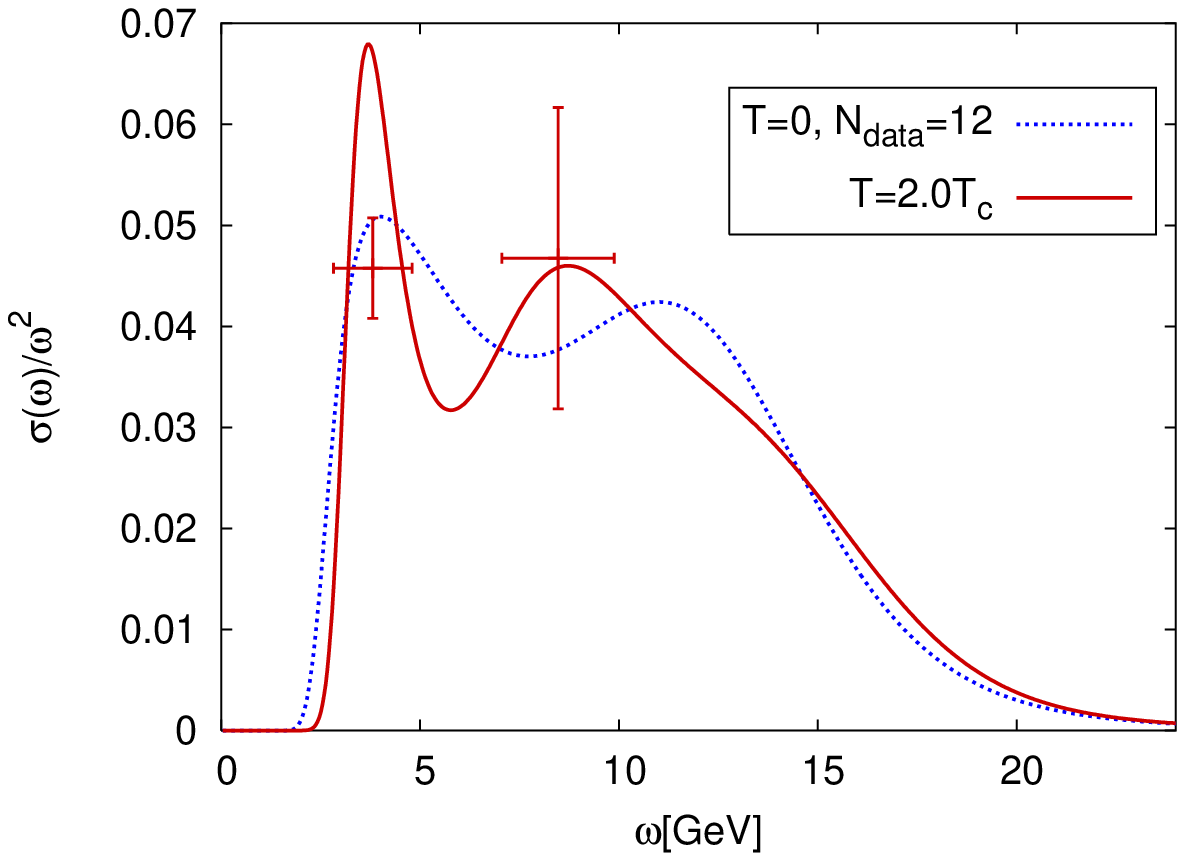}
\caption{The pseudo-scalar spectral function
for $\beta=6.5$ and $N_t=40,~32,20$ corresponding to temperatures
$1.2T_c,~1.5T_c$ and $2.0T_c$. 
In the analysis the default model $m(\omega)=0.01$ has been
used for $1.2T_c$ and $1.5T_c$, while for $2.0Tc$ we used $m(\omega)=1$.
}
\label{spf_ps_Tne0}
\end{figure}

\begin{figure*}
\includegraphics[width=0.48\columnwidth]{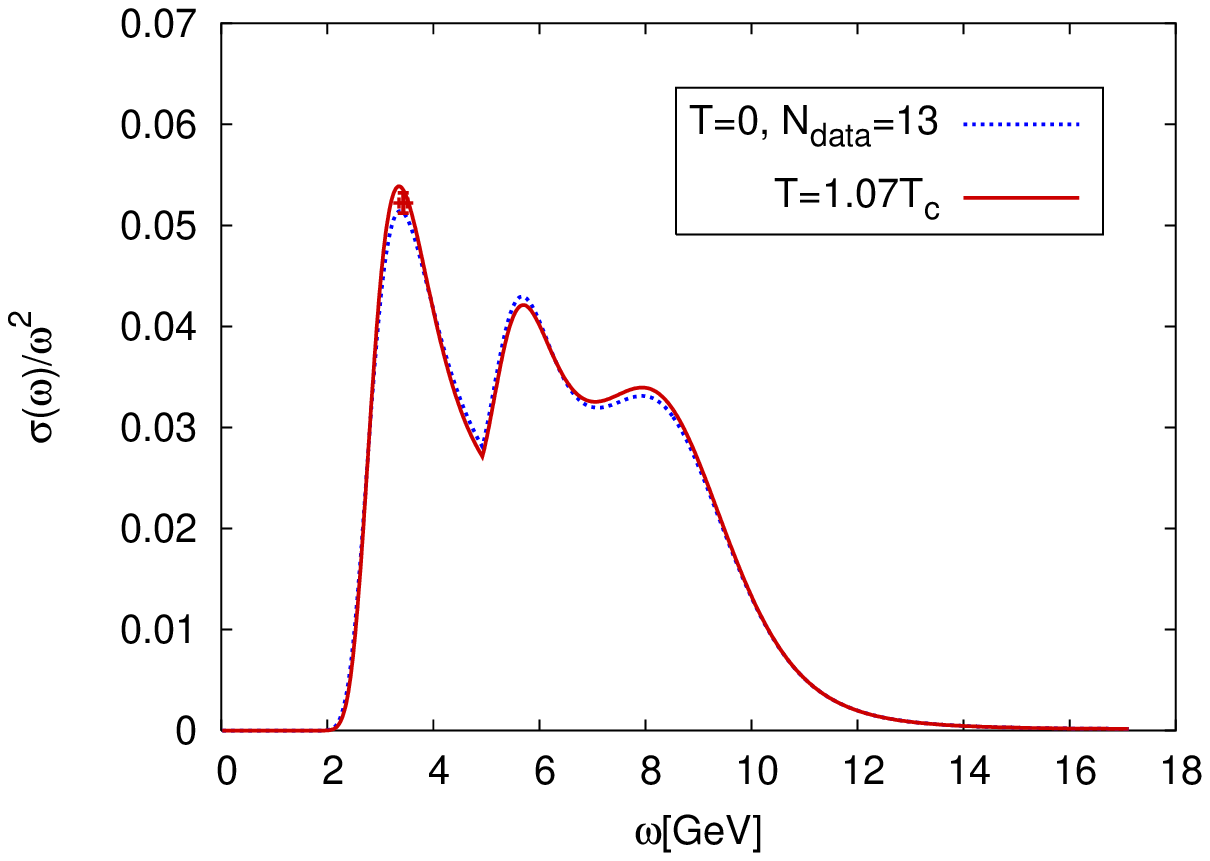}\hfill
\includegraphics[width=0.48\columnwidth]{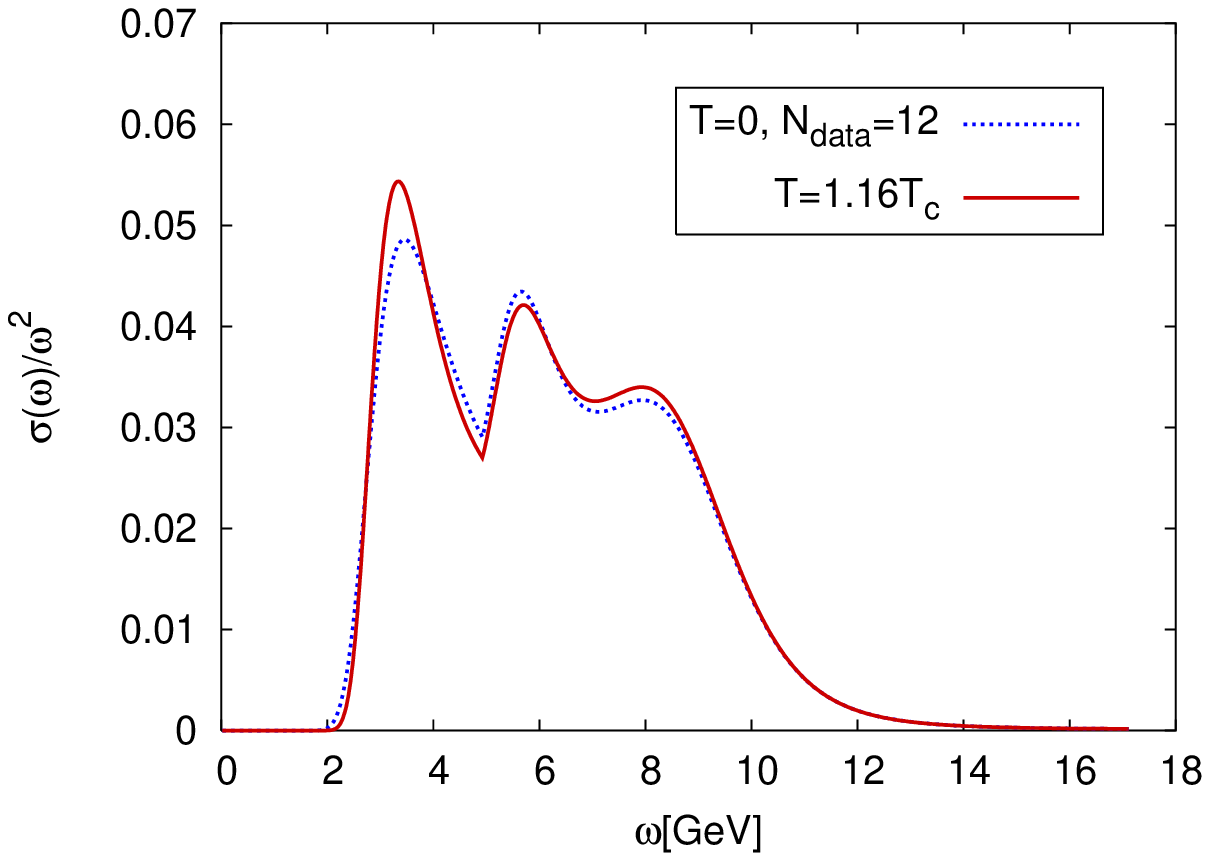}
\includegraphics[width=0.48\columnwidth]{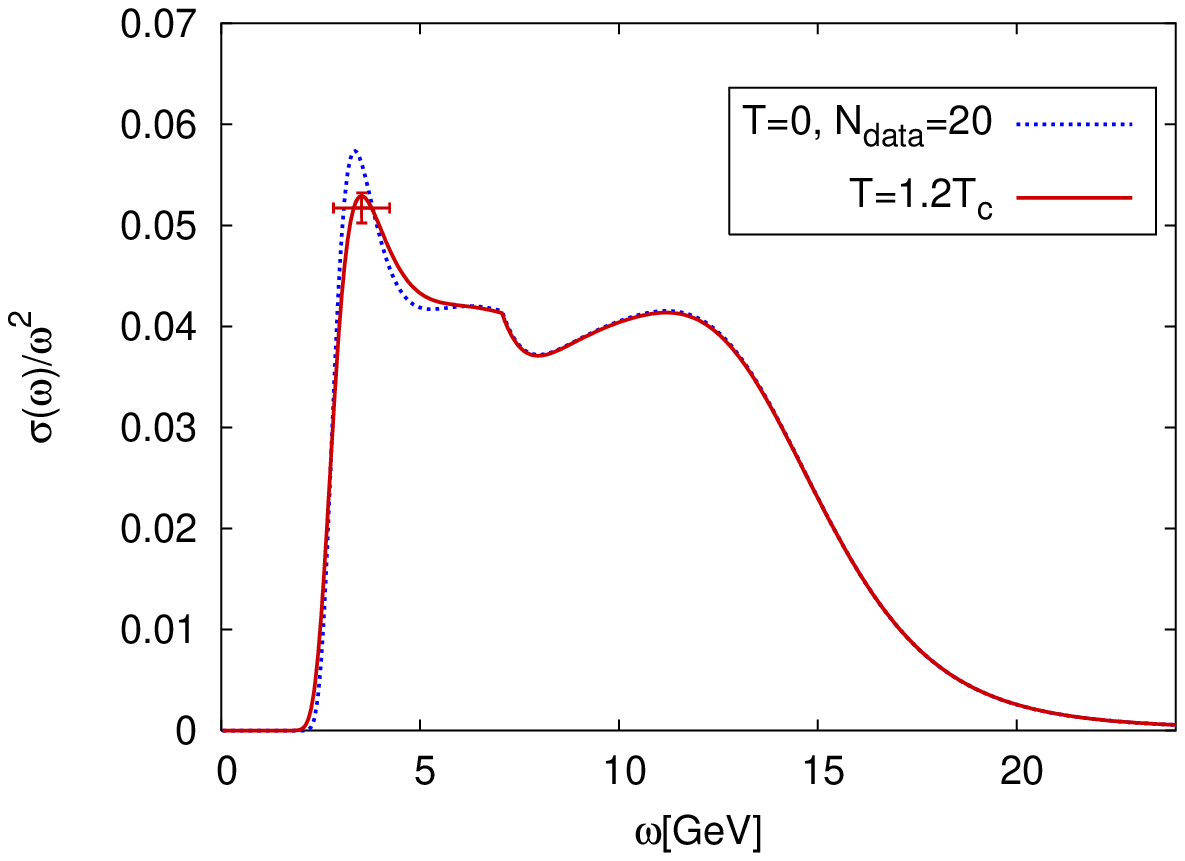}\hfill
\includegraphics[width=0.48\columnwidth]{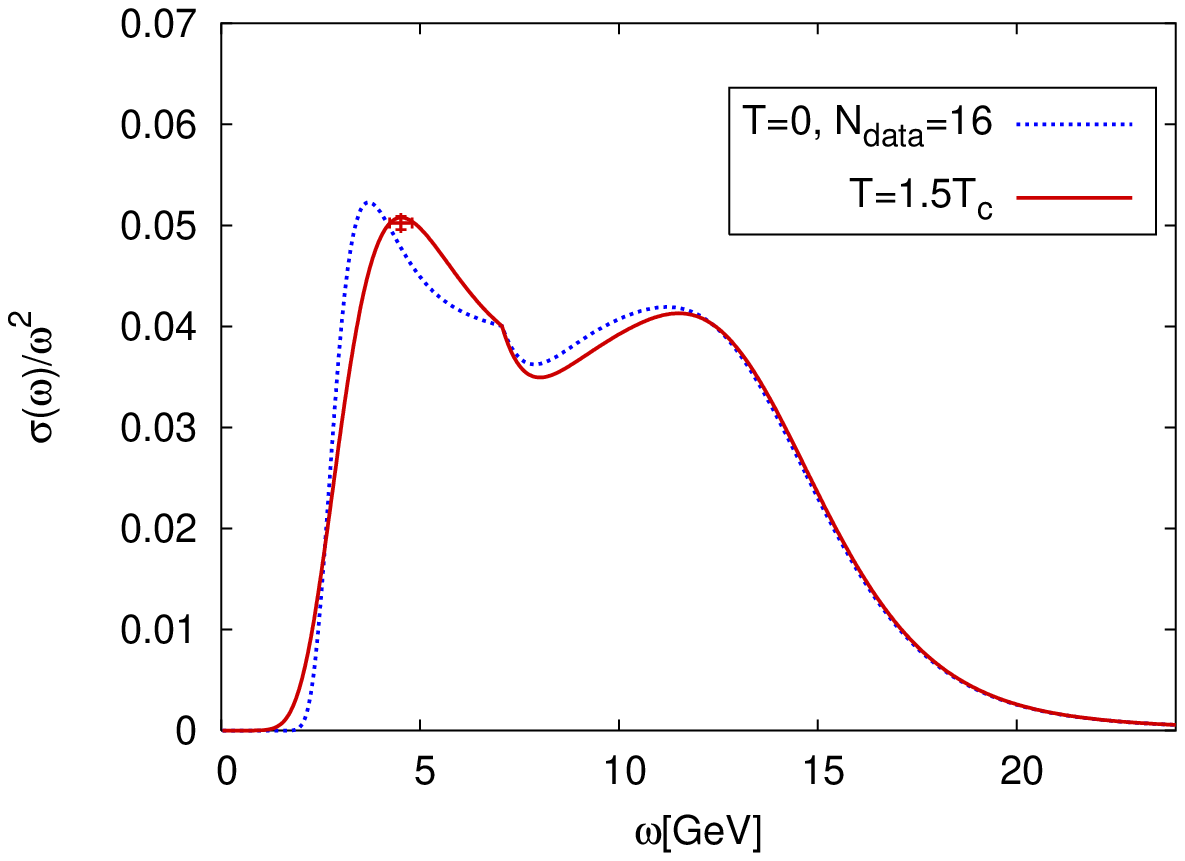}
\caption{
The pseudo-scalar spectral function at different temperatures together with the zero temperature spectral functions 
reconstructed using 
default model coming from the high energy part of the zero temperature spectral function. 
}    
\label{fig:psdefamax} 
\end{figure*}

In Fig. \ref{spf_ps_Tne0} the spectral functions at different temperatures are shown
together with the zero temperature spectral functions\cite{Jakovac:2006sf}. As a default model
for $T=1.2T_c$ and $T=1.5T_c$ $m(\omega)=0.01$ is used. For $T=2.0T_c$   $m(\omega)=1$ is used since the use of  $m(\omega)=0.01$ 
resulted in numerical problems in the MEM analysis.
The figure shows that the pseudo-scalar spectral function is not modified till $1.5T_c$ within the
errors of the calculations.
This is consistent with the conclusions of Ref. \cite{Asakawa:2003re,Datta:2003ww}.
One should note, however, that it is difficult to make any conclusive statement based on the shape
of the spectral functions as this was done in the above mentioned works. The dependence of
the reconstructed spectral functions on the default model $m(\omega)$ is much stronger
at finite temperature. The spectral functions were reconstructed using different types of default models.
For all 
temperatures  $T \le 1.5T_c$ the difference between the finite temperature spectral function
and the zero temperature one is very small compared to the statistical errors for all default model considered here. 
In particular a use is made of the default models constructed from the high energy part
of the lattice spectral functions calculated at zero temperature as this was done in Ref. \cite{Datta:2003ww}.
The idea is that at sufficiently high energy the spectral function is dominated by the continuum and
is temperature independent. Therefore it is suitable to provide the prior knowledge, i.e. the default model. 
With this
default model the spectral functions were calculated at $T=(1.07-1.5) T_c$ 
\cite{Jakovac:2006sf}. Very
little temperature dependence of the spectral functions was found, see Fig. \ref{fig:psdefamax}.
Note, however, that for this choice of the default model no clear peak can be identified  in the spectral functions 
if $T \ge 1.2T_c$.

\begin{figure}
\centering
\includegraphics[width=0.7\columnwidth]{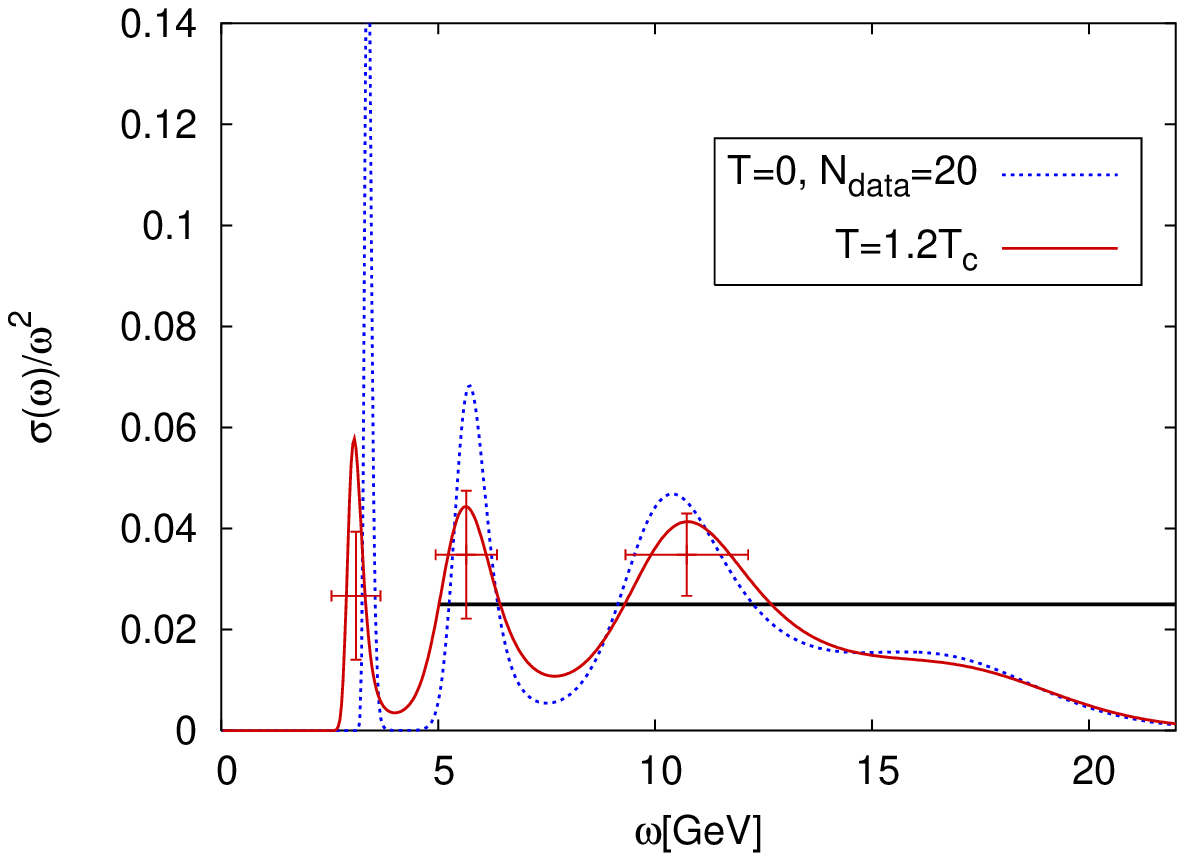}
\includegraphics[width=0.7\columnwidth]{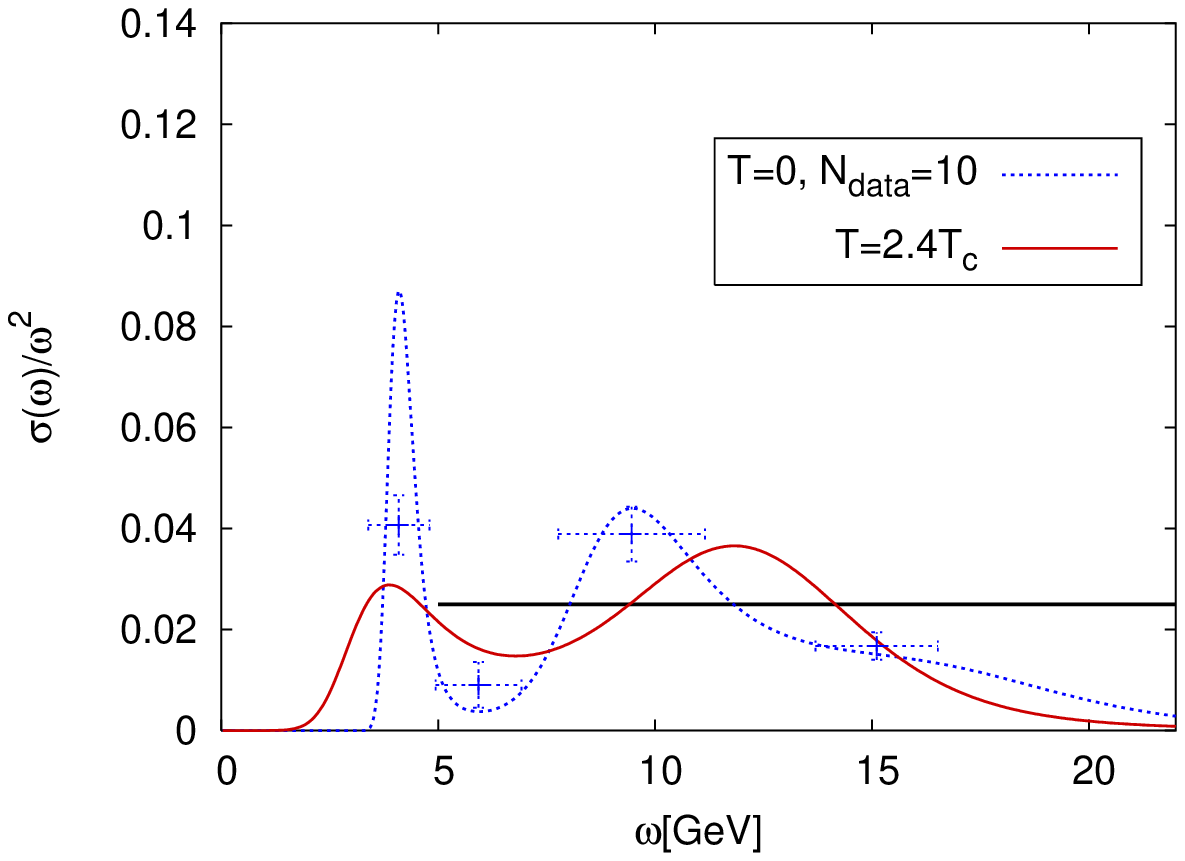}
\caption{The vector spectral function
for $\beta=6.5$ and $N_t=40,~20$ corresponding to temperatures
$1.2T_c$ and $2.4T_c$. 
In the analysis the default model $m(\omega)=0.01$ has been
used.
}
\label{spf_vc_Tne0}
\end{figure}

The spectral function in the vector channel is also calculated\cite{Jakovac:2006sf}. The results are shown in Fig. \ref{spf_vc_Tne0}
for the default model $m(\omega)=0.01$ .
As this was already discussed in the previous section the basic difference between the pseudo-scalar
and vector spectral functions at finite temperature is the presence of the transport peak at $\omega\simeq 0$.
The difference of the temperature dependence of the vector and pseudo-scalar correlators is consistent
with this assumption. The vector spectral function reconstructed with MEM shows no evidence of the transport
peak at $\omega \simeq 0$. On the other hand the spectral function at $1.2T_c$ differs from the zero temperature
spectral function, in particular the first peak is shifted to smaller $\omega$ values. We believe that 
this is a problem of the MEM analysis
which cannot resolve the peak at $\omega \simeq 0$ but instead mimics its effect by shifting the $J/\psi$ peak to smaller
$\omega$.  Also at $2.4T_c$ the spectral function extends to smaller $\omega$ values than in the pseudo-scalar
correlator which again indicates some structure at $\omega \simeq 0$. The analysis of the vector spectral functions
using other choices for the default model always indicates that the spectral functions at finite temperature differs from
the zero temperature spectral functions and extend to significantly smaller $\omega$ values.

\subsection{Charmonium correlators and spectral functions at finite momenta}
\begin{figure}[ht]
\includegraphics[width=0.49\textwidth]{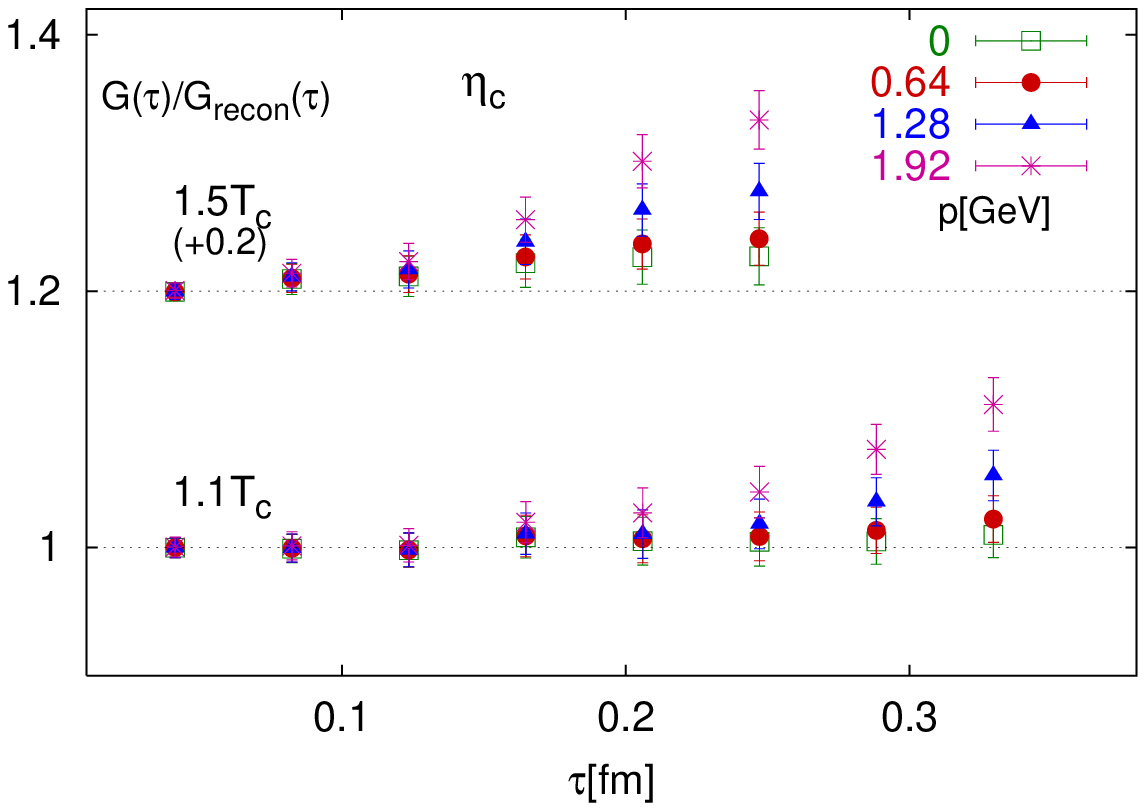}
\includegraphics[width=0.49\textwidth]{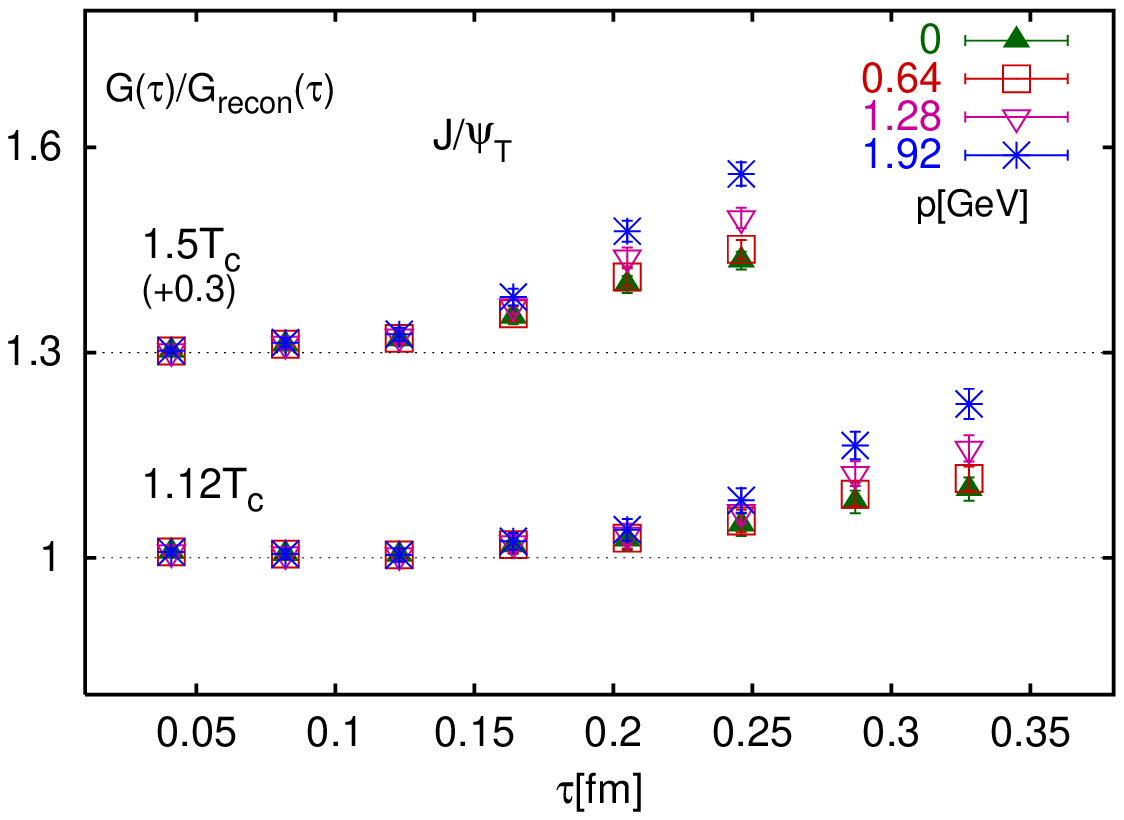}
\caption{Comparison of correlators at 1.1 $\tc$ and 1.5
$\tc$ for pseudoscalar and (transverse) vector charmonia
with those reconstructed from spectral function at 0.75
$\tc$.}
\label{fig:fin_mom}
\end{figure}

So far we reviewed charmonia at zero spatial momentum, i.e. charmonia at rest in the
heatbath's rest frame. It is certainly of interest to study the temperature dependence of
correlators and spectral functions at non-zero spatial momentum. Such a study has been
done using isotropic lattices with lattice spacing $a^{-1}=4.86$GeV and $9.72$ GeV \cite{Datta:2004js}.
It has been found that the pseudo-scalar correlators are enhanced compared to the zero temperature
correlators for non-vanishing spatial momenta, see Fig. \ref{fig:fin_mom}. Furthermore, the enhancement of vector correlator at
finite spatial momentum is larger than at zero spatial momentum.

In Ref. \cite{Jakovac:2006sf}  the finite momentum pseudo-scalar correlators  are calculated on anisotropic lattices at
$\beta=6.1$, $\xi=4$ for different temperatures.  The differences in $G/G_{recon}$ calculated in this work and in Refs. \cite{Datta:2003ww,Datta:2004js} are
present already at zero momentum and are presumably due to finite lattice spacing errors. Apart from this 
the momentum dependence of the pseudo-scalar correlators is similar to the findings of Refs.  \cite{Datta:2003ww,Datta:2004js}.
It would be interesting to see if the difference in the temperature
dependence of the correlators at zero and finite spatial momenta is due to a contribution to the spectral functions
below the light cone at finite temperature \cite{Karsch:2003wy,Aarts:2005hg}.

\subsection{Bottomonium spectral functions at zero temperature}

The use of Fermilab formulation described in the previous sections allows for a study of bottomonium for the same
range  of lattice spacings. Usually bottomonium is studied using lattice NRQCD (see e.g Ref. \cite{Davies:1994mp}).
First study of bottomonium within the relativistic framework
was presented in Ref. \cite{Liao:2001yh}. More recently it was studied in \cite{Jakovac:2006sf}, where
as before the lattice spacing is fixed by the Sommer scale $r_0$, assuming
$r_0=0.5$fm.  Note that the lattice spacings determined from $r_0$ are about $20\%$ smaller than in Ref. \cite{Liao:2001yh}
where it was determined from bottomonium $^1P_1-\overline{1S}$ mass splitting. As the result the new estimates of the
$\Upsilon$ mass are smaller than those in Ref. \cite{Liao:2001yh}. However, one finds good  agreement if  the values of
the lattice spacing quoted in Ref. \cite{Liao:2001yh} are used to calculate the physical masses.

\begin{figure}
\centering
\includegraphics[width=0.7\columnwidth]{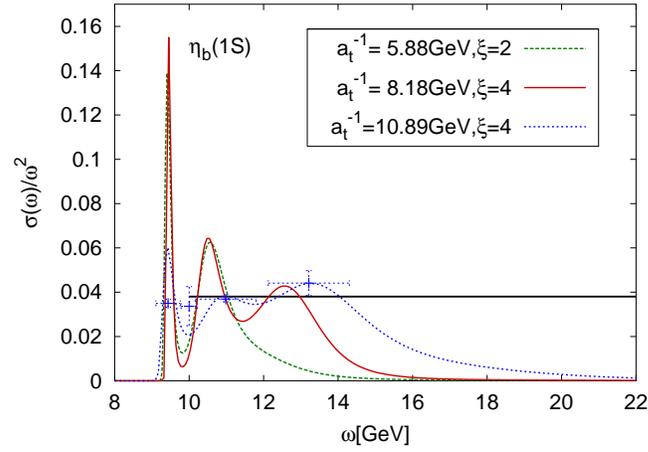}
\caption{The pseudo-scalar bottomonium spectral function at zero temperature for
different lattice spacings.}
\label{spf_ps_bot}
\end{figure}

Using MEM the spectral functions in different channels for three lattice spacings were analyzed.
In Fig. \ref{spf_ps_bot} we show the spectral functions in the 
pseudo-scalar channel. Since the physical quark mass is different at different lattice spacings
the horizontal scale was shifted by the difference of the calculated $\Upsilon$-mass and the 
corresponding experimental value. We can see that the first peak in the spectral function corresponds to
the $\eta_b(1S)$ state and its position is independent of the lattice spacing. The remaining details
of the spectral functions are cut-off dependent and we cannot distinguish the excited states
from the continuum. 
The position and the amplitude of the first peak in the spectral functions is in good agreement with
the results of simple exponential fit. As in the charmonium case the maximal energy $\omega_{max}$ for which the spectral function is 
non-zero scales approximately as $a_s^{-1}$.  Similar results have been obtained in the vector channel.

\begin{figure}
\centering
\includegraphics[width=0.7\columnwidth]{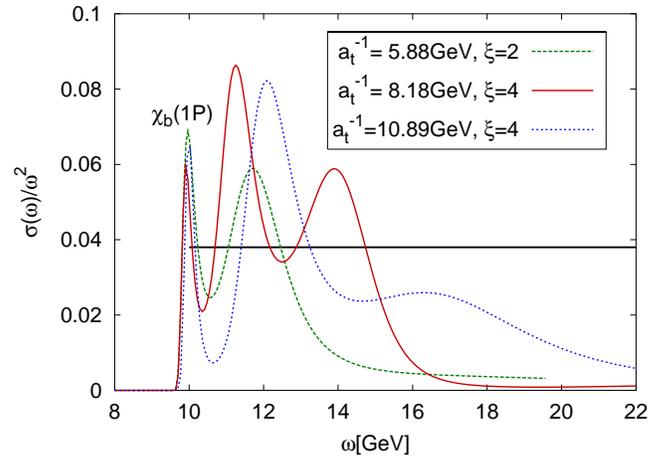}
\caption{The scalar bottomonium spectral function at zero temperature for
different lattice spacings.}
\label{spf_sc_bot}
\end{figure}

The spectral function in the scalar channel is shown in Fig.  \ref{spf_sc_bot}.
As it was the case for charmonium the correlators in this channel are more noisy
than in the pseudo-scalar channel and as a result it is more difficult to reconstruct
the spectral function. Nevertheless we are able to reconstruct the $\chi_{b0}$ state
which is the first peak in the spectral function. The peak position and the amplitude
are in reasonable agreement with the result of the simple exponential fit.

\subsection{Bottomonium at finite temperature}
\begin{figure}
\centering
\includegraphics[width=0.7\columnwidth]{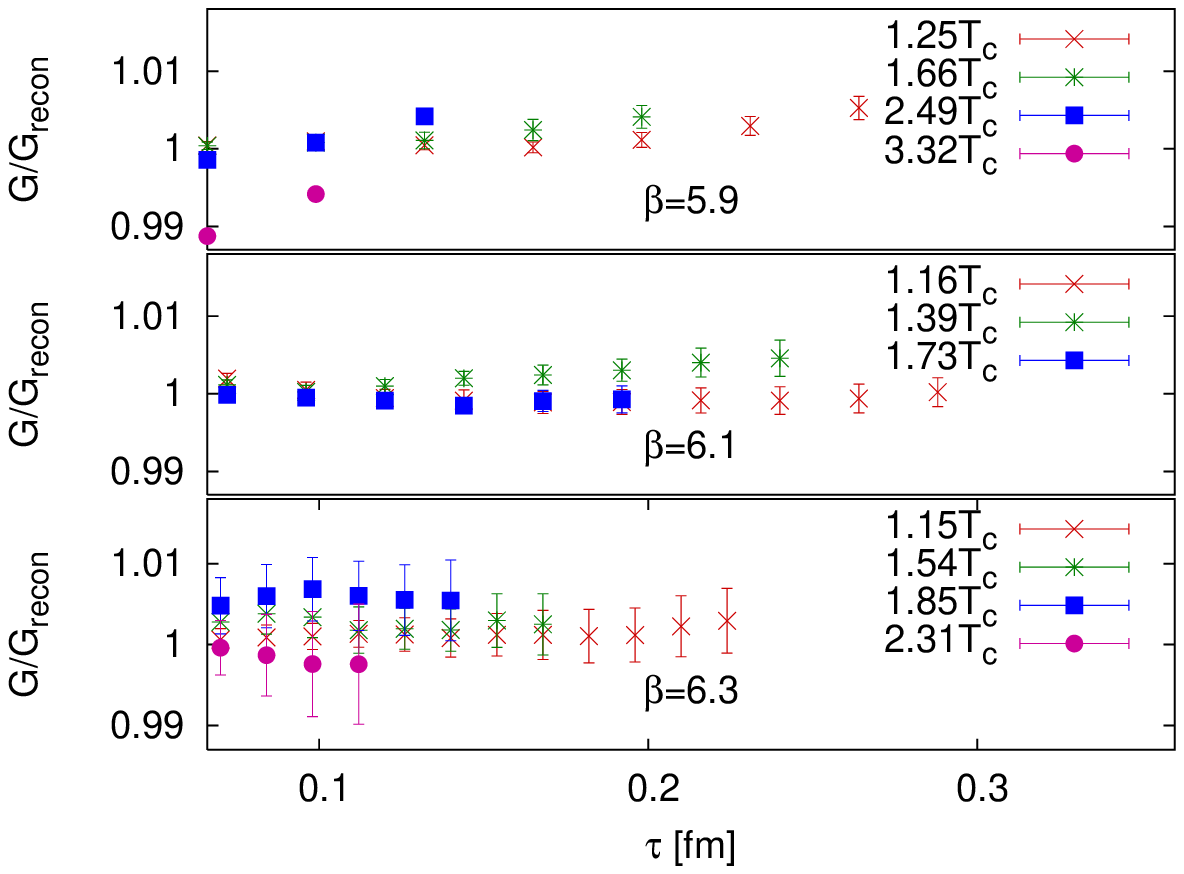}
\includegraphics[width=0.7\columnwidth]{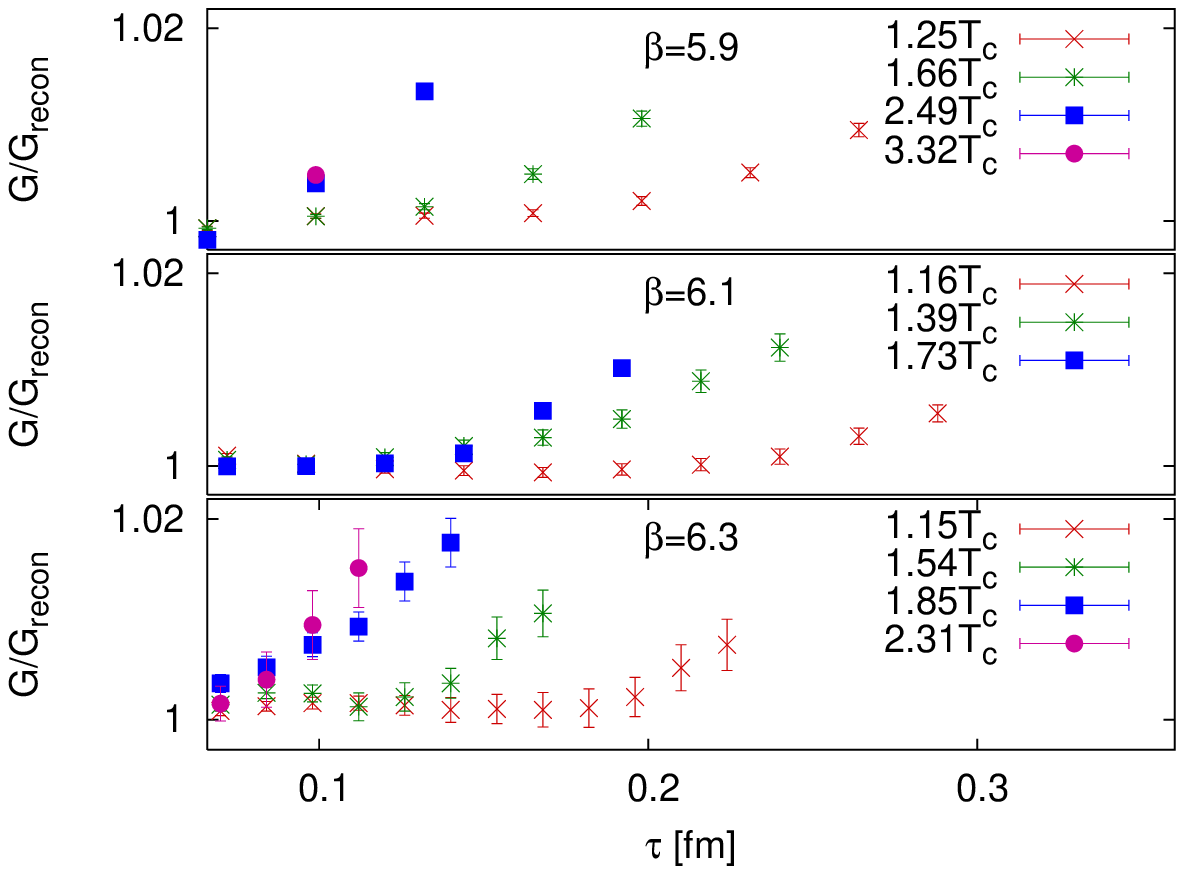}
\caption{The ratio  $G/G_{recon}$ in the pseudo-scalar (top) and vector channels at different lattice
spacings.}
\label{bot_1s}
\end{figure}

Besides calculating the bottomonium spectral function at 
zero temperature Ref. \cite{Jakovac:2006sf} presents a study of  the temperature dependence of bottomonium
correlators to see medium modification of bottomonia properties.
In Fig. \ref{bot_1s} we show $G/G_{recon }$ for vector and pseudo-scalar channel at different lattice spacings. 
This ratio appears to be temperature independent and very close to unity up to quite high temperatures.
This is consistent with the expectation that 1S bottomonia are smaller than 1S charmonia and thus are less
effected by the medium. They could survive till significantly higher temperatures.
Compared to charmonium case the  difference between the pseudo-scalar and vector channels
is smaller. This
is also expected as the transport contribution which is responsible for this difference is proportional to $\sim \exp(-m_{c,b}/T)$,
and thus is much smaller for bottom quarks (see the discussion in the next section).
Similar temperature dependence of the pseudo-scalar bottomonium correlator has been found in calculations with
isotropic clover action \cite{Datta:2006ua}.

\begin{figure}
\centering
\includegraphics[width=0.7\columnwidth]{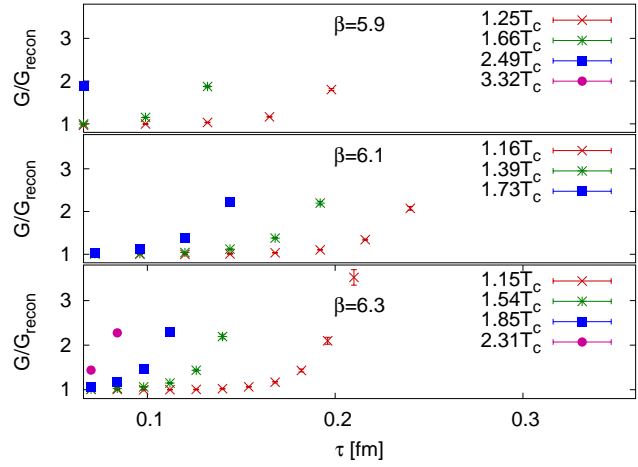}
\caption{The ratio  $G/G_{recon}$ in the scalar channel for different lattice spacings.}
\label{bot_1p}
\end{figure}

The temperature dependence of the scalar correlator is shown in Fig. \ref{bot_1p}. Contrary to the pseudo-scalar and vector 
correlators it shows strong temperature dependence and $G/G_{recon}$ is significantly larger than unity already at $1.1T_c$.
Again, similar enhancement in   $G/G_{recon}$ has been observed in isotropic lattice calculations \cite{Datta:2006ua}.

\begin{figure}
\centering
\includegraphics[width=0.7\columnwidth]{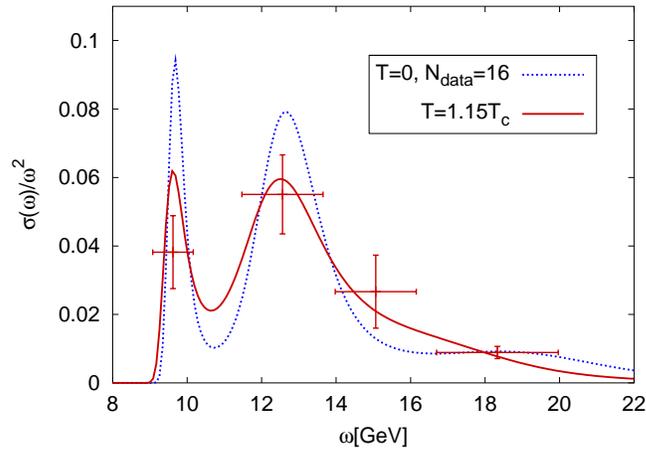}
\caption{The pseudo-scalar bottomonium spectral function at finite temperature.}
\label{spf_bot_ft}
\end{figure}

The spectral functions were reconstructed at finite temperature for $\beta=6.3$. For the pseudo-scalar channel
the results are shown in Fig. \ref{spf_bot_ft}. As in the charmonium case we compare the finite temperature spectral function
with the zero temperature spectral function obtained with the same number of data points and time interval.  As expected
the spectral function shows no temperature dependence within errors.  On the other hand it was not possible to reliably reconstruct the
scalar spectral function at finite temperature due to numerical problems.
Presumably much more statistics is needed to get some information about the scalar spectral function.

\subsection{
\label{sec_zero_modes}
Zero modes contribution
}
At finite temperature quarkonium spectral functions contain information about states containing a 
quark anti-quark pair as well as scattering of the external probe off a heavy quark from the medium. 
The later gives a contribution to the spectral function below the light cone
($\omega<k$). In the limit of zero momentum it becomes $\chi^i(T)\omega\delta(\omega)$ in the free theory. 
The generalized susceptibilities $\chi^i(T)$ were calculated in Ref. \cite{Aarts:2005hg} in the free theory. 
Interaction with the medium leads to the broadening of the delta function, which becomes a Lorentzian 
with a small width \cite{Petreczky:2005nh}. 
Because the quark anti-quark pair contributes to the spectral function at energies $\omega> 2m$ 
it is reasonable to separate quarkonium spectral function into two terms \cite{Petreczky:2008px}
\begin{equation}
\sigma^i(\omega,T)=\sigma^i_{\rm high}(\omega,T)+\sigma^i_{\rm low}(\omega,T).
\end{equation}
Here $\sigma^i_{\rm high}(\omega,T)$ is the high energy part of the spectral functions which is
non-zero only for $\omega>2 m$ and
describes the propagation of bound or unbound quark anti-quark pairs. On the other hand
$\sigma^i_{\rm low}(\omega,T)$
receives the dominant contribution
at $\omega \simeq 0$.
Because the width of the peak at $\omega \simeq 0$ is small the later gives
an almost constant contribution to the Euclidean correlator, which is called the zero mode contribution.
We can write an analogous decomposition for the Euclidean correlator
\begin{equation}
G^i(\tau,T)=G^i_{\rm high}(\tau,T) + G^i_{\rm low}(\tau,T).
\end{equation}
To a very good approximation $G^i_{\rm low}(\tau,T)=\chi^i(T) T$, i.e. constant.

In the previous sections we used the ratio $G(\tau,T)/G_{recon}(\tau,T)$ to study the temperature dependence of the correlators.
This dependence comes separately from the high energy part and low energy part, which gives the zero mode contribution. 
The zero mode contribution is absent in the derivative
of the correlator with respect to $\tau$. Therefore one can study the
temperature dependence of the correlators induced by change of bound state properties
and/or its dissolution by considering the ratio of the derivatives of the correlators
$G'(\tau,T)/G_{recon}'(\tau,T)$.
\begin{figure}                                                       
\includegraphics[width=0.48\textwidth]{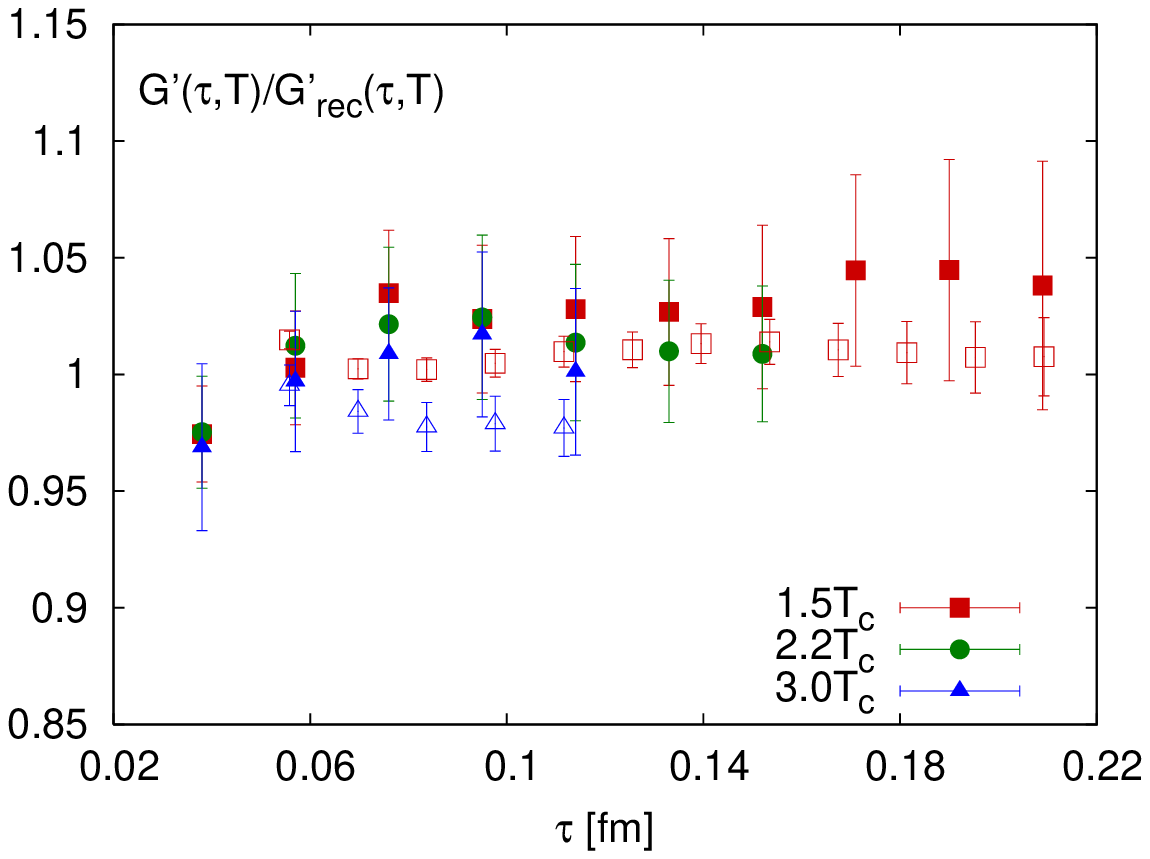}\hfill
\includegraphics[width=0.48\textwidth]{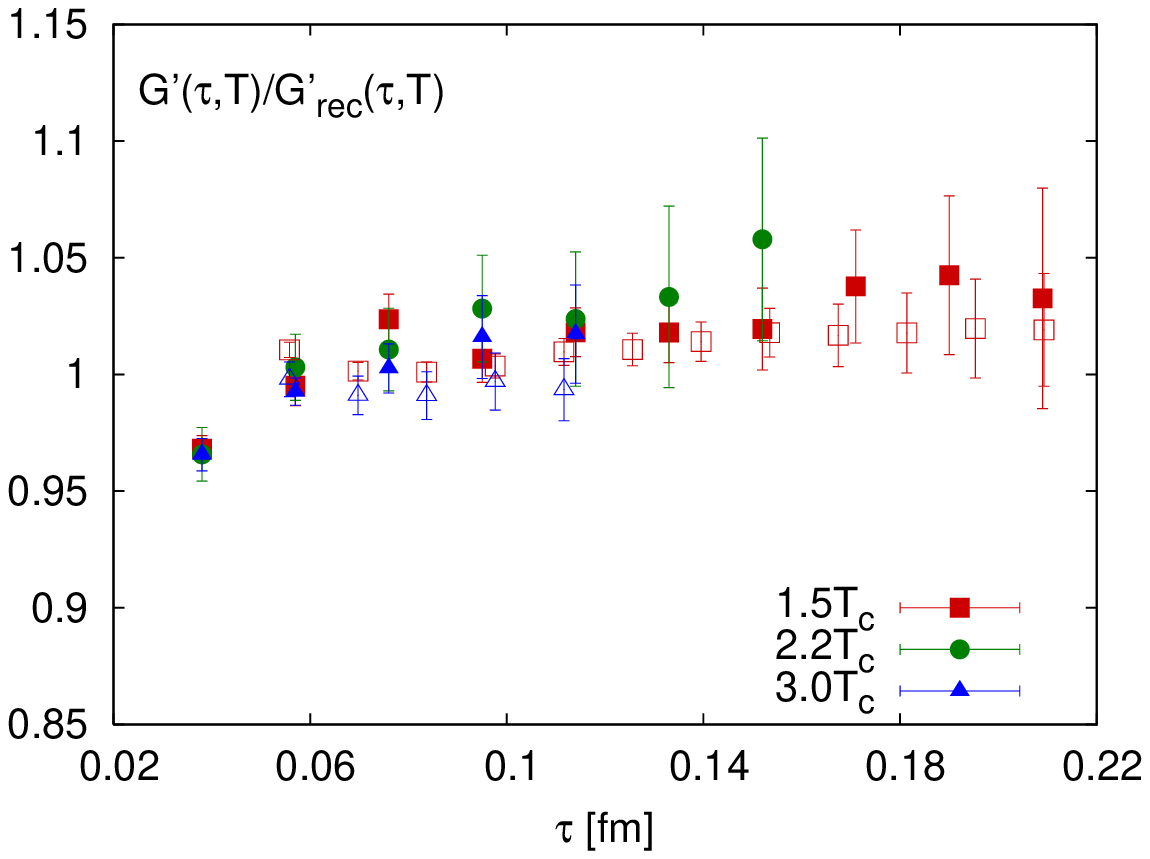}
\caption{The ratio of the derivatives $G'(\tau,T)/G'_{recon}(\tau,T)$ in the scalar
channel (left) and axial-vector channel (right) calculated at $\beta=7.192$.
The results from anisotropic lattice calculations at $\beta=6.5$ \cite{Jakovac:2006sf}
are also shown (open symbols).}
\label{fig:ratP}
\end{figure}

The temperature dependence of scalar  (\ref{scxi4}) and axial-vector correlators (\ref{axxi4}) has been presented in previous sections
in terms of $G/G_{recon}$.
It is temperature independent in the confined phase and shows
large enhancement in the deconfined phase. This large enhancement
is present both in charmonium and bottomonium correlators. To eliminate the zero mode contribution
the derivative of the correlators  and the corresponding ratio $G'(\tau,T)/G'_{recon}(\tau,T)$ have been calculated in \cite{Petreczky:2008px}
using isotropic lattices.
The numerical results for this ratio at $\beta=7.192$ are shown in Fig. \ref{fig:ratP}.
The results from anisotropic lattices \cite{Jakovac:2006sf} are also shown. There is good agreement between
the results obtained from isotropic and anisotropic lattices.
As one can see
from Fig. \ref{fig:ratP}  $G'(\tau,T)/G'_{recon}(\tau,T)$ shows very little temperature dependence and is close
to unity. This means that almost the entire temperature dependence of the scalar and axial-vector
correlators is due to zero mode contribution and $G_{\rm high}(\tau,T)$ is temperature independent.
The temperature dependence of the S-states seen in previous sections
is also greatly reduced and the agreement between isotropic and anisotropic calculations is
better for the ratio of derivatives.
For the pseudo-scalar channel this behavior can be explained by the presence of a small negative zero mode \cite{Petreczky:2008px}. 
\begin{figure}
\centering
\includegraphics[width=0.48\columnwidth]{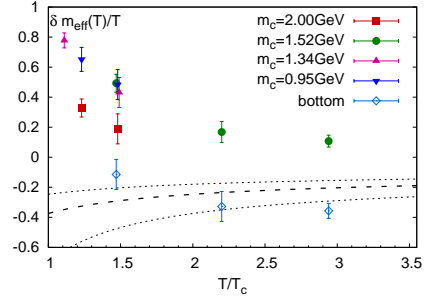}
\caption{
The thermal mass
correction  as function of the temperature.
The dashed line and the band correspond to the
perturbative prediction of the thermal mass correction.
}
\label{fig:meff}
\end{figure}
\begin{figure}
\includegraphics[width=0.48 \columnwidth]{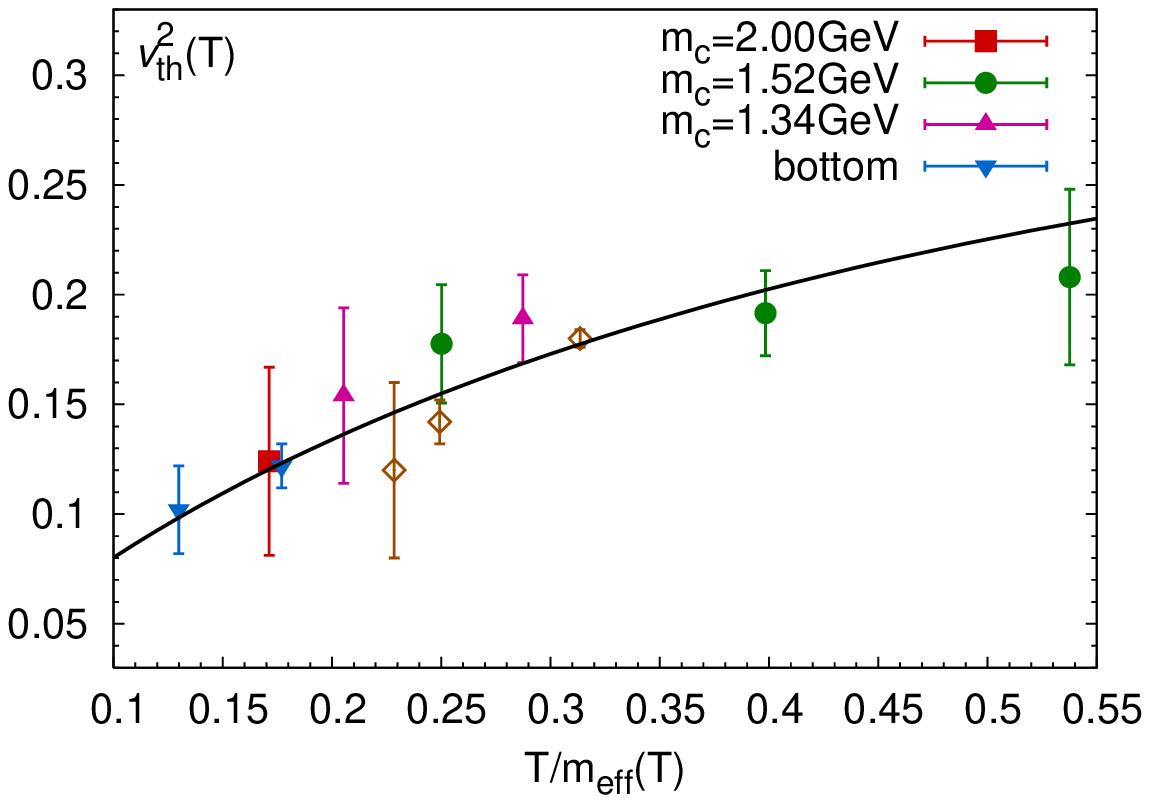}\hfill
\includegraphics[width=0.48 \columnwidth]{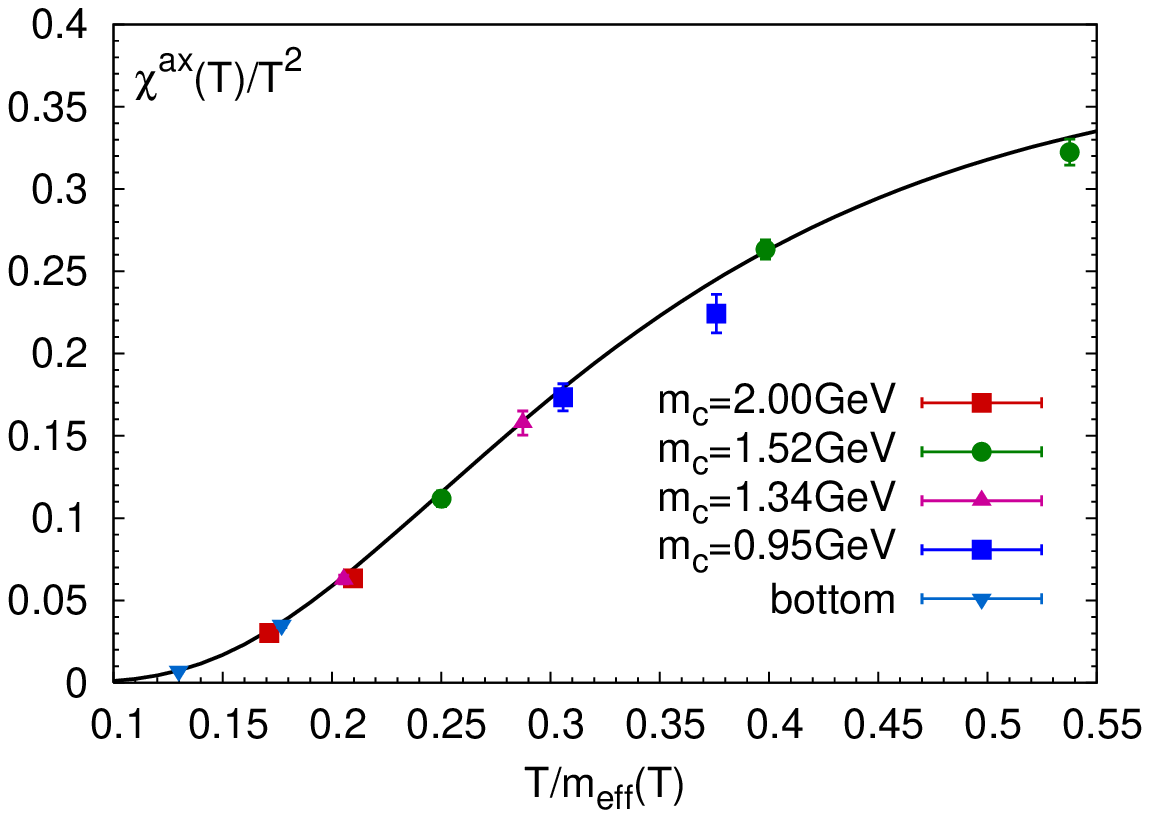}
\caption{
The thermal velocity of heavy quarks (left) and the zero mode
contribution to the axial-vector correlator $G^{ax}_{\rm low}/T^3=\chi^{ax}/T^2$ (right). The lines show
the prediction of the quasi-particle model with $m_{\rm eff}(T)$.The open symbols show the thermal velocity squared estimated on anisotropic lattice
s \cite{Jakovac:2006sf}.
}
\label{fig:zeroaxvc}
\end{figure}

Since the high energy part of the quarkonium
correlators turns out to be temperature independent to very good approximation
one can assume that  $G^i_{\rm high}(\tau,T) \simeq G^i_{recon}(\tau, T)$.
Then the low energy part of the correlators, i.e. the zero mode contribution can
be evaluated as $G^i_{\rm low}(\tau,T) = G^i(\tau,T)-G^i_{recon}(\tau,T)$ \cite{Petreczky:2008px}. 
Current lattice data are not precise enough to see a clear $\tau$ dependence 
in $G_{\rm low}^i(\tau,T)$ and it is compatible with being a constant. 
Therefore, one could take the value of low energy part of the correlator at the midpoint as an estimate for
the zero mode contribution, $G_{\rm low}^i(\tau=1/(2T),T)=T \chi^i(T)$.

The zero mode contribution is related to the propagation
of single (unbound) heavy quark in the medium. Therefore, it is natural to 
describe its temperature dependence in terms of a  quasi-particle model with effective temperature
dependent heavy quark masses.
Since the temporal component of the vector correlator has no high energy part, 
i.e. $G^{vc0}(\tau,T)=-T\chi^{vc0}(T)=-T \chi(T)$ it is
most suitable for fixing the effective quark mass. 
Matching the lattice data to the free theory expression for $\chi(T)$ one can determine the effective quasiparticle masses 
$m_{\rm eff}$ \cite{Petreczky:2008px}. 
The results of this analysis are shown in Fig. \ref{fig:meff} for different values of the constituent heavy quark mass,
including the bottom quarks, in terms of thermal mass correction $\delta m_{\rm eff}(T)=m_{\rm eff}(T)-m$. 
As one can see from the figure the thermal mass correction decreases monotonically with increasing the constituent quark mass
and increasing temperature for bottom quarks it is not incompatible with the leading-order perturbative
prediction 
\begin{equation}
\delta m_{\rm eff}=-\frac{4}{3} \frac{g^2(T)}{4 \pi} m_D,
\end{equation}
with $m_D=g(T) T$ being the perturbative Debye mass ($N_f=0$ because we work in the quenched approximation).

Having determined the effective heavy quark mass we can study the zero mode contribution in other channels.
If the quasi-particle model is correct the zero mode contribution should be a function of $m_{\rm eff}/T$ only.
Therefore in Fig. \ref{fig:zeroaxvc} the temperature dependence of zero mode contribution for the vector and axial-vector channel 
is shown as function of $T/m_{\rm eff}$. Indeed, all the lattice data seem to fall on one curve within errors, which agrees with the 
quasi-particle model prediction shown as the black lines. For the vector channel we show the data in terms
of the ratio $G_{\rm low}^{vc}(T)/G^{vc0}$, where $G_{\rm low}^{vc}(T)$ is the sum over all spatial components.
This is because this quantity does not depend on the renormalization \cite{Petreczky:2008px} and has simple physical interpretation
in terms of averaged thermal velocity squared $v_{\rm th}^2$. The later follows from the fact that due to the large quark mass the
Boltzmann approximation can be used and we have  
\begin{equation}
\frac{G_{\rm low}^{vc}(T)}{G^{vc0}(T)} \simeq
\left(\int d^3p \frac{p^2}{E_p^2} e^{-E_p/T}\right)/\left( \int d^3 p e^{-E_p/T} \right)=v_{\rm th}^2.
\end{equation}
The zero mode contribution in the vector channel has also
been studied on anisotropic lattices \cite{Jakovac:2006sf} and in Fig. \ref{fig:zeroaxvc} we also show the corresponding
results for thermal velocity squared.

\section{Potential models at finite temperature}
\label{sec_pot_model}

Quarkonium properties at finite temperature have been studied in potential models 
since the famous paper of Matsui and Satz \cite{Matsui:1986dk} (for recent review
see \cite{Mocsy:2008eg}). The basic idea behind this is that color screening will
modify the potential, which becomes short range and cannot support bound states
of heavy quarks at sufficiently high temperatures. As discussed in section 
\ref{sec_pnrqcd} this
approach can be justified if there is a separation of the scales related to binding energy
and other scales in the problem, like the inverse size of the bound states and the temperature.
Close to the QCD transition this separation of scales is not obvious, however, the lattice calculation of
static quark correlators, discussed in section \ref{sec_static_corrs} 
show that screening effects are very strong
already in the transition region. In fact, correlation functions of static quark anti-quark pair
show significant temperature modifications already at distances similar to quarkonium size.
Therefore, we may expect the most of quarkonium bound states dissolve in the deconfined phase
at temperatures close to the transition temperature. This seemingly contradicts to the 
small temperature dependence of quarkonium correlators and spectral functions discussed
in the previous section. 

\begin{figure}[htb]
\centering
\includegraphics[width=0.7\columnwidth]{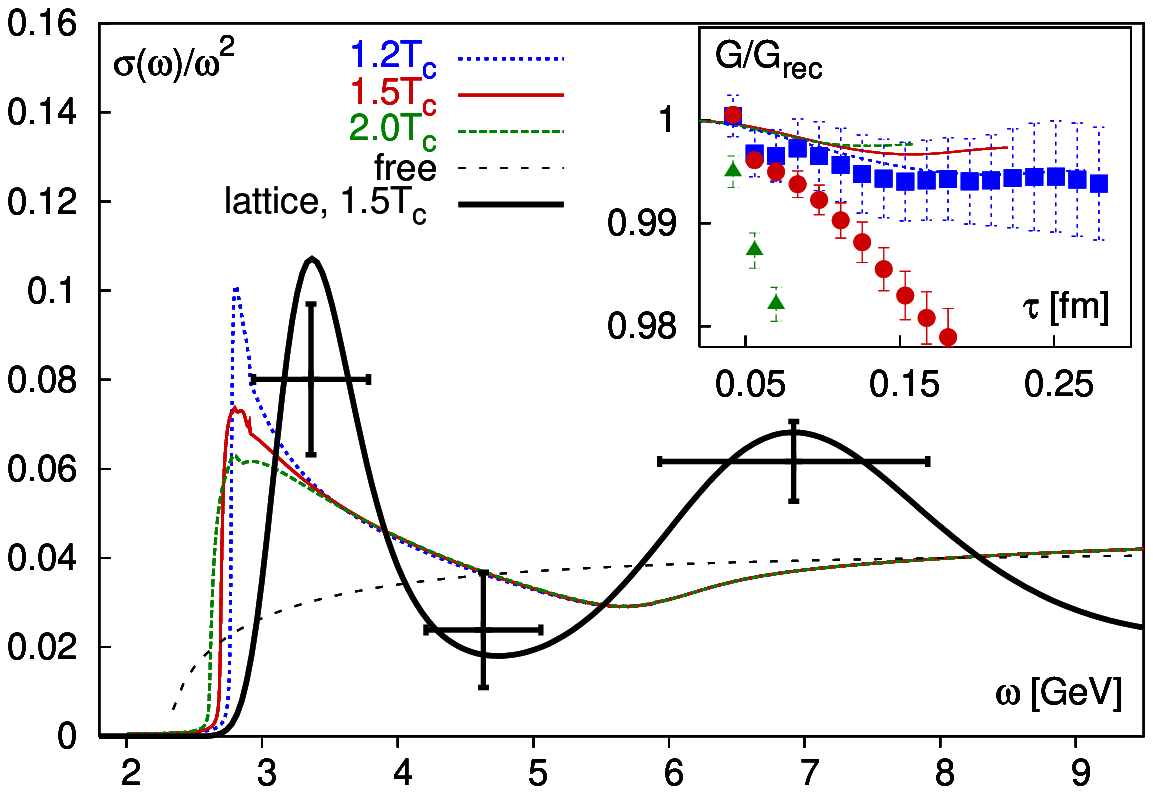}
\includegraphics[width=0.7\columnwidth]{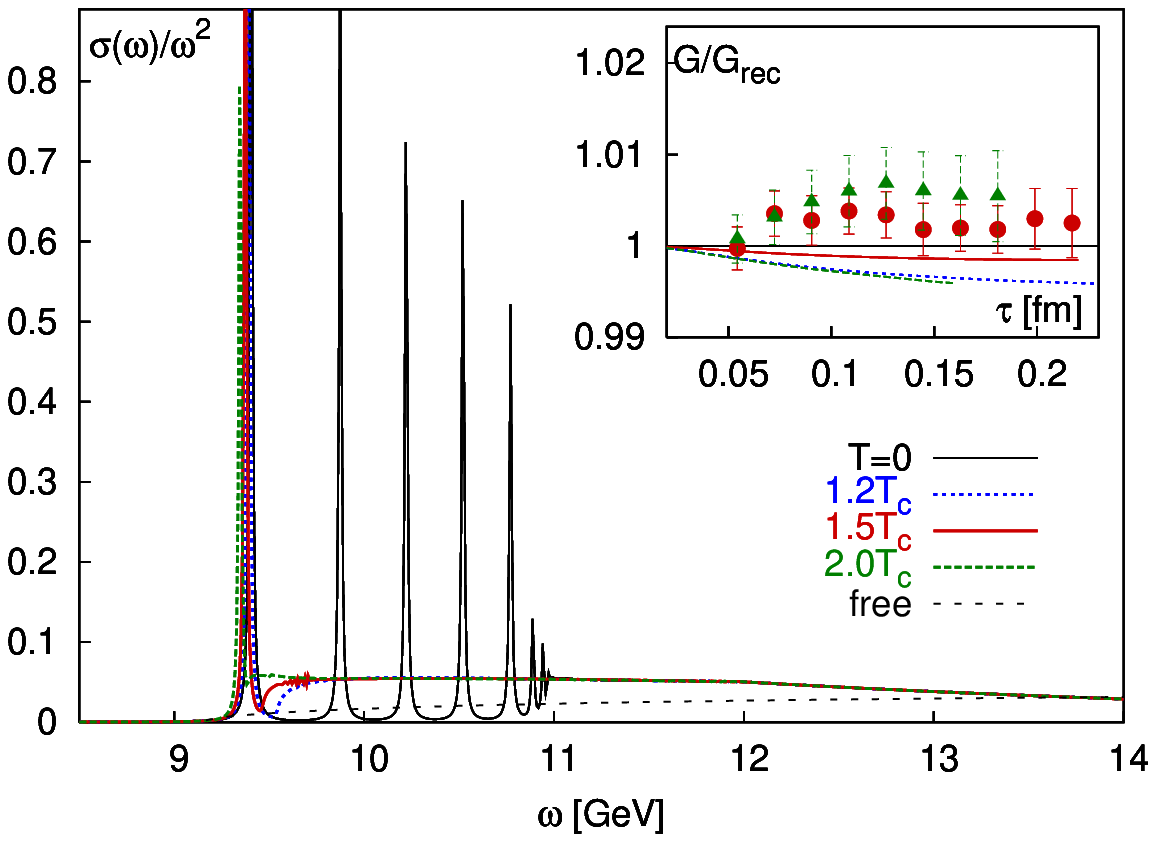}
\caption{The charmonium (top) and bottomonium (bottom) spectral
  functions at different temperatures.
  For charmonium we also show the spectral functions
  from lattice QCD obtained from the MEM at $1.5T_c$. The error-bars
  on the lattice spectral function correspond to the statistical error
  of the spectral function integrated in the $\omega$-interval
  corresponding to the horizontal error-bars.  The insets show the
  corresponding ratio $G/G_{recon}$ together with the results from
  anisotropic lattice calculations \cite{Jakovac:2006sf}.  For charmonium,
  lattice calculations of $G/G_{recon}$ are shown for $T=1.2T_c$
  (squares), $1.5T_c$ (circles), and $2.0T_c$ (triangles). For
  bottomonium lattice data are shown for $T=1.5T_c$ (circles) and
  $1.8T_c$ (triangles).  }
\label{fig:pot_spf}
\end{figure}

Due to the heavy quark mass quarkonium spectral functions can be calculated in the
potential approach by relating the spectral functions in the threshold region to
the non-relativistic Green function \cite{Mocsy:2007yj,Burnier:2007qm,Laine:2007gj}
\begin{eqnarray}
&
\displaystyle
\sigma(\omega)=K \frac{6}{\pi} {\rm Im} G^{nr}(\vec{r},\vec{r'},E)|_{\vec{r}=\vec{
r'}=0}\, ,\\[2mm]
&
\displaystyle
\sigma(\omega)=K \frac{6}{\pi}\frac{1}{m_c^2} {\rm Im} \vec{\nabla}
\cdot \vec{\nabla'} G^{nr}(\vec{r},\vec{r'},E)|_{\vec{r}=\vec{r'}=0}\, ,
\label{green_sc}
\end{eqnarray}
for $S$-wave, and $P$-wave quarkonia, respectively. Here $E=\omega-2 m$.
The pre-factor $K$ accounts for relativistic and radiative corrections.
The non-relativistic Green function is calculated from the Schr\"odinger
equation with a delta-function on the right hand side. Away from the threshold
the spectral function is matched to the perturbative result. It has been
shown that this approach can provide a fair description of quarkonium
correlators at zero temperature \cite{Mocsy:2007py}.
Other approaches to calculate quarkonium spectral functions in potential
models were proposed in Refs. 
\cite{Mocsy:2004bv,Mocsy:2005qw,Alberico:2007rg,Cabrera:2006wh,Ding:2009se}.
Close to the transition temperature the perturbative calculations of the potential
are not applicable. Therefore in Ref. \cite{Mocsy:2007yj} the singlet free energy
calculated in quenched lattice QCD has been used to construct the potential. The
drawback of this approach is that there is no one to one correspondence between
the singlet free energy and the potential in the non-perturbative domain.
Therefore in Ref. \cite{Mocsy:2007yj} several possible forms of the potential
compatible with the lattice data have been considered. Furthermore, lattice
calculations do not have much information about the imaginary part of the potential.
Therefore the imaginary part of the potential has been approximated by a small constant
term. The results of the calculations of the charmonium and bottomonium spectral
functions for $S$-wave are shown in Fig.\ref{fig:pot_spf}.
In the case of charmonium all bound states are melted at temperatures higher than $1.2T_c$.
We see, however a significant threshold enhancement, i.e. near the threshold the spectral function is
much larger than in the free case.
In the bottomonium case only the ground state survives in the deconfined phase. At temperatures
above $2T_c$ we see the melting of the ground state as well. 

Quarkonium spectral functions 
calculated using the perturbative potential, where the imaginary part is fully taken into account,
show similar qualitative features \cite{Burnier:2007qm,Laine:2007gj}. 
Here too a significant threshold enhancement is seen. The imaginary part of the potential plays
an important role in weakening the bound state peak or transforming it to mere threshold enhancement.
Let us note that the next-to-leading order
perturbative correction to the quarkonium spectral functions also give rise to significant enhancement in
the threshold region \cite{Burnier:2008ia}.

From the spectral functions we can calculate the quarkonium correlation functions in Euclidean time $G(\tau,T)$
and compare them to the available lattice data. This comparison is shown in Fig. \ref{fig:pot_spf}
for the ratio $G(\tau,T)/G_{recon}(\tau,T)$.
As one can see from the figure the melting of bound states does not lead to large change in the Euclidean
correlation functions. The ratio $G(\tau,T)/G_{recon}(\tau,T)$ calculated in the potential model 
is flat and temperature independent in agreement with lattice calculations.
This means that threshold enhancement can compensate for melting
of bound states in terms of the Euclidean correlators. Spectral functions of $P$-wave quarkonium have
been also calculated \cite{Mocsy:2007yj,Burnier:2007qm} and show significant threshold enhancement as
well. As the consequence the ratio of the derivatives  $G'(\tau,T)/G'_{recon}(\tau,T)$ is temperature
independent and close to unity \cite{Mocsy:2007yj} in agreement with lattice calculations shown in the
previous section.

The analysis discussed above has been done in quenched QCD. This is because only in quenched QCD
we have sufficiently precise lattice calculations of quarkonium correlators. 
Potential model calculations of the spectral functions have been extended to 2+1 flavor QCD using the lattice
data discussed in section \ref{sec_static_corrs}. These calculations show that 
all quarkonium states except the ground state
bottomonium dissolve in the quark gluon plasma \cite{Mocsy:2007jz}. The upper limits on the dissociation temperatures
for different quarkonium states obtained in this analysis are given in Table \ref{tab:diss}.

\begin{table}[hftb]
\renewcommand{\arraystretch}{0.81}
\begin{center}
\begin{minipage}{8.5cm} \tabcolsep 5pt
\begin{tabular}{ccccccc}
\hline
 state&$\chi_c$&$\psi'$&$J/\psi$&$\Upsilon'$&$\chi_b$&$\Upsilon$\\ 
\hline
 $T_{dis}$&$\le T_c$& $\le T_c$ &$1.2T_c$&$1.2T_c$&$1.3T_c$&$2T_c$\\
\hline
\end{tabular}
\end{minipage}
\caption{Upper bound on the dissociation temperatures for different quarkonium states in 2+1 flavor QCD \cite{Mocsy:2007jz}. }
\end{center}
\label{tab:diss}
\end{table}

\section{Conclusion}
\label{sec_concl}

In this paper we discussed quarkonium properties in quark gluon plasma.
We discussed how color screening can be studied non-perturbatively on the
lattice using spatial correlation functions of static quark and anti-quark.
We discussed the importance  of singlet and adjoint (triplet for $SU(2)$ or
octet for $SU(3)$) degrees of freedom for 
understanding the temperature dependence of static meson (quark-antiquark) correlators. 
It has been shown how color singlet and adjoint degrees of freedom can be defined in the effective 
field theory framework, the so-called thermal pNRQCD.

We have seen that the singlet correlators of static quark and 
anti-quark show strong screening effects at distances comparable to quarkonium
size. Thus it is natural to expect that most quarkonium states melt in
quark gluon plasma. Quite surprisingly lattice calculations of quarkonium
correlators in Euclidean time show very little temperature dependence.
We reviewed the current status of these calculations in quenched QCD. 
It turns out that the spatial lattice spacing should be smaller than
$(4{\rm GeV})^{-1}$ to have full control over the discretization errors. 
Therefore the extension of these calculations to full QCD pioneered in 
Ref. \cite{Aarts:2007pk} will be quite difficult.

The only source of significant temperature dependence of quarkonium correlators 
is the zero mode contribution which is not related to bound states but to transport properties 
of heavy quarks in the medium \cite{Petreczky:2005nh}. We have shown
that this contribution can be well described by a quasi-particle model
with temperature dependent heavy quark mass. 
We have presented the calculations of quarkonium spectral functions
in potential model and have shown that all quarkonium states, except
the ground state bottomonium melt in the deconfined phase. We have
also shown how the seemingly existing contradiction of strong color
screening leading to quarkonium melting and very weak temperature dependence 
of quarkonium correlators can be resolved within potential models.
It turns out that strong threshold enhancement of quarkonium spectral functions
can compensate the absence of bound state and result in Euclidean correlarors,
which are almost temperature independent.

The fact that no charmonium bound states can exist in quark gluon plasma
has important consequences for describing charmonium production in heavy
ion collisions. In absence of bound states heavy quarks in the plasma
can be treated quasi-classically using Langevin dynamics. In this scenario
the residual correlation between the heavy quark and anti-quark, visible in
the threshold enhancement and the
finite life time of the plasma play an important role \cite{Young:2008he}. 
It turns out 
that using these ideas and the interactions of heavy quark determined in
lattice QCD it is possible to uderstand the $J/\psi$ suppression pattern 
at RHIC \cite{Young:2008he}.

\section*{Acknowledgements}
This work was supported by U.S. Department of Energy under
Contract No. DE-AC02-98CH10886.
The work of A.B. was supported by grants DOE DE-FC02-06ER-41439 and NSF 0555397.
A.V. work was supported by the Joint Theory Institute funded together by
Argonne National Laboratory and the University of Chicago,
and in part by the U.S. Department of Energy,
Division of High Energy Physics and Office of Nuclear Physics, under Contract DE-AC02-06CH11357.

\bibliographystyle{ws-rv-van-mod}
\bibliography{qgp4_bpv}

\end{document}